%% file: CKM-GUT.tex
\documentclass[11pt,a4paper]{article}
\pdfoutput=1
\usepackage{amssymb,amsmath,amsfonts,mathrsfs,enumerate,wrapfig}
\usepackage[utf8]{inputenc}

\usepackage{amsmath}
\usepackage{slashed}
\usepackage{jheppub}
\usepackage{latexsym}
\usepackage{multirow}
\usepackage[usenames,dvipsnames,table]{xcolor}
\usepackage{graphicx}
\usepackage{epsfig}  
\usepackage{epsf}    
\usepackage{dcolumn}
\usepackage{bm}
\usepackage{dcolumn}
\usepackage{textcomp}
\usepackage{float}
\usepackage{subfig}
\usepackage{hypcap}
\usepackage[]{hyperref}
\usepackage{makecell}
\usepackage{color}
\usepackage{pifont}
\usepackage{appendix}
\usepackage{url}
\usepackage{cancel}
\allowdisplaybreaks[1]
\usepackage{import}
\usepackage{subfiles}
\usepackage{filecontents}
\usepackage{threeparttable, tablefootnote}
\usepackage{slashed}
\usepackage{lineno}
\usepackage{placeins}

\let\Oldsection\section
\renewcommand{\section}{\FloatBarrier\Oldsection}

\let\Oldsubsection\subsection
\renewcommand{\subsection}{\FloatBarrier\Oldsubsection}

\let\Oldsubsubsection\subsubsection
\renewcommand{\subsubsection}{\FloatBarrier\Oldsubsubsection}

\DeclareMathAlphabet      {\mathbfit}{OML}{cmm}{b}{it}

\usepackage[normalem]{ulem}
\definecolor{darkgreen}{cmyk}{1,0,1,0.4}

\definecolor{pink}{cmyk}{0.4,1,0.3,0}

\def\com2#1{\textcolor{red}{\it{#1}}}
\hypersetup{
  bookmarks=true,         
  unicode=false,          
  pdftoolbar=true,        
 pdfmenubar=true,        
 pdffitwindow=true,     
 pdfstartview={FitH},    
 pdfsubject={GUT and Topological defects},   
 pdfnewwindow=true,      
 pdfcreator={RevTeX},
 colorlinks=true,       
 linkcolor=red,          
 citecolor=blue,        
 filecolor=black,      
 urlcolor=blue,           
  }

\makeatletter
\renewcommand{\fnum@table}{\textbf{\tablename~\thetable}}
\renewcommand{\fnum@figure}{\textbf{\figurename~\thefigure}}
\makeatother
\title
{Unification, Proton Decay and Topological Defects in non-SUSY GUTs with Thresholds 
}
\author[a]{Joydeep Chakrabortty,}  
\author[b]{Stephen F. King,}
\author[a]{Rinku Maji}

  \affiliation[a]{Department of Physics, Indian Institute of Technology, Kanpur-208016, India} 
   \affiliation[b]{School of Physics and Astronomy, University of Southampton,
   	SO17 1BJ Southampton, United Kingdom}

\emailAdd{joydeep@iitk.ac.in}
\emailAdd{S.F.King@soton.ac.uk}
\emailAdd{mrinku@iitk.ac.in}

\abstract
{We calculate the proton lifetime and discuss topological defects in a wide class of 
	non-supersymmetric (non-SUSY) $SO(10)$ and $E(6)$ Grand Unified Theories (GUTs), broken
	via left-right subgroups with one or two intermediate scales (a total of 9 different scenarios with and without D-parity), including the important effect of threshold corrections.
	By performing a goodness of fit test for unification using the two-loop renormalisation group evolution equations (RGEs),
	we find that the inclusion of threshold corrections significantly affects the proton lifetime, allowing several scenarios,
	which would otherwise be excluded, to survive. Indeed we find that the threshold corrections are a saviour for many non-SUSY GUTs. For each scenario we analyse the homotopy of the vacuum manifold to estimate the possible emergence of topological defects. }

\begin{document}
\maketitle
\flushbottom
\newpage
\section{Introduction}

Grand Unified Theories (GUTs) are theoretical frameworks which aim to unify the fundamental forces described by strong, weak, and electromagnetic interactions correspond to the Standard Model (SM) of particle physics described by $SU(3)_C\otimes SU(2)_L\otimes U(1)_Y \equiv \mathcal{G}_{3_C2_L1_Y}$ gauge theory.  These unified theories are associated with a simple unified gauge group $\mathcal{G}_{U}$ and a single gauge coupling $g_{U}$ at some high energy scale $M_{U}$. However in minimal $SU(5)$, without supersymmetry (SUSY), gauge coupling unification is 
not readily achievable. Nevertheless, non-SUSY GUTs such as $SO(10)$ or $E(6)$ with one or two intermediate scales
remain viable in principle.
However, aside from the requirement of coupling unification at $M_{U}$, the main prediction of most GUTs
is that of proton decay. But 
proton decay is yet to be observed \cite{Miura:2016krn, PhysRevLett.113.121802, PhysRevD.90.072005, Takhistov:2016eqm}, and the proton decay lifetime ($\tau_p \geq 1.6 \times 10^{34}$) only serves to put a stringent constraint on the unification scale $M_{X} \geq 10^{16}$~GeV,
which threatens to exclude many of the non-SUSY GUTs. However, a detailed study of proton decay in such theories,
including the effect of threshold corrections, is required in order to address this question, and to make reliable predictions for the next generation of proton decay experiments such as Hyperkamiokande \cite{Yokoyama:2017mnt} and DUNE \cite{Acciarri:2015uup}.

In this paper, we estimate the proton lifetime in a wide class of 
non-supersymmetric GUTs, broken
via left-right subgroups with one or two intermediate scales
For the one intermediate scale breaking, we suppose that the GUT groups break into their maximal subgroups of the form $SU(N)_L\otimes SU(N)_R \otimes \mathcal{G}$, see~\cite{Chakrabortty:2017mgi}. This restricts our choice of GUT groups to be $SO(10)$, $E(6)$, with certain breaking patterns. Due to the $SU(N)_L\otimes SU(N)_R$ structure, we encounter two possibilities -- D-parity conserved and broken \cite{Mohapatra:1974gc, Senjanovic:1975rk, Senjanovic:1978ev, Chang:1983fu, Chang:1984uy}. We consider a total of 9 different scenarios with and without D-parity.
For each such breaking pattern, we compute the beta-functions up to two-loop level and find the unification solutions in terms of unification and intermediate scales.
By performing a goodness of fit test for unification using the two-loop renormalisation group evolution equations (RGEs),
we find that the inclusion of threshold corrections significantly affects the proton lifetime, allowing several scenarios,
which would otherwise be excluded, to survive. For each scenario, we also analyse the homotopy of the vacuum manifold to estimate the possible emergence of topological defects. We then go on to consider a general analysis of the two intermediate scale cases.
To understand the status of the one intermediate scale case, we have recalled our earlier work \cite{Chakrabortty:2017mgi} and computed the same for those breaking chain as well. This gives us a clear notion to understand the present status of one and two intermediate GUT scenarios. 
The various breaking patterns we assume 
are achieved through the suitable choice of the scalar representations and the orientations of their vacuum expectation values (VEVs) \cite{Fritzsch:1974nn,  Lazarides:1980nt,  Clark:1982ai, Aulakh:1982sw, Hewett:1985ss, Deshpande:1992au,  Amaldi:1991cn, Hung:2006kd, Howl:2007zi,  Chakrabortty:2008zk, Bertolini:2009es, Chakrabortty:2010az, DeRomeri:2011ie, Arbelaez:2013nga,Chakrabortty:2010xq}. Also, 
the different breaking patterns lead to different phenomenological models at low energy,
as discussed in \cite{Lazarides:1980nt, Chakrabortty:2010az, DeRomeri:2011ie, Patra:2015bga, Bandyopadhyay:2015fka, Babu:2016cri, Bandyopadhyay:2017uwc} for $SO(10)$  and   \cite{Gursey:1975ki,Achiman:1978vg,Hewett:1985ss, Hung:2006kd, Howl:2007zi, Chakrabortty:2013voa, Miller:2014jza, Chakrabortty:2015ika, Gogoladze:2014cha,Younkin:2012ui,Calmet:2011ic,Wang:2011eh,Biswas:2010yp,Atkins:2010re,King:2017cwv} for $E(6)$.  The neutrino and charged fermion mass and mixing generation in the context of unified theories are discussed in \cite{Mohapatra:1979ia, Bajc:2002iw, Goh:2003hf, King:2003rf, Goh:2003sy, Mohapatra:2003tw, Bertolini:2004eq, Dev:2009aw, Joshipura:2009tg, Chakrabortty:2010az, Patel:2010hr, Blanchet:2010kw, BhupalDev:2011gi, Joshipura:2011nn, BhupalDev:2012nm, Meloni:2014rga, Babu:2016bmy, Meloni:2016rnt, FileviezPerez:2018dyf}. In \cite{Babu:2015bna, Nagata:2015dma, Garcia-Cely:2015quu, Brennan:2015psa, Boucenna:2015sdg, Mambrini:2015vna, Parida:2016hln, Nagata:2016knk, Sahoo:2017cqg, Arbelaez:2017ptu, Ernst:2018bib, Ahmed:2018jlv, Parida:2018apw, Garg:2018trf, Ferrari:2018rey} different cosmological aspects and dark matter scenarios are discussed. 
An important result of the present paper, using 
the goodness of fit test for unification with two-loop renormalisation group evolution equations (RGEs),
is the extent to which the inclusion of threshold corrections significantly affects the proton lifetime, allowing several scenarios,
which would otherwise be excluded, to survive.

The layout of the remainder of the paper is as follows. Section~\ref{prelim} is a preliminary section in which we discuss the important aspects of the unified scenarios which are used repeatedly in our analysis, e.g., (i) renormalisation group evolutions of the gauge couplings, (ii) matching conditions, and threshold corrections, and (iii)  emergence of topological defects -- at different stages of symmetry breaking. In section~\ref{proton}, we focus on the computation of proton decay  lifetime, including a detailed discussion of the following topics: (i) dimension-6 proton decay operators, (ii) anomalous dimension matrix to perform the RG of the related Wilson coefficients, and (iii) prediction of proton decay lifetime.  In section~\ref{patterns}, we analyse the breaking of  GUT symmetry groups (in our case $SO(10)$, and $E(6)$) to the Standard Model gauge group via two intermediate scales.  We have considered only those breaking chains where the first intermediate group is of the form of $SU(N)_L\otimes SU(N)_R \otimes \mathcal{G}$. We also analyse the topological structure of the vacuum manifold for each such scenario, and note the emergence of topological defects in the subsequent process of symmetry breaking.  
In section~\ref{results}, we present our results using a goodness of fit test in order to find unification solutions which are compatible with low energy data.
We compute the proton decay lifetime predicted for each two intermediate breaking chain along with the unification solutions in the presence and absence of threshold corrections. We also discuss the impact of threshold corrections in detail.  
Section~\ref{conclusions} summarises and concludes the paper.
In a series of Appendices, we provide all the details related to the threshold corrections and group theoretic informations used in this paper.


\section{Preliminaries}
\label{prelim}
\subsection{RGEs of gauge couplings}
\label{subsec:RGEs}

The renormalisation group evolutions (RGEs) of the gauge couplings can be written in terms of the group-theoretic invariants as suggested in 
\cite{Caswell:1974gg, Jones:1974mm, Jones:1981we, Slansky:1981yr, Machacek:1983tz, Machacek:1983fi, Machacek:1984zw}. 
The gauge coupling $\beta$-functions for a product group, like $\mathcal{G}_i\otimes \mathcal{G}_j\otimes \mathcal{G}_k..$ upto two-loop can be recast as :
\begin{align}
\mu \frac{dg_i}{d\mu}&= \frac{g_i^3}{(4\pi)^2} \left[ \frac{4\kappa}{3}T(F_i)D(F_j)
+\frac{1}{3}T(S_i)D(S_j) - \frac{11}{3} C_2(G_i) \right] + \frac{1}{(4\pi)^4} g_i^5  \nonumber \\
& \times \left[  \left(\frac{10}{3} C_2(G_i)+2C_2(F_i)\right) T(F_i)D(F_j)
+ \left(\frac{2}{3} C_2(G_i)+4C_2(S_i)\right) T(S_i)D(S_j)  \right. \nonumber  \\
& \left. - \frac{34}{3} (C_2(G_i))^2 \right] 
+ \frac{1}{(4\pi)^4} g_i^3 g_j^2  \left[ 2C_2(F_j)T(F_i)D(F_j)
+ 4 C_2(S_j)T(S_i)D(S_j) \right],
\end{align}
following the conventions of \cite{Jones:1981we} where $F_i$, and $S_i$ are the representations  under group $\mathcal{G}_i$ for the scalar and fermion fields respectively.  Here, $T(R)$, $D(R)$, and $C_2(R)$ the normalisation of generators, dimensionality of representation and the quadratic Casimir for the representation $R$.

%

\subsection{Matching conditions and Threshold corrections}
\label{subsec:matching}

In the process of symmetry breaking we encounter different possibilities: (i) a single group is broken to a product group, (ii) a product group is broken to a single group, (iii) a product group is broken to a product group. Now for every such scenario, we need to encapsulate  the redistributions of the gauge couplings correspond to the broken and unbroken gauge groups. This has been done through the suitable choice of matching conditions which depends on the pattern of symmetry breaking  \cite{Hall:1980kf, WEINBERG198051, Chang:1984qr, Binger:2003by, Bertolini:2009qj}.
At this point one needs to recall that there exist some heavy modes at different scales, and they need not to be always degenerate. So their presence may affect the matching conditions as well in the form of threshold corrections.  In the absence of these threshold corrections, the detailed matching conditions for different scenarios are discussed in \cite{Chakrabortty:2017mgi}.  These conditions get modified in the presence of threshold corrections \cite{Langacker:1992rq, PhysRevD.47.R4830, PhysRevD.47.264, PhysRevD.49.3711, Parida:1995td, Li:2009fq, Bertolini:2012im, Bertolini:2013vta, Babu:2015bna, Schwichtenberg:2018cka}.

%

In this section, we have estimated the impact  of different heavy degrees of freedom on the unification in the form of threshold corrections. Till now we have assumed that all the superheavy particles that do not contribute to the renormalization group evolution of the gauge couplings are degenerate with  the symmetry breaking.

At any symmetry breaking scale, $\mu$, the gauge couplings ($1/\alpha_d$) of the daughter gauge group ($\mathcal{G}_d$) are given by the suitable linear combinations of the gauge couplings ($1/\alpha_p$) of the parent one ($\mathcal{G}_p$) along with the threshold corrections after integrating out the superheavy fields. The gauge coupling matching condition reads as
\begin{align}\label{matching}
\frac{1}{\alpha_d(\mu)}-\frac{C_2(\mathcal{G}_d)}{12\pi}=\left(\frac{1}{\alpha_p(\mu)} -\frac{C_2(\mathcal{G}_p)}{12\pi}\right) - \frac{\Lambda_d(\mu)}{12\pi} , 
\end{align}
where, 
\begin{align}\label{lamda}
\Lambda_d(\mu)=-21\; {\mathrm{Tr}} (t_{dV}^2 \ln\frac{M_V}{\mu})+2\;\eta\; {\mathrm{Tr}} (t_{dS}^2 \ln\frac{M_S}{\mu}) + 8\;\kappa\; {\mathrm{Tr}} (t_{dF}^2 \ln\frac{M_F}{\mu}),
\end{align}
is the measure of one-loop threshold correction \cite{WEINBERG198051, Hall:1980kf,Bertolini:2009qj,Bertolini:2013vta}. Here, $t_{dV}$, $t_{dS}$ and $t_{dF}$ are the generators for the representations under $\mathcal{G}_d$ of the superheavy vector, scalar and fermion fields respectively, $M_V$, $M_S$ and $M_F$ are their respective masses. In  Eqn.~\ref{lamda}, $\eta = \frac{1}{2}(1)$ for the real(complex) scalar fields and, $\kappa = \frac{1}{2}(1)$ for Weyl(Dirac) fermions. Here, all the scalars are the physical scalars. 

\vskip 0.2cm
To analyse the impact of the threshold correction we have adopted  a conservative approach where all the superheavy gauge bosons ($M_{X,Y}$) are degenerate with the symmetry breaking scale ($\mu$). The scalars $(S)$ and fermions $(F)$ are assumed to be nondegenerate and the mass ratio ${M_i}/{M_{X,Y}}$ ($i\equiv S, F$)  {\it w.r.t.} those gauge bosons are varied within $[1/2:2]$, and $[1/10:10]$.  We have first computed the total threshold corrections at the unification scale $M_X$ and intermediate scale(s) $M_I$ in terms of $\Lambda_d$.  This can be expressed as a linear combination of $\ln\frac{M_{j}}{\mu}$, see Eqn.~\ref{lamda},  with positive and negative coefficients. To maximize the unification scale $M_X$ or intermediate scale(s) $M_I$, we need to assign the maximum(minimum) value to the terms containing  coefficients with +ve(-ve) sign. We have designed our methodology to capture the impact of the threshold correction based on the following scenarios :
\begin{itemize}
	\item[I.] All the superheavy degrees of freedom have the same mass as the breaking scale. In this case we only have the contribution from the gauge bosons which is incorporated within the matching condition with the $C_2$ (quadratic Casimir) of Eqn.~\ref{matching}.
	
	\item[II.] All the superheavy multiplets have different masses within the given range of mass. We can maximize the threshold corrections at  $M_X$, $M_{I,II}$ scales adopting the methodology stated earlier. In this paper,  we have noted the maximum possible value of the partial proton decay lifetime $\tau_p$ varying  the ratio ($R$)  within following ranges: $[1/10:10]$ and $[1/2:2]$.
\end{itemize}
Then using the solutions of  two-loop RGEs and goodness of fit test, we have computed the proton decay lifetime for all the breaking patterns considered in this analysis.



\subsection{Topological defects associated with spontaneous symmetry breaking}
\label{subsec:TD}

In a spontaneously broken gauge theories within the unified framework it is important to analyse the topological structure of the vacuum manifold \cite{Lazarides:1980cc, Lazarides:1981fv, Kibble:1982ae, Pijushpani-Hill, Davis:1994py, Davis:1995bx, Jeannerot:2003qv}. In these cases, one can certainly predict the emergence of topological defects just by studying the homotopy of the vacuum manifold. In a mathematical framework this can be stated as: say a group $\mathcal{G}$ is broken spontaneously to another group $\mathcal{H}$, then the vacuum manifold is identified as $\mathcal{M}=\mathcal{G}/\mathcal{H}$. Now one needs to check whether $\Pi_{k}[\mathcal{M}]\neq \mathcal{I}$, i.e., non-trivial or not. If this is non-trivial then there will be some topological defects determined by the index $k$, e.g.,  domain walls ($k=0$), cosmic strings ($k=1$), monopoles ($k=2$), and textures ($k=3$).
Even this allows to understand which of them are stable ones. The topological defects that we are interested in are domain walls, cosmic strings, monopoles \cite{Lazarides:1980va, Kibble:1982dd, Weinberg:1983bf, Vachaspati:1997rr} as textures are very unstable and decays immediately. 

For product group we can use the following identities: (i) $\Pi_k(\mathcal{G}_i \otimes \mathcal{G}_j)=\Pi_k(\mathcal{G}_i) \otimes \Pi_k(\mathcal{G}_j)$, and (ii) $\Pi_k(\mathcal{G} / (\mathcal{G}_i \otimes \mathcal{G}_j))=\Pi_{k-1} (\mathcal{G}_i \otimes \mathcal{G}_j)$ while $\Pi_{k}(\mathcal{G})=\Pi_{k-1}(\mathcal{G})=\mathcal{I}$, $\Pi_0(\mathbb{Z}_N)=\mathbb{Z}_N$.

\begin{table}[H]
	\small
	\renewcommand*{\arraystretch}{1.2}
	\begin{center}
		\begin{tabular}{| c | c | c | c | c |  }
			\hline
			Lie   &  zeroth Homotopy  &  Fundamental &  2nd homotopy  &  3rd homotopy     \\
			Group       &  $(\Pi_0)$  &  group $(\Pi_1)$ &   group $(\Pi_2)$ &   group $(\Pi_3)$    \\
			\hline
			$U(1)$  & $\mathcal{I}$ & $\mathbb{Z}$ & $\mathcal{I}$  & $\mathcal{I}$    \\
			\hline
			$SU(2)$  & $\mathcal{I}$ & $\mathcal{I}$ & $\mathcal{I}$  & $\mathbb{Z}$    \\
			\hline
			$SU(3)$  & $\mathcal{I}$ & $\mathcal{I}$ & $\mathcal{I}$  & $\mathbb{Z}$    \\
			\hline
			$SU(4)$ & $\mathcal{I}$ & $\mathcal{I}$  & $\mathcal{I}$ & $\mathbb{Z}$   \\
			\hline
			
			$Spin(10)$ & $\mathcal{I}$ & $\mathcal{I}$  & $\mathcal{I}$ & $\mathbb{Z}$   \\
			\hline
			$E(6)$ & $\mathcal{I}$ & $\mathcal{I}$  & $\mathcal{I}$ & $\mathbb{Z}$   \\
			\hline
		\end{tabular}
		\caption{Homotopy classification of Lie Groups.}
		\label{tab:homotopy}
	\end{center}
\end{table}


\section{Computation of the proton lifetime}
\label{proton}
Proton decay is the smoking gun signal to confirm the existence of grand unification. In the non-supersymmetric GUT scenario, the proton can decay dominantly through the exchange of lepto-quark gauge bosons which induce lepton and baryon number violation simultaneously. These lepto-quark gauge bosons gain mass through the spontaneous symmetry breaking of the GUT symmetry; thus their mass is determined by the unification scale ($M_X$). Again these exotic gauge bosons need to be very heavy  to be consistent with non-observation of the proton decay so far. This justifies why the GUT scale is very close to the Planck scale. At low energy ($< M_X$), the proton decay diagrams can be featured in terms of  effective  dimension-6 operators after integrating out the  gauge bosons.

Our plan of calculations is following: First we will construct the dimension-6 proton decay operators using the Standard Model fermions along with their respective Wilson coefficients. 
Then we will perform RG running of the effective operators  till the unification scale using the relevant anomalous dimensions. Here, we have discussed and provided the detail structure of these 
anomalous dimensions for different breaking patterns.

\subsection{Dimension-6 Proton decay operators}
The lepto-quark heavy gauge bosons that mediate the proton decay, transform under the SM gauge group $( SU(2)_L\otimes U(1)_Y \otimes SU(3)_C )$ as: $(X,Y) = (2,5/6,3)$ and $(X^\prime , Y^\prime ) = (2,-1/6,3)$ respectively \cite{Goldhaber:1980dn, Langacker:1980js}. In this work we have considered the limits for $p\to e^+\pi^0$, as this channel provides the stringent constraint $\tau_p \geq 1.6 \times 10^{34}$ yrs.

The effective Lagrangian that emerges after integrating out the heavy lepto-quark gauge bosons, contains the $(B-L)$ conserving dimension-six proton decay operators, are given as \cite{Weinberg:1979sa, Wilczek:1979hc, Weinberg:1980bf, Abbott:1980zj,Lucha:1984tv, Nath:2006ut}:
\begin{align}\label{operators_at_SM}
\mathcal{O}_{I}^{d=6}  = \Omega_1^2 \epsilon^{ijk} \epsilon^{ab} \overline{u^C_{i\alpha}}\gamma^\mu Q_{j a\alpha} \overline{e^C_\beta} \gamma_\mu Q_{k b\beta} \ ,\;\; &
\mathcal{O}_{II}^{d=6}  = \Omega_1^2 \epsilon^{ijk} \epsilon^{ab} \overline{u^C_{i\alpha}}\gamma^\mu Q_{j a \alpha} \overline{d^C_{k\beta}} \gamma_\mu L_{ b\beta} \ ,  \\
\mathcal{O}_{III}^{d=6}  = \Omega_2^2 \epsilon^{ijk} \epsilon^{ab} \overline{d^C_{i\alpha}}\gamma^\mu Q_{jb\alpha} \overline{u^C_{k\beta}} \gamma_\mu L_{a \beta} \ ,\;\; &
\mathcal{O}_{IV}^{d=6}  = \Omega_2^2 \epsilon^{ijk} \epsilon^{ab} \overline{d^C_{i\alpha}}\gamma^\mu Q_{jb \alpha} \overline{\nu^C_{\beta}} \gamma_\mu Q_{k a \beta} \ .
\end{align}
These operators are written in flavour basis. Here, $\Omega_{1,2}$ are the Wilson coefficients associated with these dimension-6 operators. In the next section their structures and necessary running using anomalous dimension matrices are discussed in detail.

The SM fermions are: $Q= \begin{pmatrix}
u \\ d
\end{pmatrix}$ and  $L= \begin{pmatrix}
\nu \\ e
\end{pmatrix}$, where $i,j,k$ are the $SU(3)_C$; $a,b$ are the $SU(2)_L$; and $\alpha,\beta$ are the generation indices for light quarks.

In the physical basis, the relevant  effective terms in the Lagrangian leading to $p\to e^+\pi^0$ decay are expressed as \cite{FileviezPerez:2004hn}: 
\begin{align}\label{operator_physical_basis}
\mathcal{O}_L^{d=6}\left( e^C, d\right) & = \mathcal{WC}_1\;  \epsilon^{ijk} \overline{u^C_i}\gamma^\mu u_j \overline{e^C} \gamma_\mu d_{k} \ , \nonumber \\
\mathcal{O}_R^{d=6}\left( e, d^C\right) & = \mathcal{WC}_2 \; \epsilon^{ijk} \overline{u^C_i}\gamma^\mu u_j \overline{d^C_{k}} \gamma_\mu e \ , 
\end{align}
with their respective  the Wilson coefficients ($\mathcal{WC}_{1,2}$) :
\begin{align}
\mathcal{WC}_1 & = \Omega_1^2\; \left[ 1 + |V_{ud}|^2\right] \ , \nonumber \\
\mathcal{WC}_2 & = \Omega_1^2 + \Omega_2^2\; |V_{ud}|^2 \ , 
\end{align}
where, $|V_{ud}| = 0.9742$ is the CKM matrix element \cite{Tanabashi:2018oca}. In our analysis, we have assumed other mixing matrices to be identity.

\subsection{Computation of the anomalous dimensions and RG of dimension-6 operators}
The running of the dimension-6 proton decay operators is considered into two steps: (i) RG evolution from mass scale of proton ($m_p\sim 1$ GeV) to $M_Z$ which is taken care of by the long distant enhancement factor $A_L$ \cite{Nihei:1994tx}, and (ii) RG evolution of the same operator from $M_Z$ to unification scale $M_X$ through the intermediate scales, if any.  The impact of second level running is captured in short range renormalisation factors $A_S$ which can be written in the presence of multiple intermediate scales as \cite{Buras:1977yy,Goldman:1980ah,Caswell:1982fx,DANIEL1983219,Ibanez:1984ni,MUNOZ198655}: 
\begin{align} \label{short_range_re_factor}
A_S = \prod_{j}^{M_Z\leq M_{j}\leq M_X} \prod_i \left[ \frac{\alpha_i \left(M_{j+1}\right)}{\alpha_i \left(M_{j}\right)} \right]^{\frac{\gamma_i}{b_i}} ,
\end{align}
where, $\alpha_i = g_i^2/4\pi $, $\gamma_i$'s are the anomalous dimensions, and $b_i$'s are the $\beta$-coefficients at different stages of the renormalisation group evolutions  from the scale $M_{j}$ to the next scale $M_{j+1}$. We have computed $\gamma_i$ for different symmetry breaking patterns and they are all summarised in Table~\ref{table_anomalous}. The one-loop $\beta$ coefficients $(b_i)$ are  given explicitly in the next section for every breaking chain.

\begin{table}[h!]
	\renewcommand*{\arraystretch}{1.2}
	\begin{center}
		\begin{tabular}{|c|c|c|}
			\hline
			\multirow{2}{*}{Gauge group} & \multicolumn{2}{c|}{Anomalous dimensions}\\
			\cline{2-3}
			& $\mathcal{O}^{d=6}_L\left(e^C , d\right)$ & $\mathcal{O}^{d=6}_R\left(e , d^C \right)$  \\ 
			\hline
			$\mathcal{G}_{2_L 1_Y 3_C}$& $\lbrace \frac{9}{4} , \frac{23}{20} , 2 \rbrace$ & $\lbrace \frac{9}{4} , \frac{11}{20} , 2 \rbrace$  \\
			\hline
			$\mathcal{G}_{2_L 2_R 3_C 1_{B-L}}$& $\lbrace \frac{9}{4} , \frac{9}{4} , 2 , \frac{1}{4} \rbrace$ & $\lbrace \frac{9}{4} , \frac{9}{4} , 2 , \frac{1}{4} \rbrace$ \\
			\hline
			$\mathcal{G}_{2_L 2_R 4_C}$& $\lbrace \frac{9}{4} , \frac{9}{4} , \frac{15}{4} \rbrace$ &  $\lbrace \frac{9}{4} , \frac{9}{4} , \frac{15}{4} \rbrace$ \\
			\hline
			$\mathcal{G}_{2_L 1_{R^\prime} 4_C}$& $\lbrace \frac{9}{4} , \frac{3}{4} , \frac{15}{4} \rbrace$ & $\lbrace \frac{9}{4} , \frac{3}{4} , \frac{15}{4} \rbrace$ \\
			
			\hline
			$\mathcal{G}_{3_L 3_R 3_C}$& $\lbrace 2 , 4 , 2 \rbrace$ & $\lbrace 4 , 2 , 2 \rbrace$ \\
			\hline
			\thead{{\large{$\mathcal{G}_{2_L 2_R 4_C 1_X}$}}\\ (flipped)}& $\lbrace \frac{9}{4} , 0, \frac{15}{8} , \frac{7}{8} \rbrace$ & $\lbrace \frac{9}{4} , 0, \frac{15}{4} , \frac{1}{2} \rbrace$ \\
			
			\hline
		\end{tabular}
		\caption{Relevant anomalous dimensions for the considered breaking chains.}
		\label{table_anomalous}
	\end{center}
	
\end{table}

\subsection{Decay width and lifetime computation for different proton decay channels}

Proton is expected to decay into mesons which are pseudo scalar mesons and leptons as follows: $p\to M + \bar{l}$, where $M$ can be $\pi^o$, $\pi^{+}$, $K^0$, $K^{+}, \eta$ and  $l$ can be $ e, \mu,\nu_{e,\mu,\tau}$ \cite{Machacek:1979tx}. The current experimental bounds on the partial proton decay lifetime suggested by the Super-Kamiokande Collaboration are $\tau(p\to\pi^0 e^+)> 1.6 \times 10^{34}$ years \cite{Miura:2016krn}, $\tau(p\to\pi^+ \bar{\nu})> 3.9 \times 10^{32}$ years \cite{PhysRevLett.113.121802} and $\tau(p\to K^+ \bar{\nu})> 5.9 \times 10^{33}$ years \cite{PhysRevD.90.072005}.

The partial decay width for such decay process can be written as:
\begin{align}
\Gamma (p\to M + \bar{l}) = \frac{m_p}{32\pi}\left[1-\left(\frac{m_{M}}{m_p}\right)^2\right]^2 A_L^2 \bigg|\sum_n A_{Sn} \mathcal{WC}_n \mathcal{F}_0^n(p\to M)\bigg|^2 .
\end{align}
Here, $m_p$, and $ m_{M}$ are the mass of the proton and Mesons respectively. $\mathcal{WC}_n$ are the Wilson coefficients of the operators that give rise to that particular decay channel of the proton ($p\to M + \bar{l}$), $A_{Sn}$'s are the short-range enhancement factors computed in the form of 
Eqn.~\ref{short_range_re_factor},  and $\mathcal{F}_0^n=\langle M \rvert\epsilon^{ijk}\;(q^{\mathrm{T}}_iCP_n q'_j) \;P_L q''_k\lvert p \rangle \equiv \langle M \rvert {(q q')}_n q''_L\lvert p \rangle$ are the form factors determined by  chiral perturbation theory (passively)\cite{CLAUDSON1982297,CHADHA1984374,PhysRevD.62.014506} and(or) directly using the lattice QCD results \cite{PhysRevD.75.014507,PhysRevD.89.014505,Aoki:2017puj}. Here, $q$, $q'$ and $q''$ are the light quark ($ u,  d,  s$) which are integral part of the dimension-6 proton decay operators. Here, $C$ is the charge conjugation operator, and $P_n$ $(n=L,R)$ is the chiral projection operator.

Now once every thing is taken care of, the lifetime or inverse of partial decay width computation for the ``golden" channel $p\to e^+ \pi^0 $ as \cite{FileviezPerez:2004hn}:
\begin{align}
\tau_p = & \Bigg[ \frac{m_p}{32\pi}\left(1-\frac{m_{\pi^0}^2}{m_p^2}\right)^2 A_L^2 \frac{g_U^4}{4 M_X^4}(1+|V_{ud}|^2)^2  \nonumber \\
& \times  \left( A_{SR}^2 |\langle \pi^0 \rvert (ud)_R u_L\lvert p \rangle |^2 + A_{SL}^2 |\langle \pi^0 \rvert (ud)_L u_L\lvert p \rangle |^2 \right) \Bigg]^{-1} ,
\end{align}
where, $A_{SL}$ and $A_{SR}$ are the short-range enhancement factors associated with the left-handed $\mathcal{O}^{d=6}_L\left( e^C, d\right)$ and right-handed $\mathcal{O}^{d=6}_R\left( d^C, e \right)$ operators respectively, see Table~\ref{table_anomalous}. In our calculation we have used the following values of the matrix elements \cite{Aoki:2017puj}: 
\begin{align*}
\langle \pi^0 \rvert (ud)_R u_L\lvert p \rangle = -0.131, \ \ \, \, \langle \pi^0 \rvert (ud)_L u_L\lvert p \rangle = 0.134 .
\end{align*}   


\section{Patterns of GUT breaking: RGEs, Matching and Topological defects}
\label{patterns}
\begin{figure}[h!]
	\centering
	\includegraphics[scale=0.6]{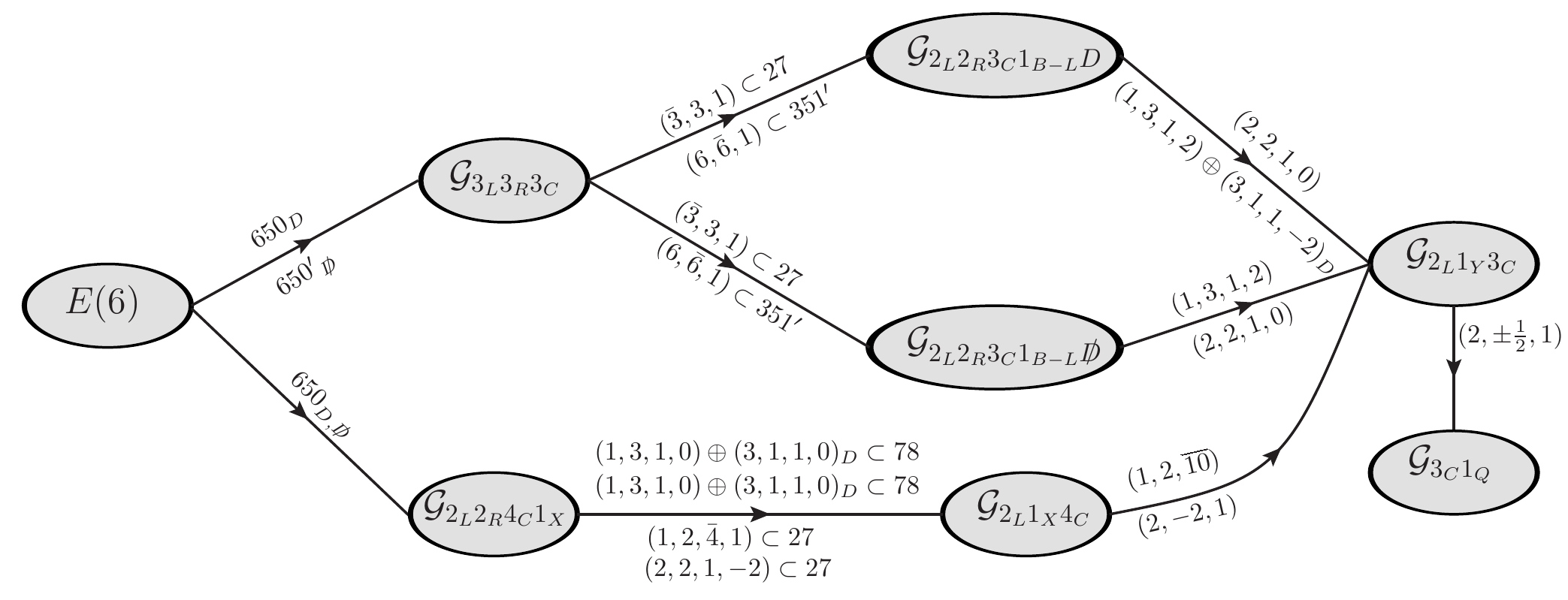}
	\caption{Two intermediate step breaking of $E(6)$ to the Standard Model.}\label{e6_sm_2step}
\end{figure}

\begin{figure}[h!]
\centering
\includegraphics[scale=0.6]{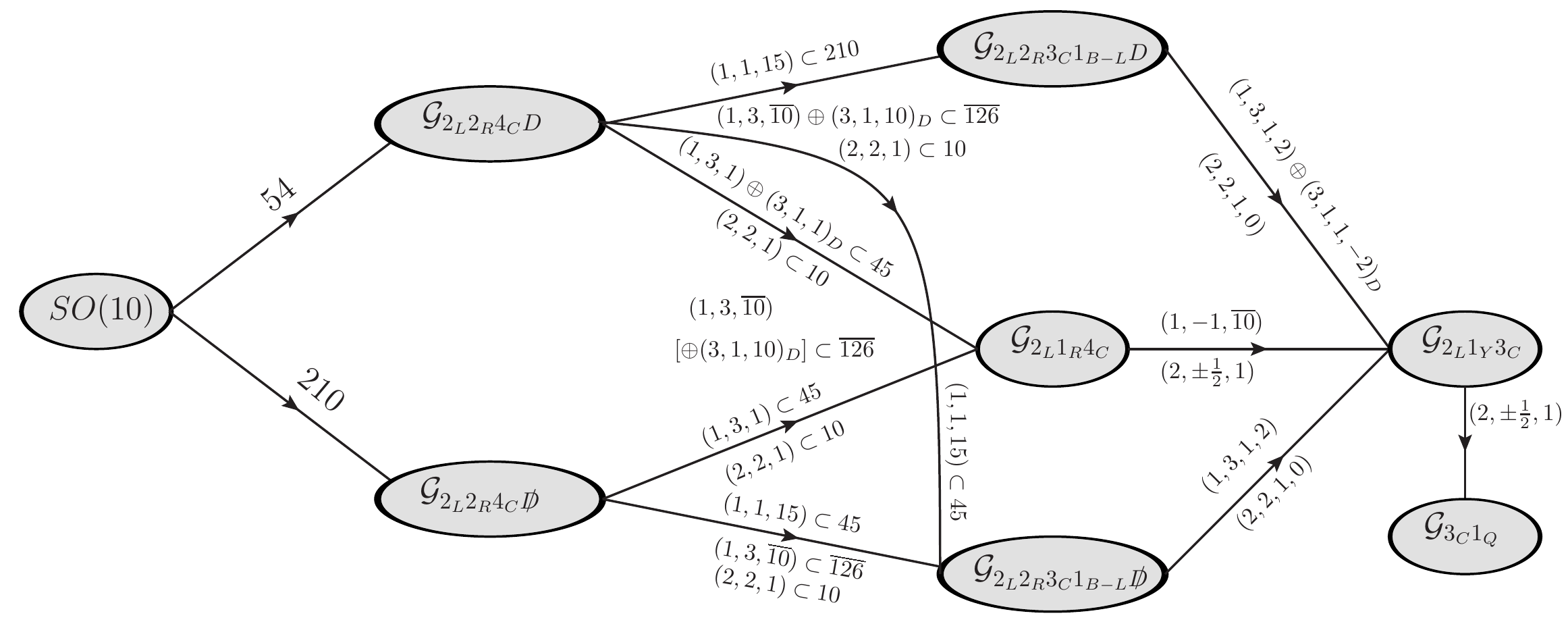}
\caption{Two intermediate step breaking of $SO(10)$ to the Standard Model.}\label{so10_sm_2step}
\end{figure}

In this section, we discuss the spontaneous breaking of $SO(10)$ and $E(6)$ GUT groups to the SM through two intermediate scales. 
 As mentioned in the earlier section,  we have chosen the first intermediate group starting from GUT is of the form $SU(N)_L\otimes SU(N)_R$. The list of such breaking patterns are encapsulated in Figs.~\ref{e6_sm_2step}, and~\ref{so10_sm_2step}. For each such breaking we have computed the $\beta$-coefficients for the gauge coupling running upto two-loop level. We have also discussed the emergence of possible topological defects at different stages of symmetry breaking. In these figures, we have also mentioned the suitable choices of the scalar representations in detail.
 To evaluate the RGEs we need to incorporate suitable matching conditions at each symmetry breaking scale. The matching conditions when all the heavy degrees of freedom are degenerate with the breaking scales are given below for each scenario. This is equivalent to the case of no threshold correction. To include the effects of threshold correction we need to modify these conditions accordingly given in Eqns.~\ref{matching}, and \ref{lamda}. The detail structures of the threshold corrections are specific to the breaking chain, and are given in the appendix.

\subsection*{I. $\mathbf{E(6)\xrightarrow{M_X} \mathcal{G}_{3_L 3_R 3_C D} \xrightarrow{M_{I}} \mathcal{G}_{2_L 2_R 3_C 1_{LR} D}\xrightarrow{M_{II}} \rm SM }$}

\vskip 0.5cm
\begin{center}
\underline{IA. $\beta$ coefficients}
\end{center}

\input{RGE/E6-G333-G2231D.tex}

\pagebreak

\begin{center}
	\underline{IB. Matching conditions}
\end{center}

At $M_{I}$ scale $SU(3)_{L,R}$ is broken to $SU(2)_{L,R}\otimes U(1)_{L,R}$, and at the same scale $U(1)_L\otimes U(1)_R$  is broken to $U(1)_{LR}$. Thus the matching conditions read as:
\begin{align}\label{matching_g333_g2231}
\frac{1}{\alpha_{2L}(M_{I})}-\frac{1}{6\pi} & = \frac{1}{\alpha_{3L}(M_{I})}-\frac{1}{4\pi} ,\nonumber \\
\frac{1}{\alpha_{2R}(M_{I})}-\frac{1}{6\pi} & = \frac{1}{\alpha_{3R}(M_{I})}-\frac{1}{4\pi} ,\nonumber \\
\frac{1}{\alpha_{1LR}(M_{I})} & = \frac{1}{2}\left(\frac{1}{\alpha_{3L}(M_{I})}-\frac{1}{4\pi}\right)+\frac{1}{2}\left(\frac{1}{\alpha_{3R}(M_{I})}-\frac{1}{4\pi}\right) ,
\end{align} 
as $\alpha_{3L}(M_{I})=\alpha_{1L}(M_{I})$, and  $\alpha_{3R}(M_{I})=\alpha_{1R}(M_{I})$.
We would like to mention that $\alpha_{3L}(M_{I})=\alpha_{3R}(M_{I})$ which is ensured by the unbroken D-parity.

\vskip 0.2cm
At $M_{II}$ scale $SU(2)_R\otimes U(1)_{LR}$ is broken to $U(1)_Y$, and the matching condition reads as:
\begin{align}\label{matching_g2231_sm}
\frac{1}{\alpha_{1Y}(M_{II})}=\frac{3}{5}\left(\frac{1}{\alpha_{2R}(M_{II})}-\frac{1}{6\pi}\right) + \frac{2}{5}\left(\frac{1}{\alpha_{1LR}(M_{II})}\right) . 
\end{align}
Here, $\alpha_{2L}(M_{II})=\alpha_{2R}(M_{II})$ as a signature of conserved D-parity.

\vskip 0.5cm

\begin{center}
\underline{IC. Topological defects}
\end{center}

\begin{itemize}
\item $E(6) \xrightarrow{M_{X}} \mathcal{G}_{3_L 3_R 3_CD}$:  Here, the non-trivial homotopy structure of the vacuum manifold is given by  $ \Pi_1(E(6)/\mathcal{G}_{3_L 3_R 3_CD})= \Pi_0(\mathcal{G}_{3_L 3_R 3_CD})= \Pi_0(D)=\mathbb{Z}_2$. Thus  $\mathbb{Z}_2$-strings are formed during this symmetry breaking.  It is important to note that $ \Pi_1(E(6)/\mathcal{G}_{2_L 2_R 3_C 1_{LR}D})=\mathbb{Z}_2$, which implies that the  strings are stable upto $M_{II}$.

\item $\mathcal{G}_{3_L 3_R 3_C} \xrightarrow{M_{I}} \mathcal{G}_{2_L 2_R 3_C 1_{LR}D}$: At this stage, stable cosmic strings are formed as we have  $ \Pi_2(\mathcal{G}_{3_L 3_R 3_C} / \mathcal{G}_{2_L 2_R 3_C 1_{LR}D})= \Pi_1(\mathcal{G}_{2_L 2_R 3_C 1_{LR}D})  =\mathbb{Z}$.
 The charge of these strings  changes  from $LR$ to $Y$ in the process of subsequent breaking to the SM.

\item $\mathcal{G}_{2_L 2_R 3_C 1_{B-L}D}\xrightarrow{M_{II}} \rm{SM} $: Here, $D$-parity is spontaneously broken leading to the formation of domain walls bounded by strings. No stable monopole and topological cosmic string are formed at this stage, though embedded strings will be generated. 
\end{itemize}

\subsection*{II. $\mathbf{E(6)\xrightarrow{M_X} \mathcal{G}_{3_L 3_R 3_C } \xrightarrow{M_{I}} \mathcal{G}_{2_L 2_R 3_C 1_{LR}\slashed{D}}\xrightarrow{M_{II}} \rm SM }$}

\vskip 0.5cm
\begin{center}
	\underline{IIA. $\beta$ coefficients}
\end{center}
\input{RGE/E6-G333-G2231.tex}

\vskip 0.5cm
\begin{center}
	\underline{IIB. Matching conditions}
\end{center}
In this case, the breaking chain is very similar to the earlier one.
Therefore, the matching conditions at $M_{I}$ and $M_{II}$ scales are same as in Eqns.~\ref{matching_g333_g2231}, and~\ref{matching_g2231_sm} respectively.
The only little departure occurs at the $M_I$ scale since the D-parity is not conserved here. Thus we have $\alpha_{2L}(M_{II})\neq \alpha_{2R}(M_{II})$ unlike the previous case.
\vskip 0.5cm
\begin{center}
	\underline{IIB. Topological Defects}
\end{center}
The formation of topological defects for this breaking scenario is very similar to the earlier case.
\begin{itemize}
\item $E(6)\xrightarrow{M_X} \mathcal{G}_{3_L 3_R 3_C }$:  As $\pi_k(E(6)/ \mathcal{G}_{3_L 3_R 3_C })=\mathcal{I}$ for $k=0,1,2$; no topological defect is created during this symmetry breaking.

\item $\mathcal{G}_{3_L 3_R 3_C } \xrightarrow{M_{I}} \mathcal{G}_{2_L 2_R 3_C 1_{LR}\slashed{D}}$: Here, $ \Pi_2(\mathcal{G}_{3_L 3_R 3_C }/\mathcal{G}_{2_L 2_R 3_C 1_{LR}\slashed{D}})= \Pi_1(U(1)_{LR})=\mathbb{Z}$ and, also $ \Pi_2(\mathcal{G}_{3_L 3_R 3_C }/\mathcal{G}_{2_L 1_Y 3_C })=\mathbb{Z}$ leading to the formation of  stable monopoles. 

\item $\mathcal{G}_{2_L 2_R 3_C 1_{LR}\slashed{D}}\xrightarrow{M_{II}} \rm SM$: At this stage only embedded cosmic strings are formed.
\end{itemize}

\subsection*{III. $\mathbf{E(6)\xrightarrow{M_X} \mathcal{G}_{2_L 2_R 4_C 1_X D} \xrightarrow{M_{I}} \mathcal{G}_{2_L 1_X 4_C }\xrightarrow{M_{II}} \rm SM }$}
In this case we have considered the flipped-$SO(10)$ scenario.
\vskip 0.5cm
\begin{center}
   \underline{IIIA. $\beta$ coefficients}
\end{center}

\input{RGE/E6-G2241D-G214.tex}

\vskip 0.5cm
\begin{center}
	\underline{IIIB. Matching conditions}
\end{center}
At $M_{I}$, the $SU(2)_R$ group completely breaks. Also, the conservation of the D-parity gives $\alpha_{2L}(M_{I})=\alpha_{2R}(M_{I})$. At the scale $M_{II}$, $SU(4)_C$ is broken to $SU(3)_C\otimes U(1)_{B-L}$, and at the same scale $U(1)_{B-L}\otimes U(1)_X$ is spontaneously broken to $U(1)_Y$. Therefore the matching condition is given by,
\begin{align}\label{matching_g2241_g214}
\frac{1}{\alpha_{1Y}(M_{II})}=\frac{1}{10}\left(\frac{1}{\alpha_{4C}(M_{II})}-\frac{1}{3\pi}\right) + \frac{9}{10}\left(\frac{1}{\alpha_{1X}(M_{II})}\right) .
\end{align}

\vskip 0.5cm
\begin{center}
	\underline{IIIC. Topological Defects}
\end{center}
\begin{itemize}
\item $E(6)\xrightarrow{M_X} \mathcal{G}_{2_L 2_R 4_C 1_X D}$: We can think of this breaking in terms of an underlying breaking pattern as $E(6)$ is spontaneously broken to $\frac{Spin(4)\otimes Spin(6)}{\mathbb{Z}_2}\otimes U(1)_X \otimes D$, where $Spin(4)\cong SU(2)_L\otimes SU(2)_R$ and $Spin(6)\cong SU(4)$. 

We find that  $ \Pi_2(E(6)/\mathcal{G}_{2_L 2_R 4_C 1_X D})=  \Pi_1({Spin(4)\otimes Spin(6)}/{\mathbb{Z}_2})\otimes  \Pi_1(U(1)_X)= \Pi_0({\mathbb{Z}_2})\otimes \mathbb{Z}={\mathbb{Z}_2}\otimes \mathbb{Z} $. These imply the formation of  topologically unstable $\mathbb{Z}_2$-monopoles and stable monopoles whose charge changes from $X$ to $Y$ in latter stage. Further, $ \Pi_1(E(6)/ \mathcal{G}_{2_L 2_R 4_C 1_X D})= \Pi_0(\mathcal{G}_{2_L 2_R 4_C 1_X D})=\mathbb{Z}_2$, thus unstable $\mathbb{Z}_2$ cosmic string is also formed.

\item $\mathcal{G}_{2_L 2_R 4_C 1_X D} \xrightarrow{M_{I}} \mathcal{G}_{2_L 1_X 4_C }$: At this stage domain walls bounded by the cosmic strings are generated.

\item $\mathcal{G}_{2_L 1_X 4_C }\xrightarrow{M_{II}} \rm SM $: Here, the  embedded strings are created.

\end{itemize}

\subsection*{IV. $\mathbf{E(6)\xrightarrow{M_X} \mathcal{G}_{2_L 2_R 4_C 1_X \slashed{D}} \xrightarrow{M_{I}} \mathcal{G}_{2_L 1_X 4_C }\xrightarrow{M_{II}} \rm SM }$}

\vskip 0.5cm
\begin{center}
	\underline{IVA. $\beta$ coefficients}
\end{center}

\input{RGE/E6-G2241-G214.tex}
\vskip 0.5cm
\begin{center}
	\underline{IVB. Matching conditions}
\end{center}
Here, the broken D-parity implies $\alpha_{2L}(M_{I})\neq\alpha_{2R}(M_{I})$. The $SU(2)_{2R}$ gauge group is completely broken at the scale $M_{I}$. The matching conditions at $M_{II}$ are the same as the Eqn.~\ref{matching_g2241_g214}.

\vskip 0.5cm
\begin{center}
	\underline{IVC. Topological Defects}
\end{center}
This breaking pattern is very similar to the earlier one apart from the absence of $D$-parity. Thus the generation of topological defects are very similar.
\begin{itemize}
\item $E(6)\xrightarrow{M_X} \mathcal{G}_{2_L 2_R 4_C 1_X \slashed{D}}$: Here, unstable $\mathbb{Z}_2$-monopoles are created, and stable monopoles with charge $X$ are generated which changes to  $Y$  in the next stage of phase transition.

\item $\mathcal{G}_{2_L 2_R 4_C 1_X \slashed{D}} \xrightarrow{M_{I}} \mathcal{G}_{2_L 1_X 4_C }$: No topological defect is formed during this phase transition since all the relevant homotopy groups are trivial for this vacuum manifold.

\item $\mathcal{G}_{2_L 1_X 4_C }\xrightarrow{M_I} \rm SM$: At this stage, only embedded strings are formed.
\end{itemize}

\subsection*{V. $\mathbf{SO(10)\xrightarrow{M_X} \mathcal{G}_{2_L 2_R 4_C D} \xrightarrow{M_{I}} \mathcal{G}_{2_L 2_R 3_C 1_{B-L}D}\xrightarrow{M_{II}} \rm SM} $}

\vskip 0.5cm
\begin{center}
	\underline{VA. $\beta$ coefficients}
\end{center}

\input{RGE/SO10-G224D-G2231D.tex}
\vskip 0.5cm
\begin{center}
	\underline{VB. Matching conditions}
\end{center}
 At $M_{I}$ scale, $SU(4)_C$ is spontaneously broken to $SU(3)_C\otimes U(1)_{B-L}$, and the matching conditions are given as,
\begin{align}\label{matching_g224_g2231}
\frac{1}{\alpha_{1(B-L)}(M_{I})} & = \frac{1}{\alpha_{4C}(M_{I})}-\frac{1}{3\pi} , \nonumber \\
\frac{1}{\alpha_{3C}(M_{I})}-\frac{1}{4\pi} & = \frac{1}{\alpha_{4C}(M_{I})}-\frac{1}{3\pi} .
\end{align}
The matching condition at the scale $M_{II}$ is dictated by,
\begin{align}\label{matching_g2231_(B-L)_sm}
\frac{1}{\alpha_{1Y}(M_{II})}=\frac{3}{5}\left(\frac{1}{\alpha_{2R}(M_{II})}-\frac{1}{6\pi}\right) + \frac{2}{5}\left(\frac{1}{\alpha_{1(B-L)}(M_{II})}\right) . 
\end{align}
D-parity remains conserved upto $M_{II}$ giving $\alpha_{2L}(M_{II})=\alpha_{2R}(M_{II})$.

\vskip 0.5cm
\begin{center}
	\underline{VC. Topological Defects}
\end{center}

Here, the GUT group  is $Spin(10)$, which is also the simply connected universal covering of $SO(10)$. $Spin(10)$ contains the maximal subgroup $\frac{Spin(4)\otimes Spin(6)}{\mathbb{Z}_2}\otimes D$, where $Spin(4)\cong SU(2)\otimes SU(2)$ and $Spin(6)\cong SU(4)$. 
\begin{itemize}
\item $SO(10)\xrightarrow{M_X} \mathcal{G}_{2_L 2_R 4_C D}$: Here, $ \Pi_2\left(SO(10)/\mathcal{G}_{2_L 2_R 4_C D}\right)= \Pi_1\left(\frac{Spin(4)\otimes Spin(6)}{\mathbb{Z}_2}\otimes D\right)= \Pi_0(\mathbb{Z}_2) = \mathbb{Z}_2 $ \cite{Lazarides:1980va,Lazarides:1980cc}. These imply that $\mathbb{Z}_2$-monopoles are but they are unstable. Again $ \Pi_0(\mathcal{G}_{2_L 2_R 4_C D})= \Pi_0(D)=\mathbb{Z}_2$. Thus $\mathbb{Z}_2$-strings are formed at the scale $M_X$ which are  stable till the next phase transition takes place at  $M_I$ scale where $D$-parity is spontaneously broken.
\item $\mathcal{G}_{2_L 2_R 4_C D} \xrightarrow{M_{I}} \mathcal{G}_{2_L 2_R 3_C 1_{B-L}D}$: At this stage only non-trivial homotopy of the vacuum manifold  is 
$ \Pi_2(\mathcal{G}_{2_L 2_R 4_C D} / \mathcal{G}_{2_L 2_R 3_C 1_{B-L}D})= \Pi_1(U(1)_{B-L})=\mathbb{Z}$. Thus topologically stable monopoles are formed as we further have $ \Pi_1(U(1)_Y)=\mathbb{Z}$. Their topological charge change from $(B-L)$ to $Y$ due to latter stage of symmetry breaking.

\item $\mathcal{G}_{2_L 2_R 3_C 1_{B-L}D}\xrightarrow{M_{II}} \rm SM $: As the $D$-parity is spontaneously broken, and string-bounded domain walls are formed.  There will be no topological cosmic string, but embedded strings are formed.
\end{itemize}

\subsection*{VI. $\mathbf{SO(10)\xrightarrow{M_X} \mathcal{G}_{2_L 2_R 4_C D} \xrightarrow{M_{I}} \mathcal{G}_{2_L 2_R 3_C 1_{B-L}\slashed{D}}\xrightarrow{M_{II}} \rm SM }$}
\vskip 0.5cm
\begin{center}
	\underline{VIA. $\beta$ coefficients}
\end{center}

\input{RGE/SO10-G224D-G2231.tex}
\vskip 0.5cm
\begin{center}
	\underline{VIB. Matching conditions}
\end{center}
The matching conditions at the scales $M_{I}$ and $M_{II}$ are given by the Eqns.~\ref{matching_g224_g2231} and~\ref{matching_g2231_(B-L)_sm} respectively. Here D-parity is broken at $M_{I}$. Therefore, $\alpha_{2L}(M_{I})=\alpha_{2R}(M_{I})$, whereas, $\alpha_{2L}(M_{II})\neq\alpha_{2R}(M_{II})$

\vskip 0.5cm
\begin{center}
	\underline{VIC. Topological Defects}
\end{center}
This breaking pattern is very similar to the earlier one apart from the breaking of $D$-parity at the first intermediate scale. Thus the formation of topological defects is very similar.

\begin{itemize}
\item $SO(10)\xrightarrow{M_X} \mathcal{G}_{2_L 2_R 4_C D}$: At this stage only $\mathbb{Z}_2$-monopoles, and $\mathbb{Z}_2$-strings are generated. Though  both of them are topologically unstable.

\item $\mathcal{G}_{2_L 2_R 4_C D} \xrightarrow{M_{I}} \mathcal{G}_{2_L 2_R 3_C 1_{B-L}\slashed{D}}$: Owe to the spontaneous breaking of $D$-parity, the domain walls bounded by the cosmic strings are formed. Along with that stable monopoles are also formed 

\item $\mathcal{G}_{2_L 2_R 3_C 1_{B-L}\slashed{D}}\xrightarrow{M_{II}} \rm SM$: At this stage only embedded cosmic strings are formed.
\end{itemize}

\pagebreak
\subsection*{VII. $\mathbf{SO(10)\xrightarrow{M_X} \mathcal{G}_{2_L 2_R 4_C D} \xrightarrow{M_{I}} \mathcal{G}_{2_L 1_R 4_C }\xrightarrow{M_{II}} \rm SM }$}
\vskip 0.5cm
\begin{center}
	\underline{VIIA. $\beta$ coefficients}
\end{center}

\input{RGE/SO10-G224D-G214.tex}
\vskip 0.5cm
\begin{center}
	\underline{VIIB. Matching conditions}
\end{center}
The $SU(2)_R$ gauge group is spontaneously broken to $U(1)_R$ at the scale $M_{I}$ with the matching condition : 
\begin{align}\label{matching_g224_g214}
\frac{1}{\alpha_{1R}(M_{I})} = \frac{1}{\alpha_{2R}(M_{I})}-\frac{1}{6\pi} .
\end{align}
Also, we have $\alpha_{2L}(M_{I})=\alpha_{2R}(M_{I})$ as a result of D-parity conservation.
At $M_{II}$, $SU(4)_C$ is broken to $SU(3)_C\otimes U(1)_{B-L}$, and at the same scale, $U(1)_{B-L}\otimes U(1)_R$ is spontaneously broken to $U(1)_Y$. Thus the matching condition at $M_{II}$ is stated as
\begin{align}\label{matching_g214_sm}
\frac{1}{\alpha_{1Y}(M_{II})}=\frac{3}{5}\left(\frac{1}{\alpha_{1R}(M_{II})}\right) + \frac{2}{5}\left(\frac{1}{\alpha_{4C}(M_{II})}-\frac{1}{3\pi}\right) .
\end{align}

\vskip 0.5cm
\begin{center}
	VIIC. Topological Defects
\end{center}
The first stage of this breaking chain is exactly same as the earlier and thus true for the formation of topological defects as well.
\begin{itemize}
\item $SO(10)\xrightarrow{M_X} \mathcal{G}_{2_L 2_R 4_C D}$: Here, only topologically unstable $\mathbb{Z}_2$-monopoles and $\mathbb{Z}_2$-strings are formed.

\item $\mathcal{G}_{2_L 2_R 4_C D} \xrightarrow{M_{I}} \mathcal{G}_{2_L 1_R 4_C }$: At this stage, the walls bounded by strings, and stable monopoles are formed. The topological  charge of the monopoles changes from $R$ to $Y$ at the subsequent stage of symmetry breaking.

\item $\mathcal{G}_{2_L 1_R 4_C }\xrightarrow{M_{II}} \rm SM $: Only embedded strings are formed.
\end{itemize}
  
\subsection*{VIII. $\mathbf{SO(10)\xrightarrow{M_X} \mathcal{G}_{2_L 2_R 4_C \slashed{D}} \xrightarrow{M_{I}} \mathcal{G}_{2_L 2_R 3_C 1_{B-L}\slashed{D}}\xrightarrow{M_{II}} \rm SM }$}

\vskip 0.5cm
\begin{center}
	\underline{VIIIA. $\beta$ coefficients}
\end{center}

\input{RGE/SO10-G224-G2231.tex}

\vskip 0.5cm
\begin{center}
	\underline{VIIIB. Matching conditions}
\end{center}
Again the matching conditions are given by the Eqns.~\ref{matching_g224_g2231} and \ref{matching_g2231_(B-L)_sm}. D-parity is broken at the scale $M_X$. Thus, $\alpha_{2L}(M_{II})\neq\alpha_{2R}(M_{II})$.

\vskip 0.5cm
\begin{center}
	\underline{VIIIC. Topological Defects}
\end{center}
Here, the $D$-parity is broken at the GUT scale itself. Thus there will not be any domain wall due to the spontaneous breaking of $D$-parity in the latter stage, unlike the previous cases.
\begin{itemize}
\item $SO(10)\xrightarrow{M_X} \mathcal{G}_{2_L 2_R 4_C \slashed{D}}$: Here, only unstable $\mathbb{Z}_2$-monopoles are formed .

\item $ \mathcal{G}_{2_L 2_R 4_C \slashed{D}} \xrightarrow{M_{I}} \mathcal{G}_{2_L 2_R 3_C 1_{B-L}}$: At this stage, the topologically stable monopoles are formed whose topological charge changes from $(B-L)$ to $Y$  in  the subsequent phase transition.

\item $\mathcal{G}_{2_L 2_R 3_C 1_{B-L}\slashed{D}}\xrightarrow{M_{II}} \rm SM$:   Here, only embedded cosmic strings are formed.
\end{itemize}
\subsection*{IX. $\mathbf{SO(10)\xrightarrow{M_X} \mathcal{G}_{2_L 2_R 4_C \slashed{D}} \xrightarrow{M_{I}} \mathcal{G}_{2_L 1_R 4_C }\xrightarrow{M_{II}} \rm SM }$}
\vskip 0.5cm
\begin{center}
	\underline{IXA. $\beta$ coefficients}
\end{center}

Here are the $\beta$-coefficients :
\input{RGE/SO10-G224-G214.tex}
\vskip 0.5cm
\begin{center}
	\underline{IXB. Matching conditions}
\end{center}
And, the matching conditions are same as the Eqns.~\ref{matching_g224_g214} and \ref{matching_g214_sm}. Here, D-parity is broken at the scale $M_X$ resulting $\alpha_{2L}(M_{I})\neq\alpha_{2R}(M_{I})$.

\begin{center}
	\underline{IXC. Topological Defects}
\end{center}
\begin{itemize}
\item $SO(10)\xrightarrow{M_X} \mathcal{G}_{2_L 2_R 4_C \slashed{D}}$: Here, only topologically unstable $\mathbb{Z}_2$-monopoles are formed.

\item $\mathcal{G}_{2_L 2_R 4_C \slashed{D}} \xrightarrow{M_{I}} \mathcal{G}_{2_L 1_R 4_C }$: Again,  stable monopoles are formed whose topological charge changes from $R$ to $Y$ in the subsequent phase transition.

\item $\mathcal{G}_{2_L 1_R 4_C }\xrightarrow{M_{II}} \rm SM$: Here, only embedded cosmic strings are generated.
\end{itemize}

\begin{table}[h!]
\begin{center}
	\caption{Here, we have summarised the possible emergence of the topological defects at the different stages of symmetry breaking starting from unified groups to the SM one.}
\begin{tabular}{|c|c|c|c|}
	\hline
	\multirow{2}{*}{$\rm{GUT}\to \mathcal{G}_I\to \mathcal{G}_{II}\to \rm{SM} $} &    	\multicolumn{3}{|c|}{Topological defects} \\ 
	\cline{2-4}
	&$\rm{GUT}\to \mathcal{G}_I$& $\mathcal{G}_I\to \mathcal{G}_{II}$ & $\mathcal{G}_{II}\to \rm{SM}$ \\
	\hline
	\thead{$E(6)\to\mathcal{G}_{3_L 3_R 3_CD}\to$ \\ $\mathcal{G}_{2_L 2_R 3_C 1_{LR} D}\to\rm{SM}$} & \thead{Unstable $\mathbb{Z}_2$-strings } & \thead{Stable monopoles} & \thead{Domain walls $+$ \\ embedded strings}  \\	
	\hline
	\thead{$E(6)\to\mathcal{G}_{3_L 3_R 3_C}\to$ \\ $\mathcal{G}_{2_L 2_R 3_C 1_{LR} \slashed{D}}\to\rm{SM}$} & \thead{No defects} & \thead{Stable monopoles} & \thead{Embedded strings}  \\
	\hline
	\thead{$E(6)\to\mathcal{G}_{2_L 2_R 4_C 1_X D}\to$ \\ $\mathcal{G}_{2_L 1_X 4_C}\to\rm{SM}$} & \thead{Unstable $\mathbb{Z}_2$-strings \\ $+$ stable monopoles \\ $+$ unstable $\mathbb{Z}_2$-monopoles} & \thead{Domain walls} & \thead{Embedded strings} \\
	\hline
	\thead{$E(6)\to\mathcal{G}_{2_L 2_R 4_C 1_X \slashed{D}}\to$ \\ $\mathcal{G}_{2_L 1_X 4_C}\to\rm{SM}$} & \thead{ Stable monopoles $+$ \\ unstable $ \mathbb{Z}_2$-monopoles} & \thead{No defects} & \thead{Embedded strings} \\
	\hline
	\thead{$SO(10)\to\mathcal{G}_{2_L 2_R 4_C D}\to$ \\ $\mathcal{G}_{2_L 2_R 3_C 1_{B-L} D}\to\rm{SM}$} & \thead{ $\mathbb{Z}_2$-strings \\ (stable upto $M_{II}$)  \\ $+$ unstable $\mathbb{Z}_2$-monopoles} & \thead{Stable monopoles} & \thead{Domain walls $+$ \\ embedded strings}  \\ 
	\hline
	\thead{$SO(10)\to\mathcal{G}_{2_L 2_R 4_C D}\to$ \\ $\mathcal{G}_{2_L 2_R 3_C 1_{B-L} \slashed{D}}\to\rm{SM}$} & \thead{Unstable $\mathbb{Z}_2$-strings $+$ \\ unstable $\mathbb{Z}_2$-monopoles} & \thead{Domain walls $+$ \\ stable monopoles} & \thead{Embedded strings}  \\ 
	
	\hline
	\thead{$SO(10)\to\mathcal{G}_{2_L 2_R 4_C \slashed{D}}\to$ \\ $\mathcal{G}_{2_L 2_R 3_C 1_{B-L} \slashed{D}}\to\rm{SM}$} & \thead{Unstable $\mathbb{Z}_2$-monopoles} & \thead{Stable monopoles} & \thead{Embedded strings} \\
	\hline
	\thead{$SO(10)\to\mathcal{G}_{2_L 2_R 4_C D}\to$ \\ $\mathcal{G}_{2_L 1_R 4_C}\to\rm{SM}$} & \thead{Unstable $\mathbb{Z}_2$-strings $+$ \\ unstable $\mathbb{Z}_2$-monopoles} & \thead{Domain walls $+$ \\ stable monopoles} & \thead{Embedded strings} \\
	\hline
	\thead{$SO(10)\to\mathcal{G}_{2_L 2_R 4_C \slashed{D}}\to$ \\ $\mathcal{G}_{2_L 1_R 4_C}\to\rm{SM}$} & \thead{ Unstable $\mathbb{Z}_2$-monopoles} & \thead{Stable monopoles} & \thead{Embedded strings} \\
	\hline
\end{tabular}
\end{center}
\end{table}

 \section{Results}
 \label{results}
 \subsection{Test for Unification with and without threshold corrections} 
 Our aim of this analysis is to find out the unification solutions in terms of unified coupling ($g_U$), intermediate scale(s) ($M_{I/II}$), and unification scale ($M_X$) and the solution space is compatible with the low energy data, given in Table~\ref{ewpo}. To do so we have constructed a $\chi^2$-function as:
 \begin{align}
 	\chi^2 = \sum_{i=1}^3 \frac{\left({g_{i}}^2-g_{i,exp}^2\right)^2}{\sigma^2(g_{i,exp}^2)},
 \end{align}
 and we minimize this function to find the solution. 
 Here, $g_i$'s denote the SM gauge couplings at the electroweak scale $M_Z$ and can be recast in terms of the unification solutions using the renormalization group equations.  The $g_{i,exp}$'s are their experimental values at $M_Z$ scale with uncertainties $\sigma(g_{i,exp})$ which can be derived from the low energy parameters tabulated in Table~\ref{ewpo}.  
 \begin{table}[h!]
 	\begin{center}
 		\begin{tabular}{|c|c|}
 			\hline
 			Z-boson mass, $M_Z$ & $91.1876(21)$ GeV \\
 			\hline
 			Strong coupling constant, $\alpha_S$ & $0.1185(6)$ \\
 			\hline
 			Fermi coupling constant, $G_F$ & $1.1663787(6)\times 10^{-5} \ \rm{GeV}^{-2}$ \\
 			\hline
 			Weinberg angle, $\sin^2{\theta_W}$ & $0.23126(5)$ \\
 			\hline
 		\end{tabular}
 		\caption{The low energy parameters at the electroweak scale. Unification solutions are compatible with these values.}\label{ewpo}
 	\end{center}
 \end{table}
 
 In presence of one intermediate scale, we have noted solutions in terms of intermediate scale $(M_I)$, unification scale $(M_X)$ and  proton decay lifetime $(\tau_p)$, see Table~\ref{thrshldcr}, using two-loop RGEs and minimising the $\chi^2$ function. In case of two intermediate scales, similar solution space is found. But, here, due to the presence of an extra intermediate scale, the D.O.F. increases and that allows a range of solutions, unlike the one intermediate case. Here, we have considered three different choices to incorporate the  threshold corrections: (i) no threshold correction ($R=1$),  (ii) short range variation ($R \in [1/2:2]$), and  (iii) long range variation ($R \in [1/10:10]$) .

\subsection{One intermediate scale  and the proton lifetime: present status }
In this section, we have considered all possible one step breaking chain, as in Ref.~\cite{Chakrabortty:2017mgi} from $SO(10)$ and $E(6)$ having a left-right symmetric gauge group ($SU(N)_L\otimes SU(N)_R$) at the intermediate stage. 

\begin{figure}[h!]
	\centering
	{
		\includegraphics[scale=0.90]{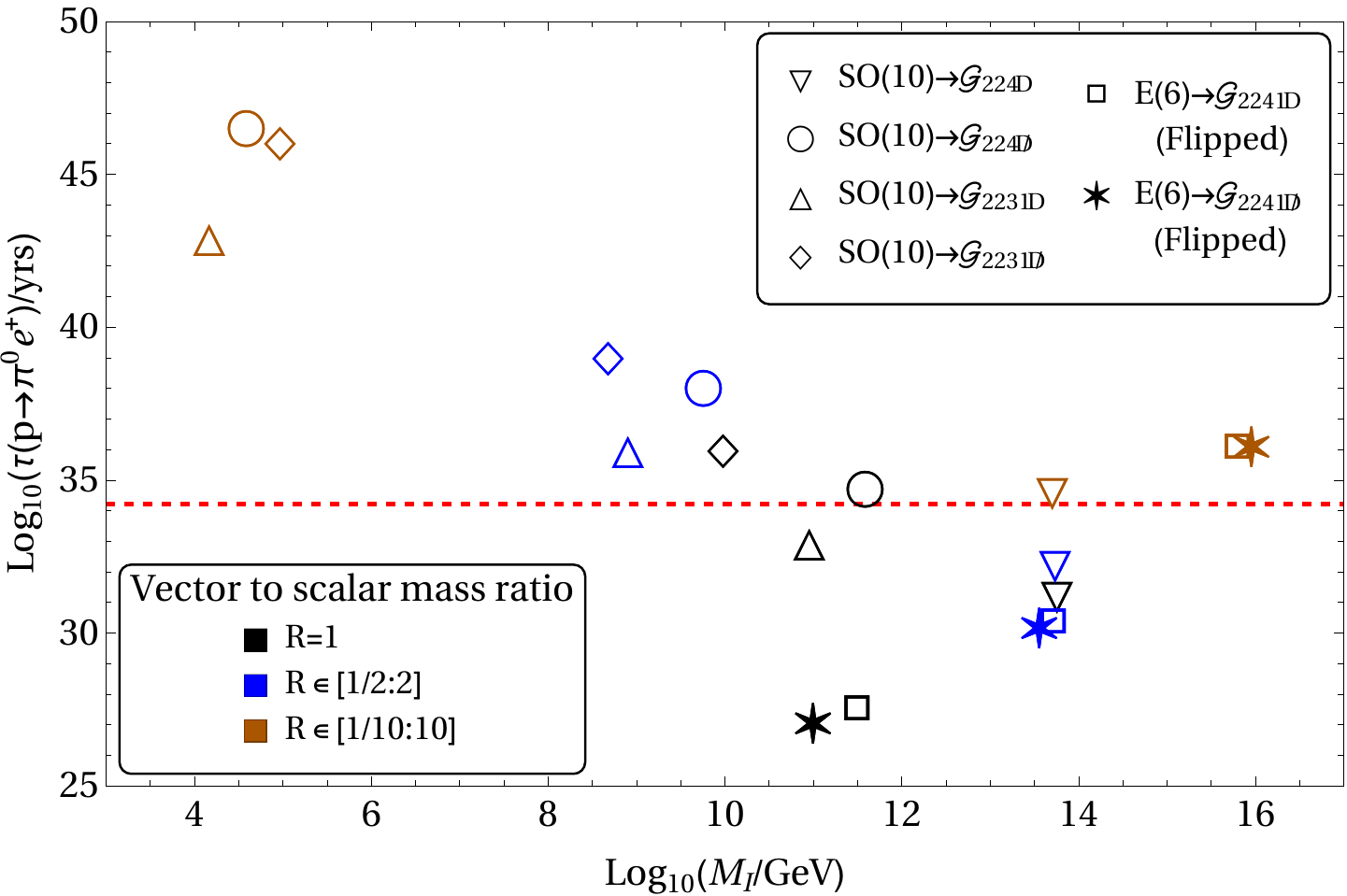}
	}
	
	\caption{GUT prediction for proton lifetime in presence of one intermediate scale.}\label{fig:one-step}
\end{figure}

We have computed the proton decay lifetime for  different scenarios: (i) no threshold correction ($R$=1), (ii) non-zero threshold correction featured through the variation of $R$ in two different ranges. Performing two-loop RGEs, we have found out unification solutions for all one step breaking chain. We have explained how the inclusion of threshold correction affects the unification solutions. The question that we want to address is whether the threshold corrections could be the saviour for this kind of theory which are ruled by proton decay lifetime constraints?


\begin{table}[h!]
	\begin{center}
		\begin{tabular}{|c|c|c|c|c|}
			\hline
			\multirow{2}{*}{Breaking chain} & \multirow{2}{*}{Observables} & \multicolumn{3}{|c|}{ $R=\frac{M_i}{M_X}$, $i=S,F$} \\
			\cline{3-5}
			& & $1$ & $[1/2 : 2]$ & $[1/10 : 10]$  \\
			\hline
			\multirow{3}{*}{$SO(10)\to \mathcal{G}_{2_L2_R4_C\slashed{D}} \to SM$} & $\log_{10}(M_I/{\rm{GeV}})$ & $11.6$ & $9.8$ & $4.6$ \\
			\cline{2-5}
			& $\log_{10}(M_X/{\rm{GeV}})$ & $15.7$ & $16.6$ & $18.7$ \\
			\cline{2-5}
			& $\tau_p$ (yrs) & $\mathbf{5.5\times 10^{34}}$ & $\mathbf{1.1\times 10^{38}}$ & $\mathbf{3.4\times 10^{46}}$ \\
			\hline
			\multirow{3}{*}{$SO(10)\to \mathcal{G}_{2_L2_R4_C D} \to SM$} & $\log_{10}(M_I/{\rm{GeV}})$ & $13.8$ & $13.7$ & $13.7$ \\
			\cline{2-5}
			& $\log_{10}(M_X/{\rm{GeV}})$ & $14.8$ & $15.1$ & $15.7$  \\
			\cline{2-5}
			& $\tau_p$ (yrs) & $1.6\times 10^{31}$ & $1.7\times 10^{32}$ & $\mathbf{4.0\times 10^{34}}$ \\
			\hline
			\multirow{3}{*}{$SO(10)\to \mathcal{G}_{2_L2_R3_C1_{B-L}\slashed{D}} \to SM$} & $\log_{10}(M_I/{\rm{GeV}})$ & $10.0$ & $8.7$ & $5.0$\\
			\cline{2-5}
			& $\log_{10}(M_X/{\rm{GeV}})$ & $16.0$ & $16.8$ & $18.6$\\
			\cline{2-5}
			& $\tau_p$ (yrs) & $\mathbf{1.0\times 10^{36}}$ & $\mathbf{1.1\times 10^{39}}$ & $\mathbf{1.2\times 10^{46}}$ \\
			\hline
			\multirow{3}{*}{$SO(10)\to \mathcal{G}_{2_L2_R3_C1_{B-L} D} \to SM$} & $\log_{10}(M_I/{\rm{GeV}})$ & $11.0$ & $8.9$ & $4.2$ \\
			\cline{2-5}
			& $\log_{10}(M_X/{\rm{GeV}})$ & $15.3$ & $16.0$ & $17.8$\\
			\cline{2-5}
			& $\tau_p$ (yrs) & $9.4\times 10^{32}$ & $\mathbf{9.4\times 10^{35}}$ & $\mathbf{8.1\times 10^{42}}$ \\
			\hline
			\multirow{3}{*}{$E(6)\to \mathcal{G}_{2_L2_R4_C1_X\slashed{D}} \to SM$} & $\log_{10}(M_I/{\rm{GeV}})$ & $11.0$ & $13.6$ & $15.9$\\
			\cline{2-5}
			& $\log_{10}(M_X/{\rm{GeV}})$ & $13.7$ & $14.5$  & $16.0$\\
			\cline{2-5}
			& $\tau_p$ (yrs) & $9.4\times 10^{26}$ & $1.3\times 10^{30}$ & $\mathbf{1.0\times 10^{36}}$ \\
			\hline
			\multirow{3}{*}{$E(6)\to \mathcal{G}_{2_L2_R4_C1_X D} \to SM$} & $\log_{10}(M_I/{\rm{GeV}})$ & $11.5$ & $13.7$ & $15.8$ \\
			\cline{2-5}
			& $\log_{10}(M_X/{\rm{GeV}})$ & $13.9$ & $14.6$ & $16.0$\\
			\cline{2-5}
			& $\tau_p$ (yrs) & $2.8\times 10^{27}$ & $1.9\times 10^{30}$ &  $\mathbf{9.7\times 10^{35}}$ \\
			\hline
		\end{tabular}
		\caption{$\log_{10}(M_X/{\rm{GeV}})$ and $\log_{10}(M_I/{\rm{GeV}})$, and proton decay lifetime $\tau_p$ (in yrs) are computed for the different one step GUT breaking scenarios. Here, $R$=1 implies the absence of threshold correction. The non-zero threshold corrections are incorporated for two different choices: $R$ varied between $[1/2:2]$, and $[1/10:10]$. The proton decay predictions which satisfy $\tau_p \geq 1.6 \times 10^{34}$ yrs constraint are boldfaced.}\label{thrshldcr}
	\end{center}
\end{table}


To answer this query, we need to understand the the Fig.~\ref{fig:one-step}. In this plot the red dotted line signifies the experimentally allowed minimum value of proton decay lifetime. Any solution below that is ruled out. The solutions correspond to $R=1$, i.e., no threshold corrections are all ruled out apart from the breaking chains $SO(10) \to \mathcal{G}_{224\slashed{D}}$ and $SO(10) \to \mathcal{G}_{2231\slashed{D}}$. Then we have varied $R$ within $[1/2:2]$ and $[1/10:10]$ to estimate the impact of threshold correction. It is evident from Fig.~\ref{fig:one-step}, that inclusion of these corrections certainly push the proton decay lifetime prediction for each model to the higher values. Thus to save these models from this constraint, these corrections may  play crucial and important role. It is interesting to note that these corrections also affect the intermediate scales, they are even brought down to much low scale in some cases. The amount of threshold corrections depends on the range of $R$. 

We have summarised the unification solutions for each breaking chain in terms of intermediate scale $M_I$, unification scale $M_X$, and computed the proton decay lifetime for three different choices of $R$ in Table~\ref{thrshldcr}. We have noted  the models that pass the proton decay lifetime constraint ($\tau_p \geq 1.6 \times 10^{34}$ yrs), and their predictions are mentioned in boldface.
This clearly shows the impact and importance of the threshold corrections.

\subsection{Two intermediate scales and the proton lifetime}
In this section, we have performed a similar analysis, as the earlier section, but for two intermediate symmetry groups. As proton decay lifetime is one of the deciding factors to rule in or out GUT models, we have discussed, first, the models which are compatible with this constraint. 

\begin{figure}[h!]
	\centering
	\subfloat[\tiny{$E(6)\to\mathcal{G}_{3_L 3_R 3_C}\to \mathcal{G}_{2_L 2_R 3_C 1_{B-L}\slashed{D}}\to \rm{SM}$}]
	{
		\includegraphics[trim={1cm 0 1.2cm 0},scale=0.54]{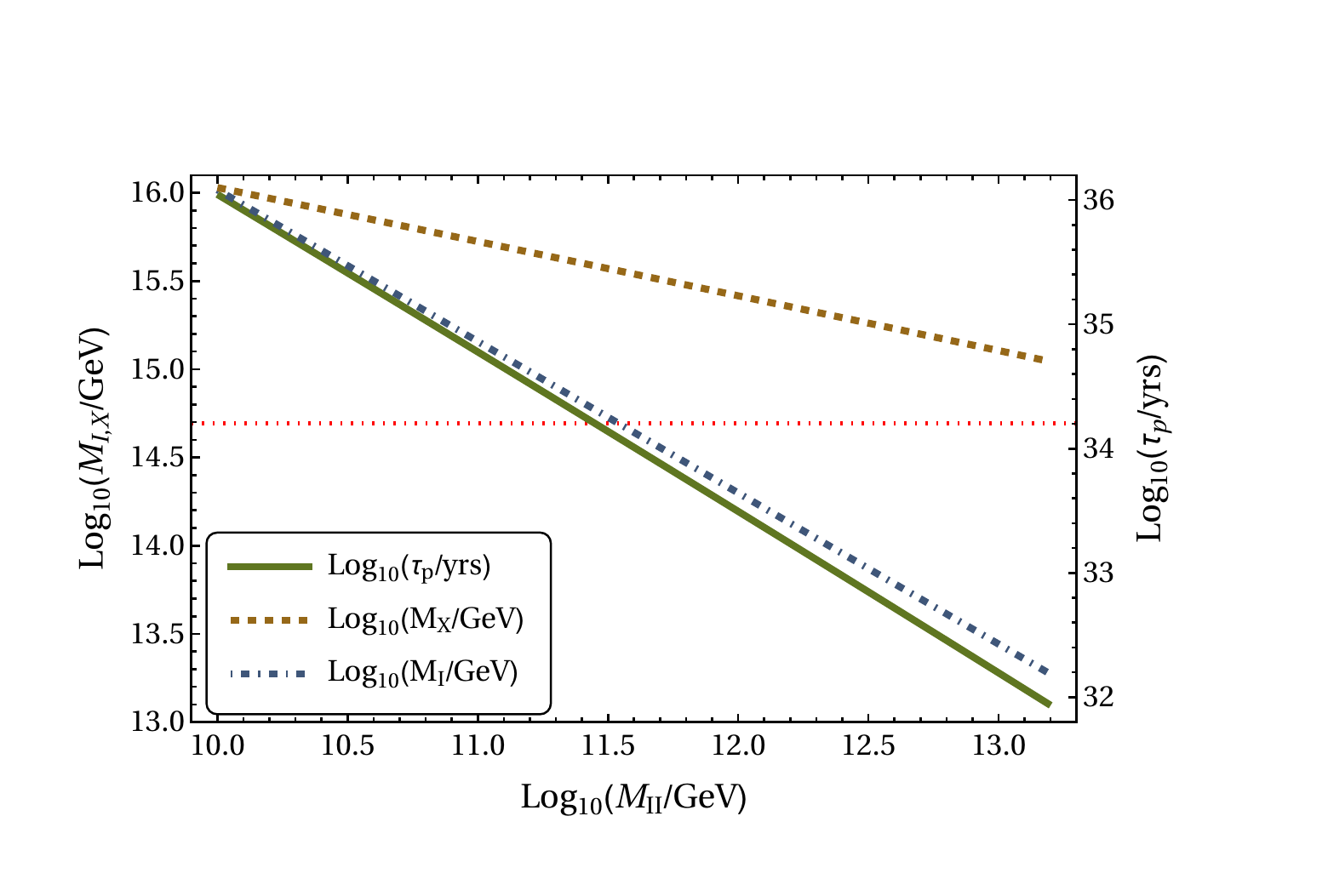}
	}
	\subfloat[\tiny{$E(6)\to\mathcal{G}_{2_L 2_R 4_C 1_XD}\to \mathcal{G}_{2_L 1_X 4_C}\to \rm{SM}$}]
	{
		\includegraphics[trim={1cm 0 1.2cm 0},scale=0.54]{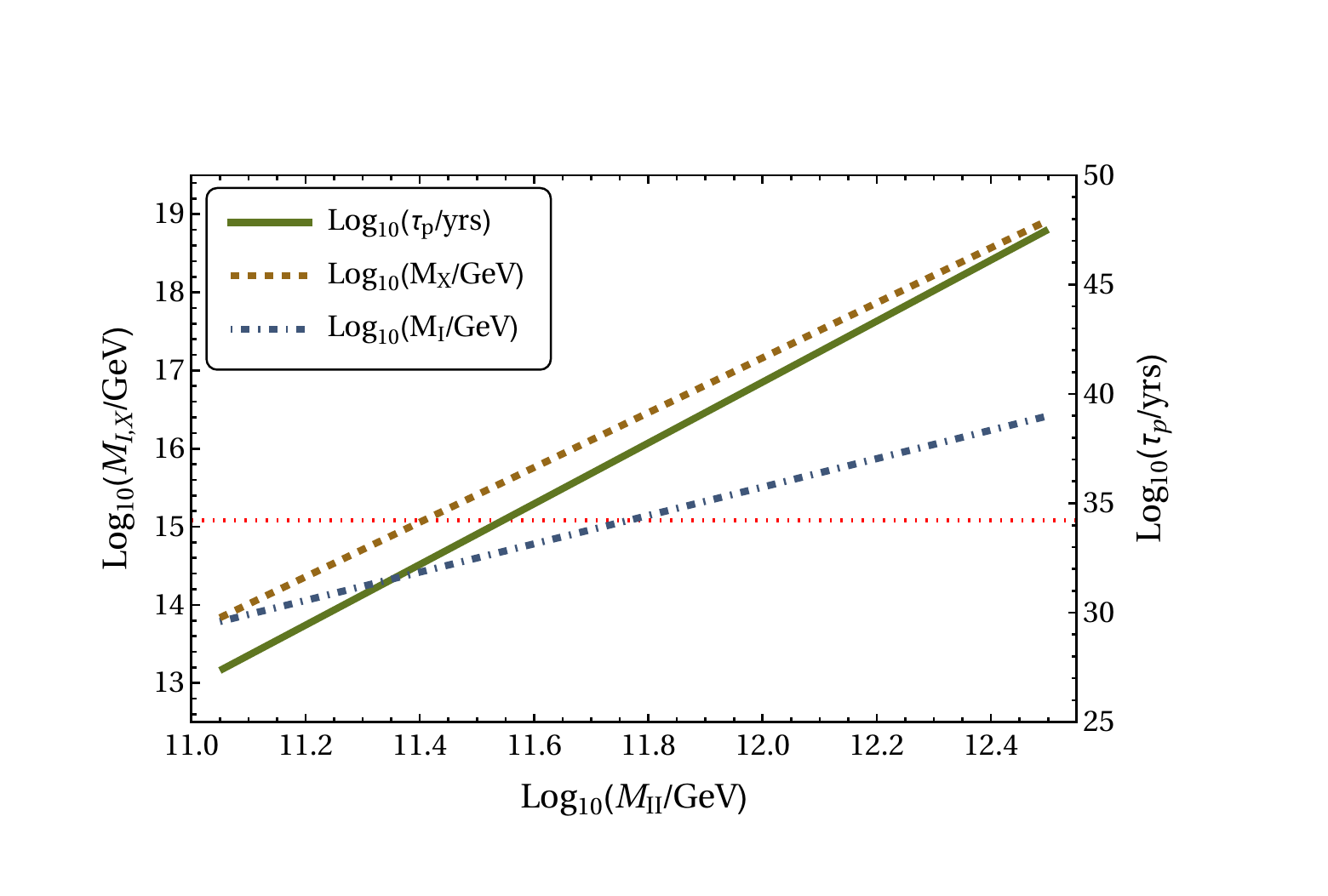}
	} \\
	\subfloat[\tiny{$E(6)\to\mathcal{G}_{2_L 2_R 4_C 1_X\slashed{D}}\to \mathcal{G}_{2_L 1_X 4_C}\to \rm{SM}$}]
	{
		\includegraphics[trim={0 0 0 0},scale=0.54]{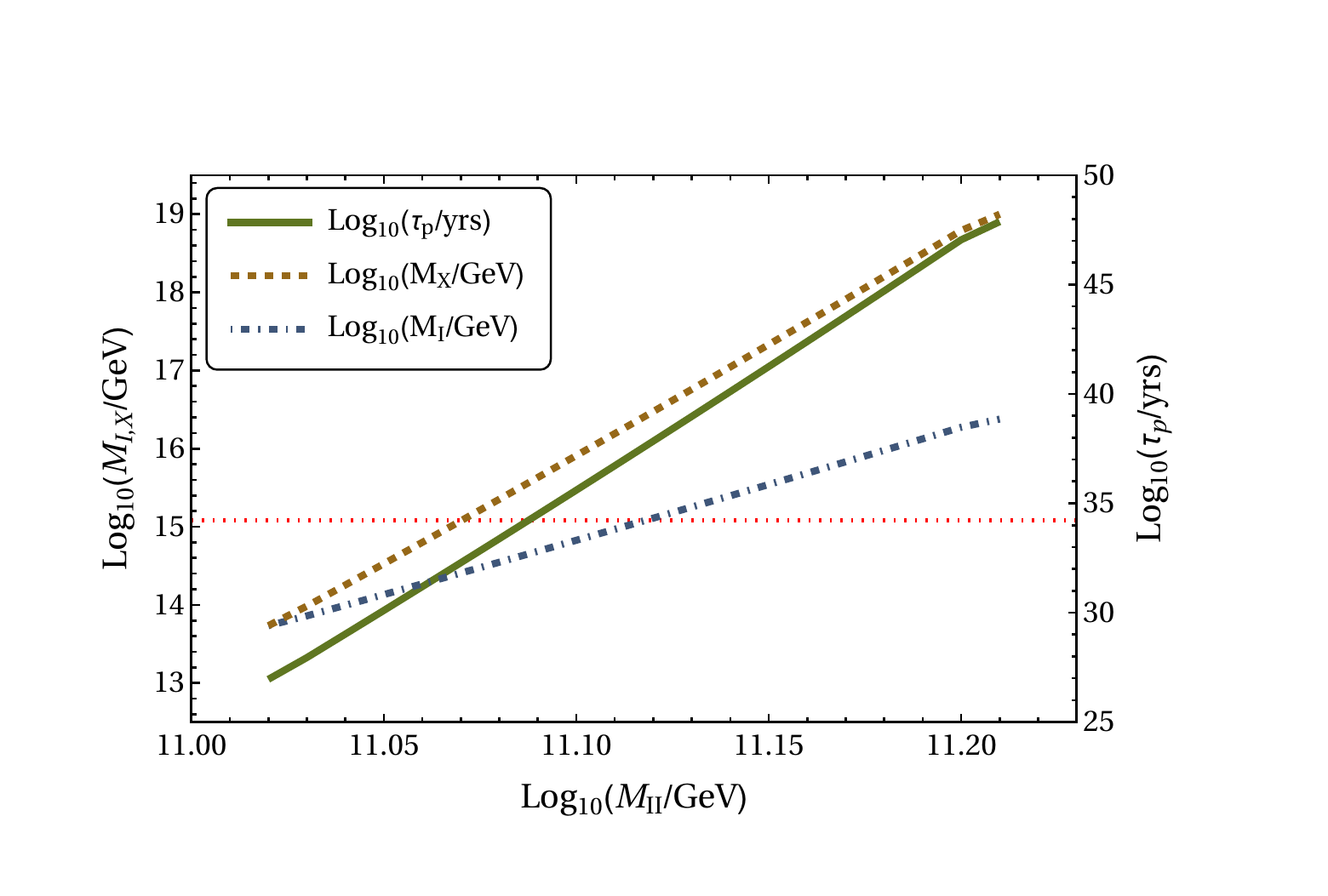}
	} 
	\caption{Proton lifetime with unification solutions for two intermediate step breaking of $E(6)$ to the SM. The horizontal dotted red line corresponds to  the proton lifetime constraint from Super-Kamiokande collaboration, below which solutions are ruled out.  }\label{fig:e6toSM}
\end{figure}

Here, we have shown the unification solutions in terms of two intermediate scales ($M_I,M_{II}$), unification scale ($M_X$) and computed the proton decay lifetime ($\tau_p$). It is evident from the plots that for each breaking chain the part of the solutions are satisfying the $\tau_p$-constraint and thus are allowed till date. But some solutions are already ruled out. Compared to the one intermediate scale, here we have more freedom due to the presence of one more intermediate symmetry group. Thus we have found a range of scales for the intermediate symmetries consistent with the unification picture, unlike the one intermediate case.  In the context of $SO(10)$, the related analysis can be found in \cite{Bertolini:2009qj,Bertolini:2013vta}. 

\begin{figure}[h!]
	\centering
	\subfloat[\tiny{$SO(10)\to\mathcal{G}_{2_L 2_R 4_C D}\to \mathcal{G}_{2_L 2_R 3_C 1_{B-L} \slashed{D}}\to \rm{SM}$}]
	{
		\includegraphics[trim={1cm 0 1.2cm 0},scale=0.54]{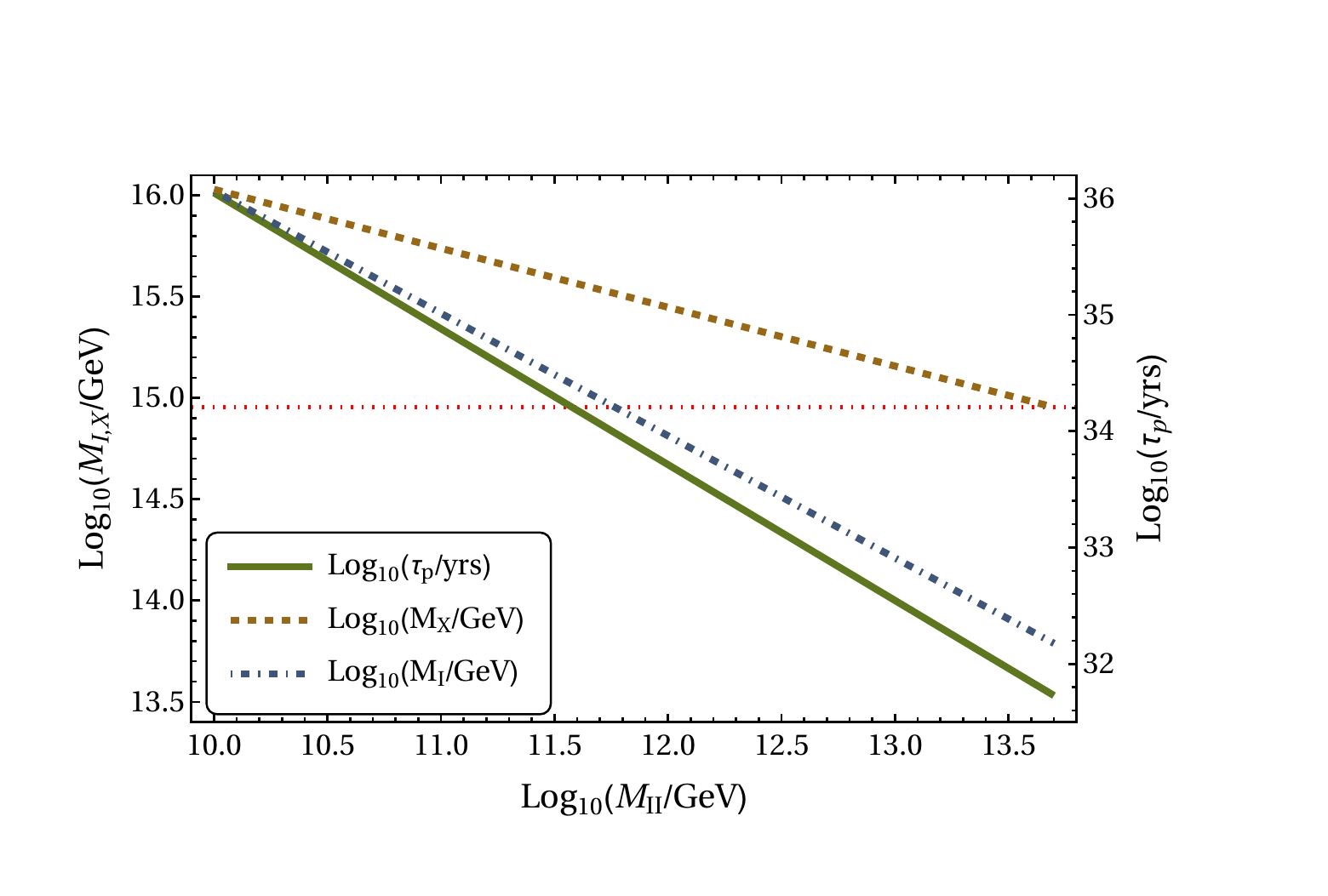}
	} 
	\subfloat[\tiny{$SO(10)\to\mathcal{G}_{2_L 2_R 4_C \slashed{D}}\to \mathcal{G}_{2_L 2_R 3_C 1_{B-L} \slashed{D}}\to \rm{SM}$}]
	{
		\includegraphics[trim={1cm 0 1.2cm 0},scale=0.54]{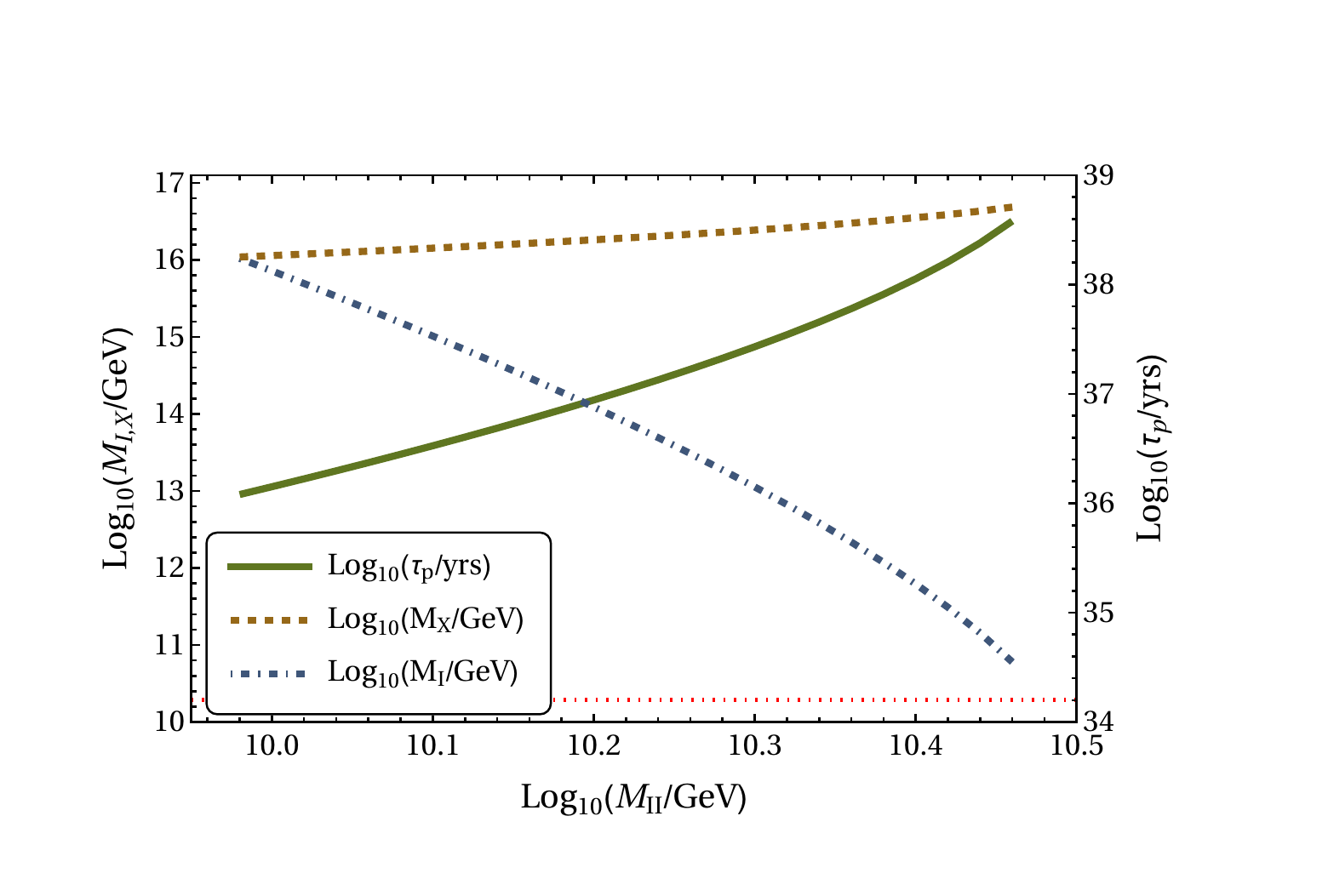}
	} 
	
	\caption{Proton lifetime prediction in case of two intermediate step breaking of $SO(10)$. The horizontal dotted red line corresponds to the lower limit bound of the proton lifetime from Super-Kamiokande collaboration.  }\label{fig:so10toSM}
\end{figure}

In this section in each plot, unification $(M_X)$, and the first $(M_{I})$ intermediate scales  starting from the unification side are depicted by the brown-dashed and blue-dot-dashed lines (see the $Y_1$-axis labelling) as a function of second intermediate scale $(M_{II})$ (see $X$-axis).  The proton decay lifetime $(\tau_p)$ for each model is shown by the green-solid line (see the $Y_2$-axis labelling). The horizontal red-dotted line represent the experimental limit on proton lifetime, i.e., $\tau_p \geq 1.6\times10^{34}$ years.

In Fig.~\ref{fig:e6toSM}, we have discussed the unification and proton decay for three different breaking chains: (a) $E(6)\to\mathcal{G}_{3_L 3_R 3_C}\to \mathcal{G}_{2_L 2_R 3_C 1_{B-L}\slashed{D}}\to \rm{SM}$, (b) $E(6)\to\mathcal{G}_{2_L 2_R 4_C 1_XD}\to \mathcal{G}_{2_L 1_X 4_C}\to \rm{SM}$, and (c) $E(6)\to\mathcal{G}_{2_L 2_R 4_C 1_X\slashed{D}}\to \mathcal{G}_{2_L 1_X 4_C}\to \rm{SM}$. Here, we have set $R=1$, i.e., no threshold correction has been injected.
We have noted that for breaking chain shown in Fig.~\ref{fig:e6toSM}(a) the solutions, allowed by proton lifetime constraint, exist only for $M_{II}$ within the range of $[10^{10.0}:10^{11.4}]$ GeV. Similarly for the models shown in Fig.~\ref{fig:e6toSM}(b), and Fig.~\ref{fig:e6toSM}(c), the unification solutions compatible with $\tau_p$ are for $10^{11.5}\,{\rm GeV} < M_{II} < 10^{12.5}\,{\rm GeV}$, and $10^{11.1}\,{\rm GeV} < M_{II} < 10^{11.2}\,{\rm GeV}$ respectively.

In Fig.~\ref{fig:so10toSM}, we have discussed the unification and proton decay for three different breaking chains: (a) $SO(10)\to\mathcal{G}_{2_L 2_R 4_C D}\to \mathcal{G}_{2_L 2_R 3_C 1_{B-L} \slashed{D}}\to \rm{SM}$, and (b) $SO(10)\to\mathcal{G}_{2_L 2_R 4_C \slashed{D}}\to \mathcal{G}_{2_L 2_R 3_C 1_{B-L} \slashed{D}}\to \rm{SM}$. Here, we have set $R=1$, i.e., no threshold correction has been incorporated.
We have noted that for breaking chain shown in Fig.~\ref{fig:so10toSM}(a), and ~\ref{fig:so10toSM}(b) the unification solutions compatible with $\tau_p >1.6\times 10^{34}$ years are for $10^{10.0}\,{\rm GeV} < M_{II} < 10^{11.6}\,{\rm GeV}$, and $10^{10.0}\,{\rm GeV} < M_{II} < 10^{10.5}\,{\rm GeV}$ respectively.

\begin{figure}[h!]
	\centering
	\subfloat[$R=1$]
	{
		\includegraphics[trim={1cm 0 1.2cm 0},scale=0.54]{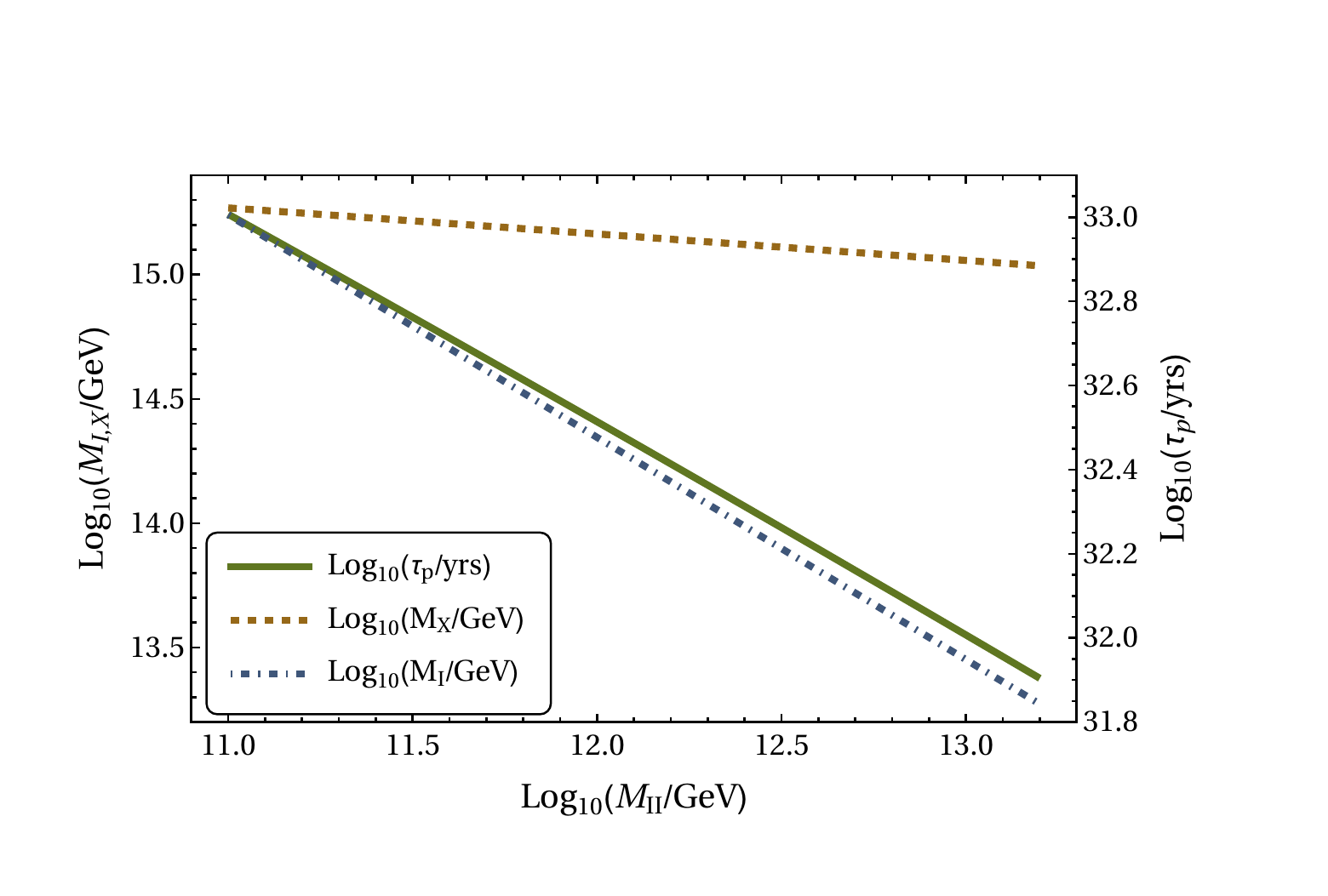}
	}
	\subfloat[${R\in \left[1/2:2\right]}$]
	{
		\includegraphics[trim={1cm 0 1.2cm 0},scale=0.54]{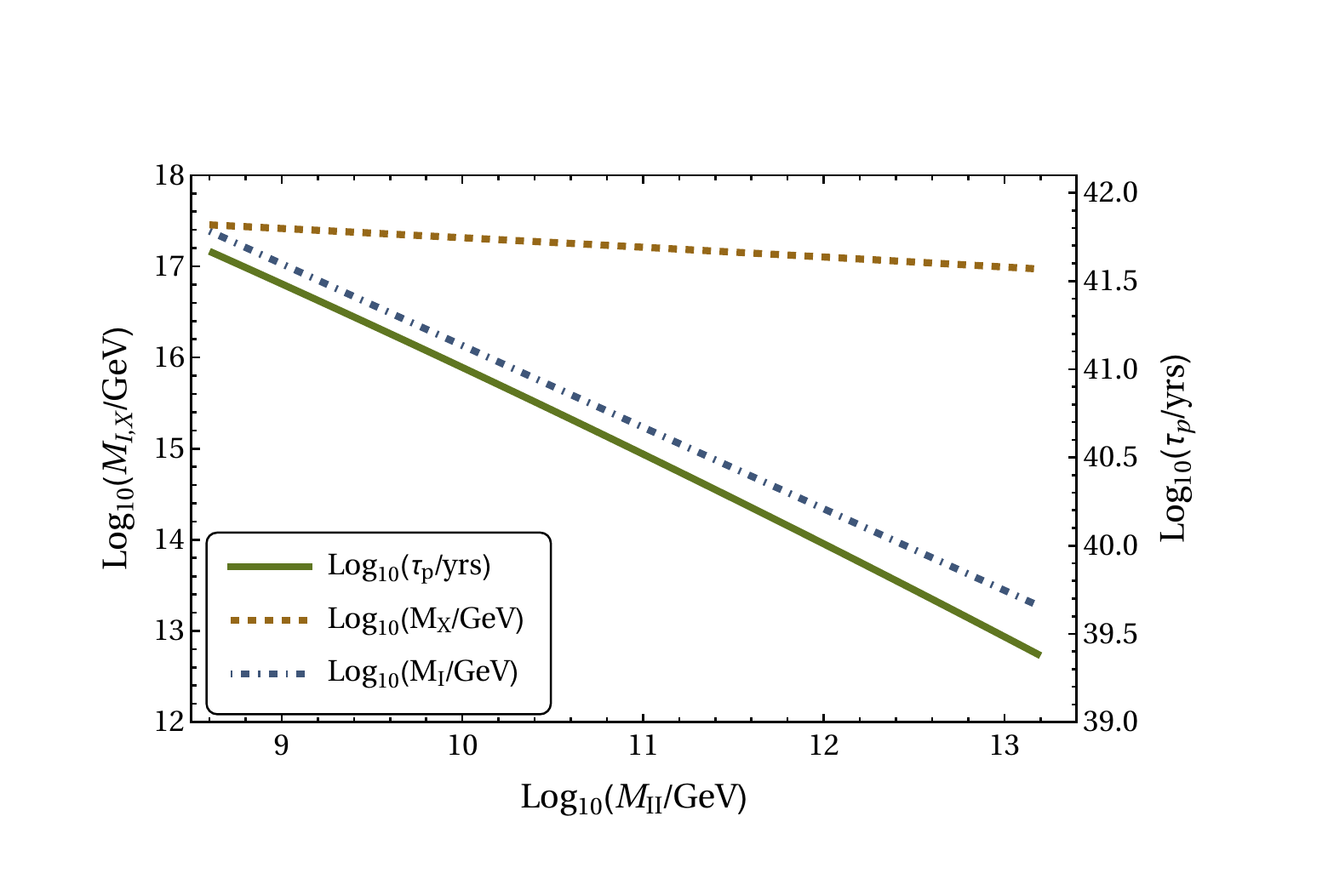}
	}
	\caption{Proton lifetime prediction in case following breaking chain: $E(6)\to\mathcal{G}_{3_L 3_R 3_CD}\to \mathcal{G}_{2_L 2_R 3_C 1_{B-L} D}\to \rm{SM}$. The proton lifetime constraint ($\geq 1.6 \times 10^{34}$ yrs) rules out the entire range of unification solutions in the absence of threshold correction ($R=1$). But once the threshold correction is incorporated and $R$ is being varied between $[1/2:2]$, a range of unification solutions are found.}
	\label{fig:e6-thc-1}
\end{figure}

\begin{figure}[h!]
	\centering
	\subfloat[$R=1$]
	{
		\includegraphics[trim={1cm 0 1.2cm 0},scale=0.54]{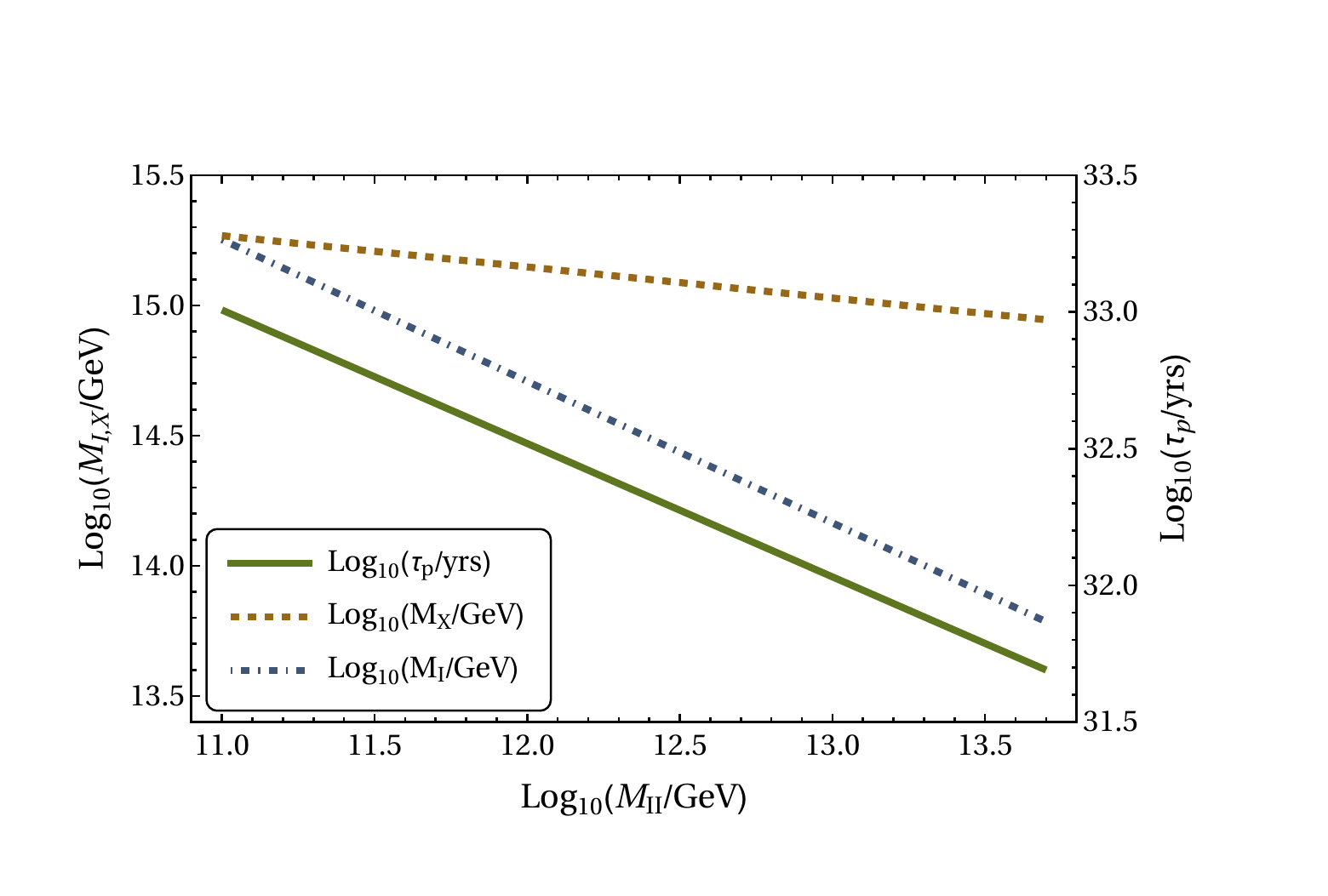}
	}
	\subfloat[ ${R\in \left[1/2:2\right]}$]
	{
		\includegraphics[trim={1cm 0 1.2cm 0},scale=0.54]{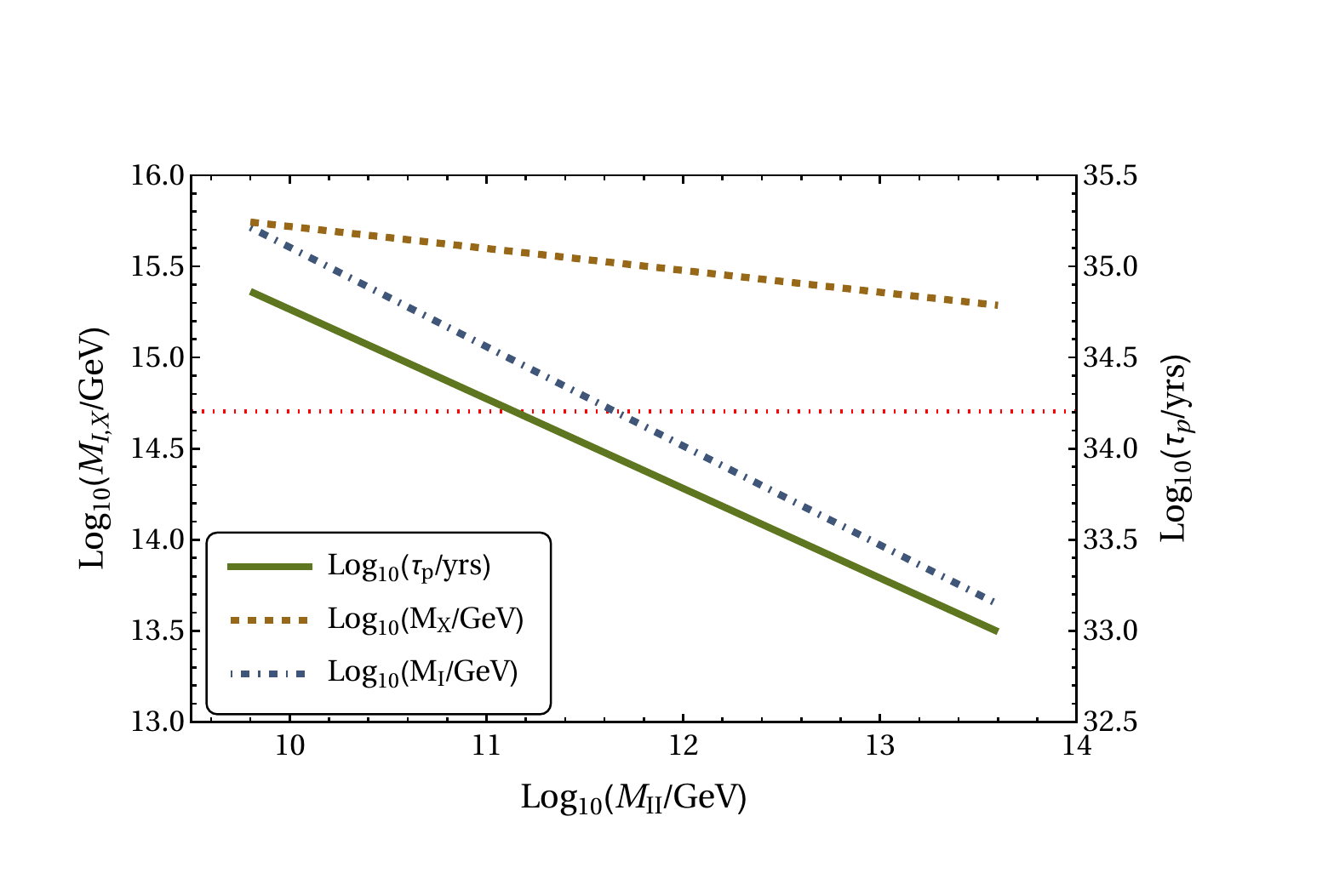}
	}
	\caption{Proton lifetime prediction in case following breaking chain: $SO(10)\to\mathcal{G}_{2_L 2_R 4_C D}\to \mathcal{G}_{2_L 2_R 3_C 1_{B-L} D}\to \rm{SM}$. The proton lifetime constraint ($\geq 1.6 \times 10^{34}$ yrs) rules out the entire range of unification solutions in the absence of threshold correction ($R=1$). But once the threshold correction is incorporated and $R$ is being varied between $[1/2:2]$, we have noted an improvement in the unification solution. This correction allows a partial range of unification solutions and revive this breaking pattern.}\label{fig:so10to224D}
\end{figure}

This implies that even in the absence of threshold corrections we have unification solutions for these models compatible with the limit on $\tau_p$. Thus we have not discussed the impact of threshold correction within these frameworks.


Now we have shifted our focus to other two intermediate breaking patterns where all most all of the 
unification solutions are ruled out bu the proton decay lifetime constraint. Our aim is to check whether the incorporation of  threshold corrections can have enough contribution to the unification program to revive some of the ruled out models. More precisely whether we can find a range of unification solutions compatible with the limit on $\tau_p$. 

\begin{figure}[h!]
	\centering
	\subfloat[$R=1$]
	{
		\includegraphics[trim={1cm 0 1.2cm 0},scale=0.54]{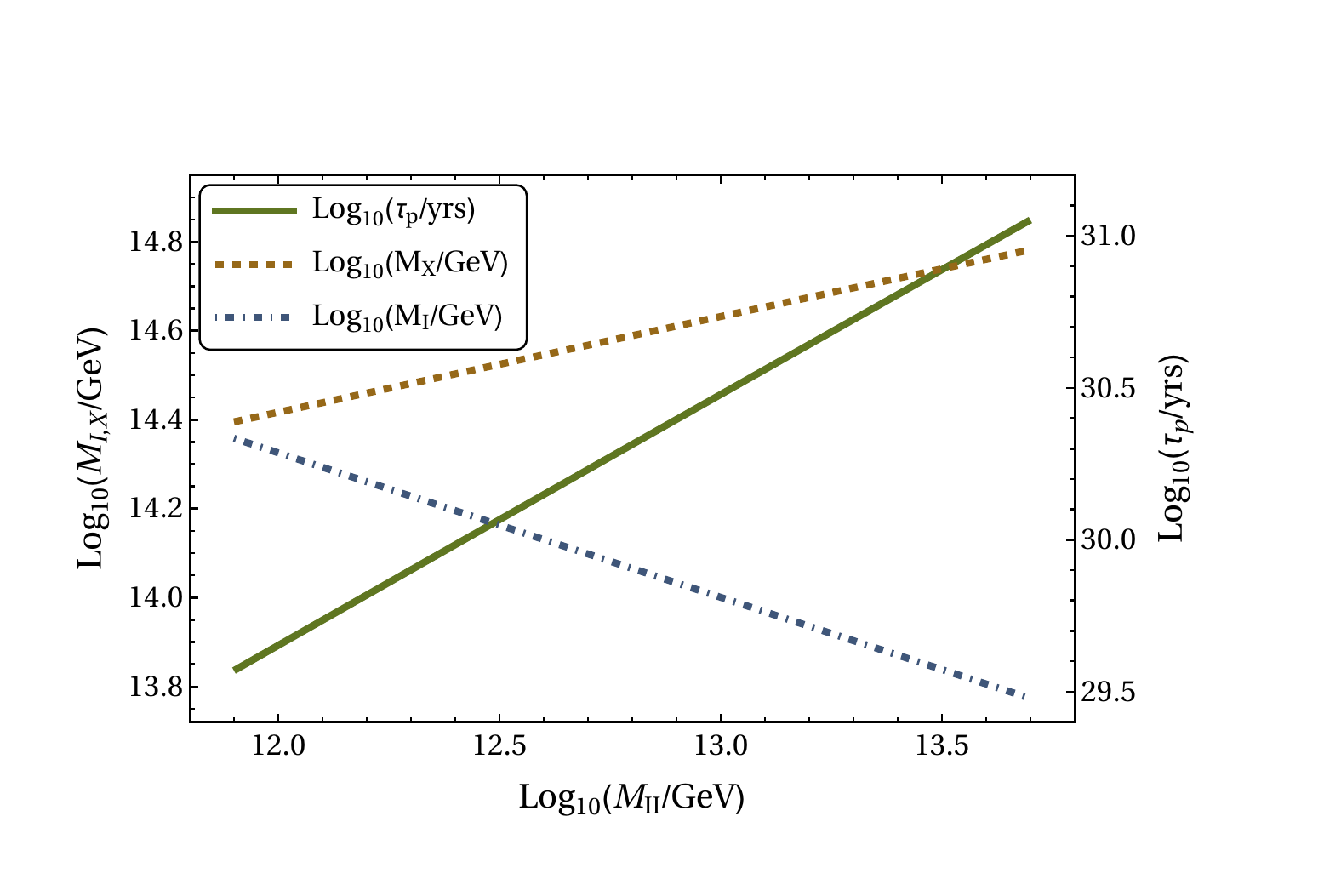}
	} 
	\subfloat[${R\in \left[ 1/10 : 10 \right] }$]
	{
		\includegraphics[trim={1cm 0 1.2cm 0},scale=0.54]{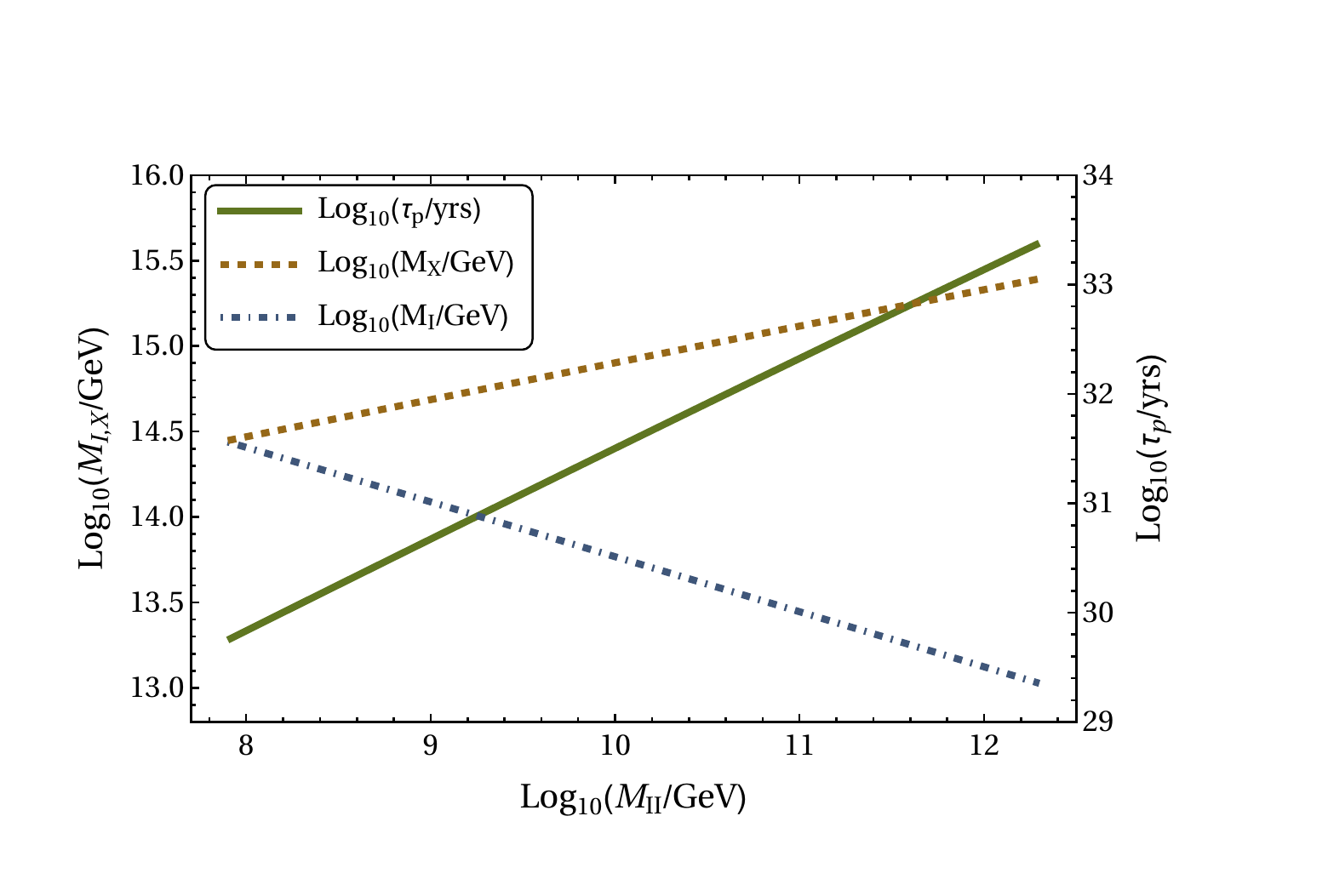}
	} 
	\caption{Proton lifetime prediction in case following breaking chain: $SO(10)\to\mathcal{G}_{2_L 2_R 4_C D}\to \mathcal{G}_{2_L 1_R 4_C}\to \rm{SM}$. The proton lifetime constraint ($\geq 1.6 \times 10^{34}$ yrs) rules out the entire range of unification solutions in the absence of threshold correction ($R=1$). Here, unlike the earlier cases where the threshold corrections are the saviour for the ruled out scenarios, fail to serve the similar purpose. Even after the incorporation of threshold correction by varying $R$ in the range of $[1/10:10]$, the full range of unification solutions are disallowed by the proton decay constraint.}
	\label{fig:so10to214}
\end{figure}

In Fig.~\ref{fig:e6-thc-1}, we have considered the breaking chain: $E(6)\to\mathcal{G}_{3_L 3_R 3_CD}\to \mathcal{G}_{2_L 2_R 3_C 1_{B-L} D}\to \rm{SM}$. The plot in Fig.~\ref{fig:e6-thc-1}(a) shows the solution space for $R=1$, i.e., in absence of threshold correction, and it is quite clear that all the solution space is below the $\tau_p$ limit and thus ruled out. Now in Fig.~\ref{fig:e6-thc-1}(b) we have noted the solution space when the minimal threshold correction (as $R$ is varied in range of $[1/2:2]$) is incorporated. This clearly shows that now we have $\tau_p$ compatible unification solution for 
$10^{8.6}\,{\rm GeV} < M_{II} < 10^{13.2}\,{\rm GeV}$.

\begin{figure}[h!]
	\centering
	\subfloat[$R=1$]
	{
		\includegraphics[trim={1cm 0 1.2cm 0},scale=0.54]{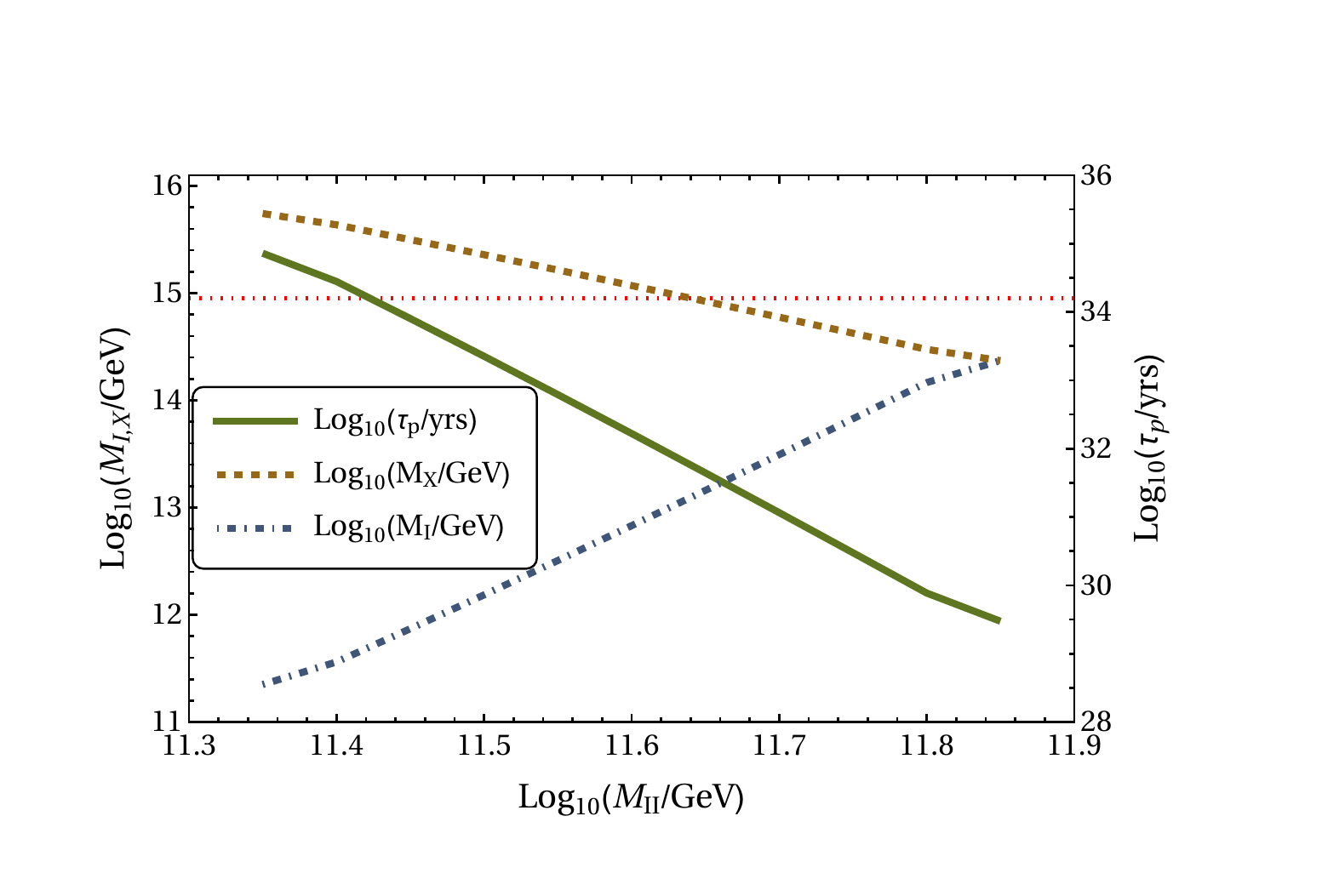}
	} 
	\subfloat[${R\in \left[1/2:2\right]}$]
	{
		\includegraphics[trim={1cm 0 1.2cm 0},scale=0.54]{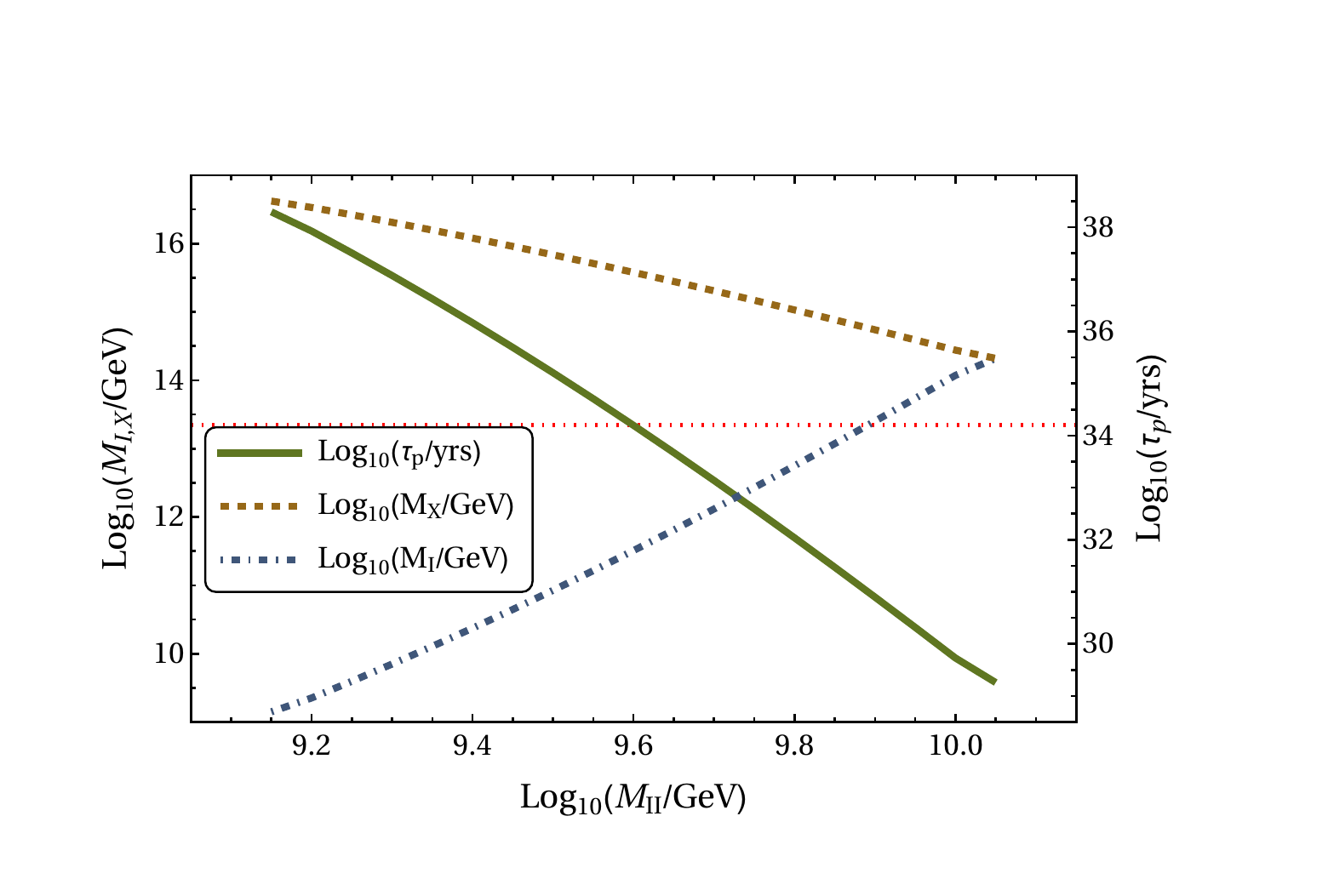}
	} 
	\caption{Proton lifetime prediction in case following breaking chain: $SO(10)\to\mathcal{G}_{2_L 2_R 4_C \slashed{D}}\to \mathcal{G}_{2_L 1_R 4_C}\to \rm{SM}$. The proton lifetime constraint ($\geq 1.6 \times 10^{34}$ yrs) allows a very small range of unification solutions in the absence of threshold correction ($R=1$). But once the threshold correction is incorporated and $R$ is being varied between $[1/2:2]$, we have noted an improvement in the unification solution. It is clearly evident that the threshold correction allow more proton lifetime compatible unification solutions.	}
	\label{fig:so10to224nD}
\end{figure}

In Fig.~\ref{fig:so10to224D}, the following  breaking chain: $SO(10) \to \mathcal{G}_{2_L 2_R 4_C D} \to \mathcal{G}_{2_L 2_R 3_C 1_{B-L}D}\to{\rm SM}$ is considered. The plot in Fig.~\ref{fig:so10to224D}(a) shows the solution space for $R=1$, i.e., in absence of threshold correction, and it is quite clear that all the solution space is below the $\tau_p$ limit and thus ruled out. Now in Fig.~\ref{fig:so10to224D}(b) we have noted the solution space when the maximal threshold correction (as $R$ is varied in range of $[1/2:2]$) is incorporated. This clearly shows that now we have $\tau_p$ compatible unification solution for 
$10^{9.8}\,{\rm GeV} < M_{II} < 10^{11.1}\,{\rm GeV}$.

In Fig.~\ref{fig:so10to214}, we have considered the breaking chain: $SO(10)\to\mathcal{G}_{2_L 2_R 4_C D}\to \mathcal{G}_{2_L 1_R 4_C}\to \rm{SM}$. The plot in Fig.~\ref{fig:so10to214}(a) shows the solution space for $R=1$, i.e., in absence of threshold correction, and it is quite clear that all the solution space is below the $\tau_p$ limit and thus ruled out. Now in Fig.~\ref{fig:so10to214}(b) we have noted the unlike the other cases even after inclusion of maximal  threshold correction (as $R$ is varied in a range of $[1/10:10]$) solution space is improved but still ruled out. Thus this model cannot be saved by this amount of threshold correction.

In Fig.~\ref{fig:so10to224nD}, we have considered the breaking chain: $SO(10)\to\mathcal{G}_{2_L 2_R 4_C \slashed{D}}\to \mathcal{G}_{2_L 1_R 4_C}\to \rm{SM}$. The plot in Fig.~\ref{fig:so10to224nD}(a) shows the solution space for $R=1$, i.e., in absence of threshold correction, and it is quite clear that most of  the solution space is below the $\tau_p$ limit and thus ruled out. Only allowed regime is $10^{11.35}\,{\rm GeV} < M_{II} < 10^{11.42}\,{\rm GeV}$. 
Now in Fig.~\ref{fig:so10to224nD}(b) we have noted the solution space when the minimal threshold correction (as $R$ is varied in range of $[1/2:2]$) is incorporated. This clearly shows that now we have $\tau_p$ compatible unification solution for 
$10^{9.2}\,{\rm GeV} < M_{II} < 10^{9.6}\,{\rm GeV}$.


\section{Summary and Conclusion}
\label{conclusions}
In this paper, we have analysed the unification scenario for non-supersymmetric $SO(10)$ and $E(6)$ GUT groups which are broken spontaneously to the Standard Model through one and two intermediate symmetries. We have focussed on those breaking chain where the GUT groups are broken in the form of $SU(N)_L\otimes SU(N)_R\otimes \mathcal{G}$, where $\mathcal{G}$ is a single or product group.
For each two-step breaking chain we have catalogued all possible topological defects which can emerge during the process of spontaneous symmetry breaking at different scales.

We have computed the two-loop beta coefficients for two intermediate scale scenarios, and performed a goodness of fit test to find out the unification solutions in terms of the unification ($M_X)$, intermediate ($M_I,M_{II}$) scales and also unified coupling. For each such case, we have estimated the proton decay lifetime by constructing the dimension-6 proton decay operators and considering their running. We have also computed the anomalous dimension matrix for each such case to perform RGEs of the proton decay operators. In the absence of any threshold correction, 
we have noted that the unification solutions in the case of non-supersymmetric GUTs in presence of one (see Ref.~\cite{Chakrabortty:2017mgi}), and two intermediate scales are mostly incompatible with the bound from proton decay lifetime. However, by including threshold corrections, we have found that many of these models can be revived.
In particular, for the models which are incompatible with bound on $\tau_p$, we have estimated the 
minimal requirement of threshold correction such that these models can be revived, in terms of 
the ratio ($R$) of the heavy scalar and fermion fields to the superheavy gauge bosons, 
assumed degenerate with the symmetry breaking scale.
Choosing two different sets of $R$ $\in [1/2:2]$, and $[1/10:10]$, we have noticed that most of the scenarios can be made safe from the  proton lifetime bound apart from $SO(10)\to\mathcal{G}_{2_L 2_R 4_C D}\to \mathcal{G}_{2_L 1_R 4_C}\to \rm{SM}$. Here, the improved solution space is still not compatible with the $\tau_p$ constraint. 

In conclusion, although most of the non-supersymmetric GUT scenarios with one, and two intermediate scales are not compatible with the proton decay lifetime in absence of threshold correction, many of these cases become viable once
threshold corrections are correctly taken into account in a consistent way. 
We conclude that threshold corrections are a saviour for many non-SUSY GUTs.
 

\subsection*{Acknowledgements}
J.C., and R.M. are supported by the Department of Science and Technology,
Government of India, under the Grant IFA12-PH-34 (INSPIRE Faculty Award); and the Science and Engineering
Research Board, Government of India, under the agreement SERB/PHY/2016348 (Early Career Research Award). R.M. acknowledges the useful discussions with Triparno Bandyopadhyay, Sunando Kumar Patra, and Tripurari Srivastava.
S.F.K. acknowledges the STFC Consolidated Grant ST/L000296/1 and the European Union's Horizon 2020 Research and Innovation programme under Marie Sk\l{}odowska-Curie grant agreements Elusives ITN No.\ 674896 and InvisiblesPlus RISE No.\ 690575.


\appendix
\numberwithin{equation}{section}
\numberwithin{table}{section}

\section*{APPENDIX}
\label{sec:appendix}
\section{Algorithm to calculate the one loop anomalous dimensions}
\begin{figure}[h!]
\centering
\subfloat[$\mathcal{O}\left( e^C_\alpha, d_\beta\right)$]
{
\includegraphics[scale=0.6]{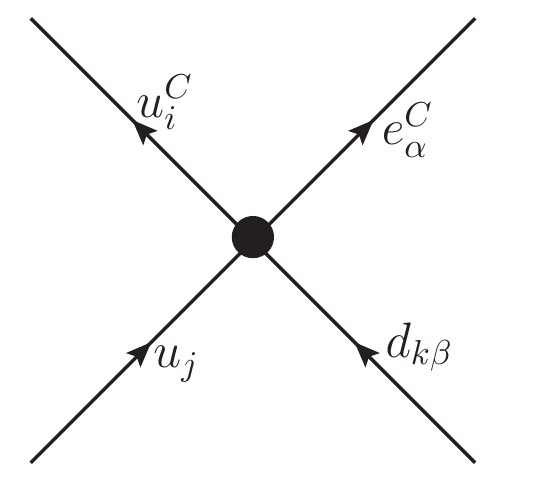}
\label{Operator_I}
}
\subfloat[$\mathcal{O}\left(e_\alpha , d_\beta^c \right)$]
{
\includegraphics[scale=0.6]{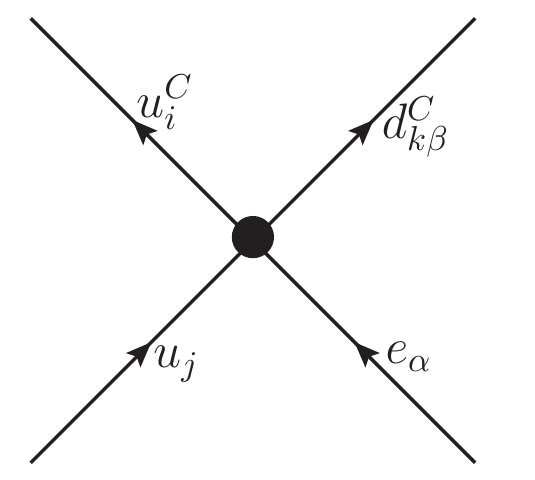}
\label{Operator_II}
}
\subfloat[$\mathcal{O}\left(\nu_l , d_\alpha , d_\beta^c\right)$]
{
\includegraphics[scale=0.6]{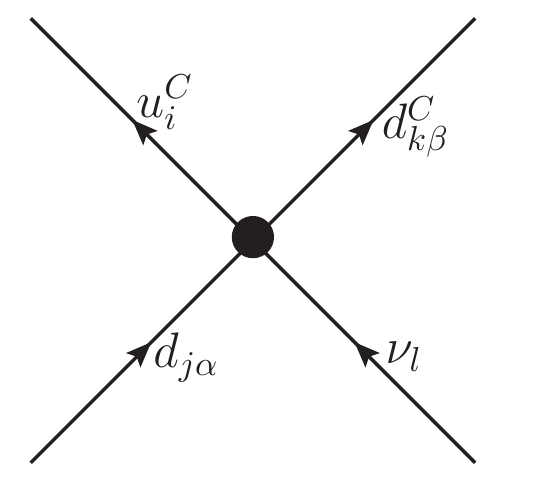}
\label{Operator_III}
}
\subfloat[$\mathcal{O}\left(\nu_l^c , d_\alpha , d_\beta^c\right)$]
{
\includegraphics[scale=0.6]{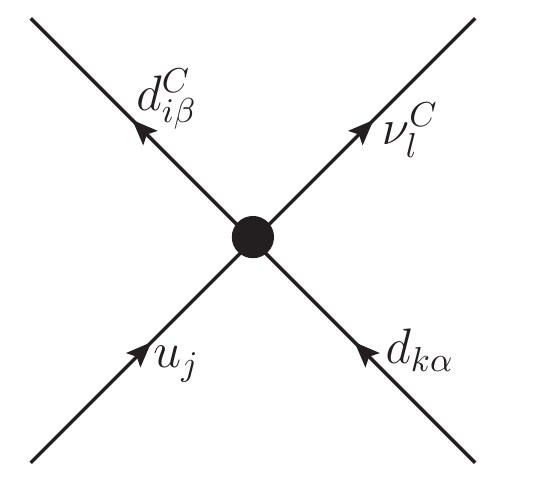}
\label{Operator_IV}
}
\caption{Proton decay operators at tree level}
\label{proton_decay_operators}
\end{figure}
\begin{figure}[h!]
\centering
\subfloat[Vertex of the type I]
{
\includegraphics[scale=0.6]{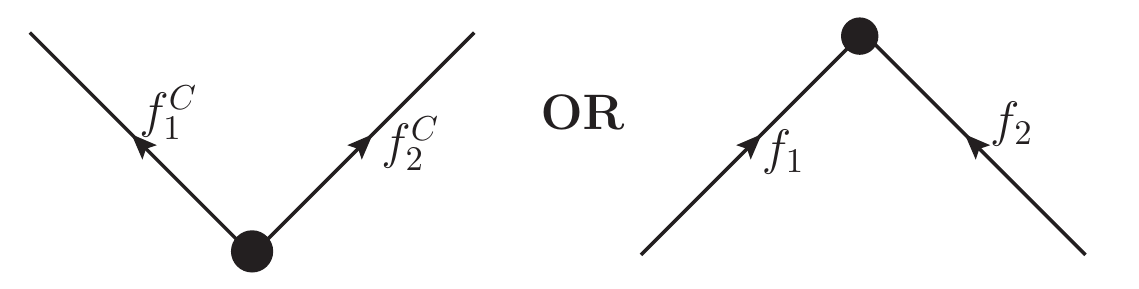}
}\hspace{5em}
\subfloat[Vertex of the type II]
{
\includegraphics[scale=0.6]{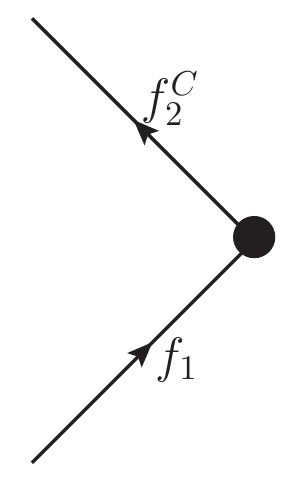}
}\hspace{5em}
\caption{Two types of vertices appearing in the operators. }\label{generic_vertices}
\end{figure}
\begin{figure}[h!]
\centering
\subfloat[Vertex correction of type I]
{
\includegraphics[scale=0.6]{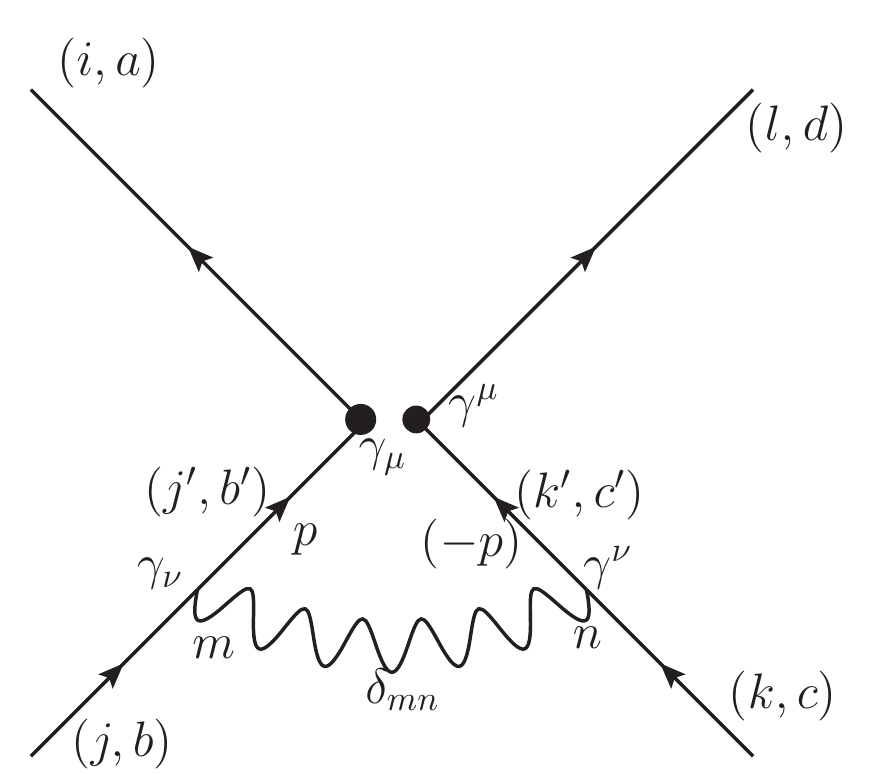}
\label{correction_I}
}
\hspace*{5em}
\subfloat[Vertex correction of type II]
{
\includegraphics[scale=0.6]{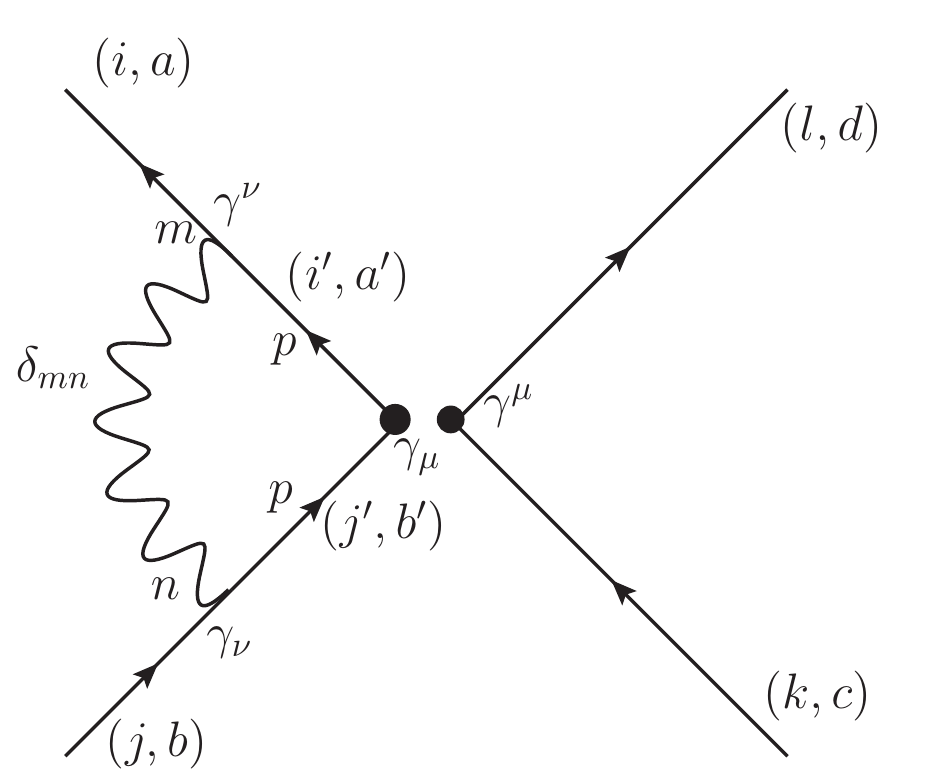}
\label{correction_II}
}
\caption{Generic vertex corrections to the operators. The two indices within the parentheses indicate the representation index and the Dirac index respectively of each fermion. $m$ and $n$ are the group indices.}\label{generic_vertex_correction}
\end{figure}

The dimension-6 effective operators that induce proton decay are listed in  Fig.~\ref{proton_decay_operators}.  These effective operators are accompanied by the relevant Wilson coefficients at low scale. But to compute the prediction for proton decay for an unified scenario we need to incorporate the renormalisation group evolutions of these Wilson coefficients. This can be done by considering quantum corrections of these operators (vertex corrections and the self-energy corrections) leading to computation of anomalous dimension matrix ($\gamma_{ij}$ in Eqn.~\ref{short_range_re_factor}) for these set of operators. To simplify the computation without loosing out any generalisation we have set external momenta and masses to be zero. The necessary vertex corrections are given in Fig.~\ref{generic_vertex_correction}. There are two different types of vertices occur here, see Fig.~\ref{generic_vertices}.

The self-energy correction is captured in $C_2(R)$ when the fields are in $R$-dimensional representation of $SU(N)$. The vertex correction is encapsulated in a combined factor due to: $\left[(\text{Dirac algebra)}\otimes (\text{color algebra})\right]$.

The Dirac algebra factor is independent of the gauge symmetry. To compute this factor for the type-I vertex (see Fig.~\ref{correction_I}) we can write
\begin{align}
d_I (\bar{f^c_1}_L\gamma_\mu f_{2L}) (\bar{f^c_4}_L\gamma^\mu f_{3L}) &= (\bar{f^c_1}_L\gamma_\mu \slashed{p} \gamma_\nu f_{2L}) \frac{1}{p^2} (\bar{f^c_4}_L\gamma^\mu (-\slashed{p}) \gamma^\nu f_{3L})\nonumber \\
&= (-4)(\bar{f^c_1}_L\gamma_\mu f_{2L}) (\bar{f^c_4}_L\gamma^\mu f_{3L}) \ .
\end{align}
Thus the Dirac algebra factor  for the type-I vertex is $d_I = -4$. Similarly, in the case of type-II vertex, we have $d_{II}=1$.

Now we will concentrate in the color factor computation part. For a given gauge group $SU(N)$, we have noted the color  factors are $\left(-\frac{N+1}{2N}\right)$, and $\left(\frac{N+1}{2N}\right)$ for the type-I, and type-II vertices.

\pagebreak

 For example in presence of $SU(N)$ gauge theory, with $n_I$ and $n_{II}$ number of  vertex of type-I, and  type-II where $n_f$ fermions receive the self-energy corrections due to the the gauge bosons,  the anomalous dimension is given as
\begin{align}
\gamma_N = & n_I \times\underbrace{ (-4)}_{\text{Dirac algebra factor}}\times \underbrace{\left( - \frac{N+1}{2N} \right)}_{\text{color algebra factor}} \nonumber \\  & + \  n_{II} \times \underbrace{(1)}_{\text{Dirac algebra factor}} \times \underbrace{\left( \frac{N+1}{2N} \right)}_{\text{color algebra factor}} - \; \frac{n_f}{2} \left( \frac{N^2-1}{2N} \right).
\end{align}

We must mention that one needs to modify the algorithm for the gauge symmetry $\mathcal{G}_{3_L 3_R 3_C}$, and the flipped $\mathcal{G}_{2_L 2_R 4_C 1_X}$. Specifically for these type of scenarios, we need to first construct the parent operators, and then calculate the color factors. Thus we prefer to provide their structures for these two cases explicitly below.

The fermion representations under the gauge group $\mathcal{G}_{3_L 3_R 3_C}$  transform as:  $(\bar{3},3,1)\equiv L_{\alpha_R}^{\alpha_L}$, $(3,1,3)\equiv Q_{\alpha_L \alpha_C}$ and, $(1,\bar{3},\bar{3})\equiv Q_C^{\alpha_R\alpha_C}$. The parent operators leading to  the proton decay   ($p\to e^+ \pi^0$) are given in flavour basis as:
\begin{align}
\mathcal{O}_L^{d=6}(e^C,d) & \subset \epsilon^{\alpha_L\beta_L\gamma_L}\delta_{\beta_R}^{\alpha_R}\epsilon^{\alpha_C\beta_C\gamma_C}\left( \overline{Q_{C}}_{\alpha_R\alpha_C}\gamma^\mu Q_{\alpha_L\beta_C}\right) \left(\overline{L}^{\beta_R}_{\beta_L} \gamma_\mu Q_{\gamma_L\gamma_C} \right), \nonumber \\
\mathcal{O}_R^{d=6}(e,d^C) & \subset \delta_{\beta_L}^{\alpha_L} \epsilon^{\alpha_R\beta_R\gamma_R}\epsilon^{\alpha_C\beta_C\gamma_C}\left( \overline{Q_{C}}_{\alpha_R\alpha_C}\gamma^\mu Q_{\alpha_L\beta_C}\right)\left( \overline{Q_{C}}_{\beta_R\gamma_C}\gamma^\mu L_{\gamma_R}^{\beta_L}\right). 
\end{align}

While for the  flipped scenario $\mathcal{G}_{2_L 2_R 4_C 1_X}$, the similar relevant parent operators for $p\to e^+ \pi^0$ decay in flavour basis are given as:
\begin{align}
\mathcal{O}_L^{d=6}(e^C,d) & \subset \epsilon^{abcd}\epsilon^{ij} \left(\bar{U}_{ab}\gamma^\mu Q_{ic}\right)\left(\bar{E}\gamma_\mu Q_{jd}\right) , \nonumber \\
\mathcal{O}_R^{d=6}(e,d^C) & \subset \epsilon^{abcd}\epsilon^{ij}\epsilon^{\alpha\beta} \left(\bar{U}_{ab}\gamma^\mu Q_{ic}\right)\left(\overline{DN}_{\alpha d}\gamma_\mu L_{j\beta}\right). 
\end{align}
Here, $\lbrace i,j \rbrace$, $\lbrace \alpha, \beta \rbrace $, and $\lbrace a,b,c,d \rbrace$ denote the $SU(2)_L$, $SU(2)_R$, and $SU(4)_C$ indices respectively. The representations of  fermion multiplet under this flipped gauge group are given as: $(1,1,1,4)\equiv E$, $(2,2,1,-2)\equiv L_{i\alpha}$, $(1,1,6,-2)\equiv U_{ab}$, $(2,1,4,1)\equiv Q_{ia}$, and $(1,2,\bar{4},1)\equiv DN^a_\alpha$.

\section{Threshold corrections ($\Lambda_i$'s) for  one  intermediate step breaking scenario }
In this section, we have enlisted the threshold corrections ($\Lambda$'s) that arise when the heavy scalars and fermions are integrated out. These particles have nondegenerate masses different from the symmetry breaking scales. These corrections modify the matching conditions, see Eqns.~\ref{matching}, and~\ref{lamda}.  We have assumed that all heavy gauge bosons have same masses degenerate with the breaking scale. The computation requires the information regarding the index  and normalizations of representations which are provided in Tables~\ref{index_repn}, and \ref{abelian_normn}.

\subsection*{$\mathbf{SO(10)\to \mathcal{G}_{2_L 2_R 4_C D} \to {\rm \bf{SM}}}$}

The threshold corrections arise after integrating out the heavy scalar fields are tabulated in the Table~\ref{SO10-G224D}.

\begin{table}[h!]
	
	\begin{center}
		
		\begin{tabular}{|c|c|c|c|}
			\hline
			& $SO(10)$ & $\mathcal{G}_{2_L2_R4_CD}$ & $\mathcal{G}_{2_L1_Y3_C}$ \\
			\hline
			Scalars & $\mathbf{10}$ & $\mathbf{(2,2,1)}$ & $\mathbf{(2,\pm{\frac{1}{2}},1)}$ \\
			&  & $(1,1,6)$ &  \\
			\cline{2-4}
		    & $\mathbf{\overline{126}}$ & $\mathbf{(1,3,\overline{10})}$ & $(1,2,1)$ \\
		    &  &  & $(1,\frac{4}{3},\overline{3})$ \\
		    &  &  & $(1,\frac{2}{3},\overline{6})$ \\
		    &  &  & $(1,1,1)_{GB}$ \\
		    &  &  & $(1,\frac{1}{3},\overline{3})$ \\
		    &  &  & $(1,-\frac{1}{3},\overline{6})$ \\
		    &  &  & $(1,0,1)_{GB}$ \\
		    &  &  & $(1,-\frac{2}{3},\overline{3})_{GB}$ \\
		    &  &  & $(1,-\frac{4}{3},\overline{6})$ \\
		    \cline{3-4}
		    &  & $\mathbf{(3,1,10)_D}$ & $(3,-1,1)$ \\
		    &  &  & $(3,-\frac{1}{3},3)$ \\
		    &  &  & $(3,\frac{1}{3},6)$ \\
		    \cline{3-4}
		    &  & $(2,2,15)$ &  \\
		    &  & $(1,1,6)$ &  \\
		    
		    \cline{2-4}
		    & $\mathbf{54}$ & $(1,1,1)$ &  \\
		    &  & $(2,2,6)_{GB}$ &  \\
		    &  & $(3,3,1)$ &  \\
		    &  & $(1,1,20')$ &  \\
			\hline
		\end{tabular}
		\caption{$SO(10)\to \mathcal{G}_{2_L2_R4_CD}\to\rm SM $. The bold multiplets contribute to the RGE and others are the heavy degrees of freedom which are integrated out. }\label{SO10-G224D}
		
	\end{center}
	
\end{table}

\begin{center}
\underline{Threshold corrections at $M_X$}
\end{center}
\input{"thrcr_I/thrcr_mx_SO10-G224D.tex"}
\begin{center}
	\pagebreak
\underline{Threshold corrections at $M_I$}
\end{center}
\input{"thrcr_I/thrcr_mr_SO10-G224D.tex"}
\subsection*{$\mathbf{SO(10)\to \mathcal{G}_{2_L 2_R 4_C \slashed{D}} \to {\rm \bf{SM}}}$}

The threshold corrections arise after integrating out the heavy scalar fields are tabulated in the Table~\ref{SO10-G224}.

\begin{table}[h!]
	
	\begin{center}
		
		\begin{tabular}{|c|c|c|c|}
			\hline
			& $SO(10)$ & $\mathcal{G}_{2_L2_R4_C\slashed{D}}$ & $\mathcal{G}_{2_L1_Y3_C}$ \\
			\hline
			Scalars & $\mathbf{10}$ & $\mathbf{(2,2,1)}$ & $\mathbf{(2,\pm{\frac{1}{2}},1)}$ \\
			&  & $(1,1,6)$ &  \\
			\cline{2-4}
		    & $\mathbf{\overline{126}}$ & $\mathbf{(1,3,\overline{10})}$ & $(1,2,1)$ \\
		    &  &  & $(1,\frac{4}{3},\overline{3})$ \\
		    &  &  & $(1,\frac{2}{3},\overline{6})$ \\
		    &  &  & $(1,1,1)_{GB}$ \\
		    &  &  & $(1,\frac{1}{3},\overline{3})$ \\
		    &  &  & $(1,-\frac{1}{3},\overline{6})$ \\
		    &  &  & $(1,0,1)_{GB}$ \\
		    &  &  & $(1,-\frac{2}{3},\overline{3})_{GB}$ \\
		    &  &  & $(1,-\frac{4}{3},\overline{6})$ \\
		    \cline{3-4}
		    &  & $(3,1,10)$ &  \\
		    
		    &  & $(2,2,15)$ &  \\
		    &  & $(1,1,6)$ &  \\
		    \cline{2-4}
		    & $\mathbf{210}$ & $(1,1,1)$ &  \\
		    &  & $(1,1,15)$ &  \\
		    &  & $(3,1,15)$ &  \\
		    &  & $(1,3,15)$ &  \\
		    &  & $(2,2,6)_{GB}$ &  \\
		    &  & $(2,2,10)$ &  \\
		    &  & $(2,2,\overline{10})$ &  \\
		    
			\hline
		\end{tabular}
		\caption{$SO(10)\to \mathcal{G}_{2_L2_R4_C\slashed{D}}\to\rm SM $. The bold multiplets contribute to the RGE and others are the heavy degrees of freedom which are integrated out. }\label{SO10-G224}
		
	\end{center}
	
\end{table}

\begin{center}
\underline{Threshold corrections at $M_X$}
\end{center}
\input{"thrcr_I/thrcr_mx_SO10-G224.tex"}
\begin{center}
	\underline{Threshold corrections at $M_I$}
\end{center}
\input{"thrcr_I/thrcr_mr_SO10-G224.tex"}
\subsection*{$\mathbf{SO(10)\to \mathcal{G}_{2_L 2_R 3_C 1_{B-L} D} \to {\rm \bf{SM}}}$}

The threshold corrections arise after integrating out the heavy scalar fields are tabulated in the Table~\ref{SO10-G2231D}.

\begin{table}[h!]
	
	\begin{center}
		
		\begin{tabular}{|c|c|c|c|}
			\hline
			& $SO(10)$ & $\mathcal{G}_{2_L2_R3_C1_{B-L}D}$ & $\mathcal{G}_{2_L1_Y3_C}$ \\
			\hline
			Scalars & $\mathbf{10}$ & $\mathbf{(2,2,1,0)}$ & $\mathbf{(2,\pm{\frac{1}{2}},1)}$ \\
			&  & $(1,1,3,-\frac{2}{3})$ &  \\
		    &  & $(1,1,\overline{3},\frac{2}{3})$ &  \\
			\cline{2-4}
		    & $\mathbf{\overline{126}}$ & $\mathbf{(1,3,1,2)}$ & $(1,2,1)$ \\
		    &  &  & $(1,1,1)_{GB}$ \\
		    &  &  & $(1,0,1)_{GB}$ \\
		    \cline{3-4}
		    &  & $\mathbf{(3,1,1,-2)_D}$ & $(3,-1,1)$ \\
		    \cline{3-4}
		    &  & $(1,3,\overline{3},\frac{2}{3})$ &  \\
		    &  & $(1,3,\overline{6},-\frac{2}{3})$ &  \\
		    
		    &  & $(3,1,3,-\frac{2}{3})$ &  \\
		    &  & $(3,1,6,\frac{2}{3})$ &  \\
		    
		    &  & $(2,2,1,0)$ &  \\
		    &  & $(2,2,3,\frac{4}{3})$ &  \\
		    &  & $(2,2,\overline{3},-\frac{4}{3})$ &  \\
		    &  & $(2,2,8,0)$ &  \\
		    &  & $(1,1,3,-\frac{2}{3})$ &  \\
		    &  & $(1,1,\overline{3},\frac{2}{3})$ &  \\
		    \cline{2-4}
		    & $\mathbf{210}$ & $(1,1,1,0)$ &  \\
		    &  & $(1,1,8,0)$ &  \\
		    &  & $(1,1,3,\frac{4}{3})_{GB}$ &  \\
		    &  & $(1,1,\overline{3},-\frac{4}{3})_{GB}$ &  \\
		    &  & $(1,3,1,0)$ &  \\
		    &  & $(1,3,3,\frac{4}{3})$ &  \\
		    &  & $(1,3,\overline{3},-\frac{4}{3})$ &  \\
		    &  & $(1,3,8,0)$ &  \\
		    &  & $(3,1,1,0)$ &  \\
		    &  & $(3,1,3,\frac{4}{3})$ &  \\
		    &  & $(3,1,\overline{3},-\frac{4}{3})$ &  \\
		    &  & $(3,1,8,0)$ &  \\
		    &  & $(2,2,3,-\frac{2}{3})_{GB}$ &  \\
		    &  & $(2,2,\overline{3},\frac{2}{3})_{GB}$ &  \\
		    &  & $(2,2,1,2)$ &  \\
		    &  & $(2,2,3,-\frac{2}{3})$ &  \\
		    &  & $(2,2,6,\frac{2}{3})$ &  \\
		    &  & $(2,2,1,-2)$ &  \\
		    &  & $(2,2,\overline{3},\frac{2}{3})$ &  \\
		    &  & $(2,2,\overline{6},-\frac{2}{3})$ &  \\
			\hline
		\end{tabular}
		\caption{$SO(10)\to \mathcal{G}_{2_L2_R3_C1_{B-L}D}\to\rm SM $. The bold multiplets contribute to the RGE and others are the heavy degrees of freedom which are integrated out. }\label{SO10-G2231D}
		
	\end{center}
	
\end{table}

\begin{center}
\underline{Threshold corrections at $M_X$}
\end{center}
\input{"thrcr_I/thrcr_mx_SO10-G2231D.tex"}
\begin{center}
\underline{Threshold corrections at $M_I$}
\end{center}
\input{"thrcr_I/thrcr_mr_SO10-G2231D.tex"}
\subsection*{$\mathbf{SO(10)\to \mathcal{G}_{2_L 2_R 3_C 1_{B-L} \slashed{D}} \to {\rm \bf{SM}}}$}

The threshold corrections arise after integrating out the heavy scalar fields are tabulated in the Table~\ref{SO10-G2231}.

\begin{table}[h!]
	
	\begin{center}
		
		\begin{tabular}{|c|c|c|c|}
			\hline
			& $SO(10)$ & $\mathcal{G}_{2_L2_R3_C1_{B-L}\slashed{D}}$ & $\mathcal{G}_{2_L1_Y3_C}$ \\
			\hline
			Scalars & $\mathbf{10}$ & $\mathbf{(2,2,1,0)}$ & $\mathbf{(2,\pm{\frac{1}{2}},1)}$ \\
			&  & $(1,1,3,-\frac{2}{3})$ &  \\
		    &  & $(1,1,\overline{3},\frac{2}{3})$ &  \\
			\cline{2-4}
		    & $\mathbf{\overline{126}}$ & $\mathbf{(1,3,1,2)}$ & $(1,2,1)$ \\
		    &  &  & $(1,1,1)_{GB}$ \\
		    &  &  & $(1,0,1)_{GB}$ \\
		    \cline{3-4}
		    &  & $(3,1,1,-2)$ &  \\
		    &  & $(1,3,\overline{3},\frac{2}{3})$ &  \\
		    &  & $(1,3,\overline{6},-\frac{2}{3})$ &  \\
		    
		    &  & $(3,1,3,-\frac{2}{3})$ &  \\
		    &  & $(3,1,6,\frac{2}{3})$ &  \\
		    
		    &  & $(2,2,1,0)$ &  \\
		    &  & $(2,2,3,\frac{4}{3})$ &  \\
		    &  & $(2,2,\overline{3},-\frac{4}{3})$ &  \\
		    &  & $(2,2,8,0)$ &  \\
		    &  & $(1,1,3,-\frac{2}{3})$ &  \\
		    &  & $(1,1,\overline{3},\frac{2}{3})$ &  \\
		    \cline{2-4}
		    & $\mathbf{45}$ & $(1,1,1,0)$ &  \\
		    &  & $(1,1,8,0)$ &  \\
		    &  & $(1,1,3,\frac{4}{3})_{GB}$ &  \\
		    &  & $(1,1,\overline{3},-\frac{4}{3})_{GB}$ &  \\
		    &  & $(1,3,1,0)$ &  \\
		    &  & $(3,1,1,0)$ &  \\
		    &  & $(2,2,3,-\frac{2}{3})$ &  \\
		    &  & $(2,2,\overline{3},\frac{2}{3})$ &  \\
		    
			\hline
		\end{tabular}
		\caption{$SO(10)\to \mathcal{G}_{2_L2_R3_C1_{B-L}\slashed{D}}\to\rm SM $. The bold multiplets contribute to the RGE and others are the heavy degrees of freedom which are integrated out. }\label{SO10-G2231}
		
	\end{center}
	
\end{table}

\begin{center}
\underline{Threshold corrections at $M_X$}
\end{center}
\input{"thrcr_I/thrcr_mx_SO10-G2231.tex"}
\begin{center}
	\underline{Threshold corrections at $M_I$}
\end{center}
\input{"thrcr_I/thrcr_mr_SO10-G2231.tex"}

\subsection*{$\mathbf{E(6)\to \mathcal{G}_{2_L 2_R 4_C 1_X D} \to {\rm \bf{SM}}}$}
\begin{table}[h!]
	
	The threshold corrections arise after integrating out the heavy scalar fields are tabulated in the Table~\ref{E6-G2241D}.
	
	\begin{center}
		
		\begin{tabular}{|c|c|c|c|}
			\hline
			& $E(6)$ & $\mathcal{G}_{2_L2_R4_C1_X D}$ & $\mathcal{G}_{2_L1_Y3_C}$ \\
			\hline
			Fermions & $\mathbf{27}$ & $\mathbf{(1,1,1,4)}$ & $\mathbf{(1,1,1)}$ \\
			&  & $\mathbf{(2,2,1,-2)}$ & $\mathbf{(2,-\frac{1}{2},1)}$ \\
			&  & $\mathbf{(1,1,6,-2)}$ & $\mathbf{(1,-\frac{2}{3},\overline{3})}$ \\
			&  &  & $(1,-\frac{1}{3},3)$ \\
			&  & $\mathbf{(2,1,4,1)}$ & $\mathbf{(2,\frac{1}{6},3)}$ \\
		    &  &  & $(2,\frac{1}{2},1)$ \\
		    &  & $\mathbf{(1,2,\overline{4},1)}$ & $\mathbf{(1,0,1)}$ \\
		    &  &  & $\mathbf{(1,\frac{1}{3},\overline{3})}$ \\
			\hline
			
		    Scalars & $\mathbf{27}$ & $\mathbf{(2,2,1,-2)}$ & $\mathbf{(2,-\frac{1}{2},1)}$ \\
		    &  & $(1,1,6,-2)$ &  \\
		    &  & $(1,1,1,4)$ &  \\
		    &  & $(2,1,4,1)$ &  \\
		    &  & $(1,2,\overline{4},1)$ &  \\
		    \cline{2-4}
		    & $\mathbf{27}$ & $\mathbf{(1,2,\overline{4},1)}$ & $(1,0,1)$ \\
		    &  &  & $(1,\frac{1}{3},\overline{3})_{GB}$ \\
		    &  & $\mathbf{(2,1,4,1)_D}$ & $(2,\frac{1}{6},3)$ \\
		    &  &  & $(2,\frac{1}{2},1)$ \\
		    &  & $(2,2,1,-2)$ &  \\
		    &  & $(1,1,6,-2)$ &  \\
		    &  & $(1,1,1,4)$ &  \\

			\hline
		\end{tabular}
		\caption{(To be continued)}
		
	\end{center}
	
\end{table}
\begin{table}[h!]\ContinuedFloat
	
	\begin{center}
		
		\begin{tabular}{|c|c|c|c|}
			\hline
			& $E(6)$ & $\mathcal{G}_{2_L2_R4_C1_X D}$ & $\mathcal{G}_{2_L1_Y3_C}$ \\
			\hline
			Scalars & $\mathbf{650}$ & $(1,1,1,0)$ &  \\
		    &  & $(2,2,1,6)$ &  \\
		    &  & $(1,1,6,6)$ &  \\
		    &  & $(2,1,4,-3)_{GB}$ &  \\
		    &  & $(1,2,\overline{4},-3)_{GB}$ &  \\
		    &  & $(1,3,1,0)$ &  \\
		    &  & $(3,1,1,0)$ &  \\
		    &  & $(1,1,15,0)$ &  \\
		    &  & $(2,2,6,0)$ &  \\
		    &  & $(1,1,1,0)_{D}$ &  \\
		    &  & $(3,3,1,0)$ &  \\
		    &  & $(1,1,20',0)$ &  \\
		    &  & $(2,2,6,0)_{GB}$ &  \\
		    &  & $(2,1,4,-3)$ &  \\
		    &  & $(1,2,\overline{4},-3)$ &  \\
		    &  & $(2,3,4,-3)$ &  \\
		    &  & $(3,2,\overline{4},-3)$ &  \\
		    &  & $(2,1,20,-3)$ &  \\
		    &  & $(1,2,\overline{20},-3)$ &  \\
		    &  & $(1,1,1,0)_{\slashed{D}}$ &  \\
		    &  & $(1,1,15,0)$ &  \\
		    &  & $(3,1,15,0)$ &  \\
		    &  & $(1,3,15,0)$ &  \\
		    &  & $(2,2,6,0)$ &  \\
		    &  & $(2,2,10,0)$ &  \\
		    &  & $(2,2,\overline{10},0)$ &  \\
			\hline
		\end{tabular}
		\caption{$E(6)\to \mathcal{G}_{2_L2_R4_C1_X D}\to\rm SM $. The bold multiplets contribute to the RGE and others are the heavy degrees of freedom which are integrated out. }\label{E6-G2241D}
		
	\end{center}
	
\end{table}

\begin{center}
\underline{Threshold corrections at $M_X$}
\end{center}
\input{"thrcr_I/thrcr_mx_E6-G2241D.tex"}
\begin{center}
\underline{Threshold corrections at $M_I$}
\end{center}
\input{"thrcr_I/thrcr_mr_E6-G2241D.tex"}
\subsection*{$\mathbf{E(6)\to \mathcal{G}_{2_L 2_R 4_C 1_X \slashed{D}} \to {\rm \bf{SM}}}$}
\begin{center}
\underline{Threshold corrections at $M_X$}
\end{center}
\input{"thrcr_I/thrcr_mx_E6-G2241.tex"}
\begin{center}
\underline{Threshold corrections at $M_I$}
\end{center}
\input{"thrcr_I/thrcr_mr_E6-G2241.tex"}

\section{Threshold corrections ($\Lambda_i$'s) for two intermediate step breaking scenario}
In this section we have quoted the threshold corrections that arise in terms of $\Lambda$'s when all the heavy scalars and fermions have nondegenerate masses different from the symmetry breaking scales in the RGE. We have assumed that all heavy gauge bosons have same masses degenerate with the breaking scale.
As a result the matching conditions in the Eqn.~\ref{matching_g224_g2231} get modified as happen for one intermediate cases also. The detailed structure of these threshold corrections for the considered breaking chains are given below.
\subsection*{$\mathbf{SO(10)\to \mathcal{G}_{2_L 2_R 4_C D}\to \mathcal{G}_{2_L 2_R 3_C 1_{B-L}D}\to {\rm \bf{SM}}}$} 

The threshold corrections arise after integrating out the heavy scalar fields are tabulated in the Table~\ref{SO10-G224D-G2231D}.

\begin{table}[h!]
	
	\begin{center}
		
		\begin{tabular}{|c|c|c|c|c|}
			\hline
			& $SO(10)$ & $\mathcal{G}_{2_L2_R4_CD}$ & $\mathcal{G}_{2_L2_R3_C1_{B-L}D}$ & $\mathcal{G}_{2_L1_Y3_C}$ \\
			\hline
			Scalars & $\mathbf{10}$ & $\mathbf{(2,2,1)}$ & $\mathbf{(2,2,1,0)}$ & $\mathbf{(2,\pm{\frac{1}{2}},1)}$ \\
			&  & $(1,1,6)$ &  &  \\
			\cline{2-5}
		    & $\mathbf{\overline{126}}$ & $\mathbf{(1,3,\overline{10})}$ & $\mathbf{(1,3,1,2)}$ & $(1,2,1)$ \\
		    &  &  &  & $(1,1,1)_{GB}$ \\
		    &  &  &  & $(1,0,1)_{GB}$ \\
		    &  &  & $(1,3,\overline{3},\frac{2}{3})$ &  \\
		    &  &  & $(1,3,\overline{6},-\frac{2}{3})$ &  \\
		    \cline{3-5}
		    &  & $\mathbf{(3,1,10)_D}$ & $\mathbf{(3,1,1,-2)_D}$ & $(3,-1,1)$ \\
		    &  &  & $(3,1,3,-\frac{2}{3})$ &  \\
		    &  &  & $(3,1,6,\frac{2}{3})$ &  \\
		    \cline{3-5}
		    &  & $(2,2,15)$ &  &  \\
		    &  & $(1,1,6)$ &  &  \\
		    \cline{2-5}
		    & $\mathbf{210}$ & $\mathbf{(1,1,15)}$ & $(1,1,1,0)$ &  \\
		    &  &  & $(1,1,8,0)$ &  \\
		    &  &  & $(1,1,3,\frac{4}{3})_{GB}$ &  \\
		    &  &  & $(1,1,\overline{3},-\frac{4}{3})_{GB}$ &  \\
		    \cline{3-5}
		    &  & $(3,1,15)$ &  &  \\
		    &  & $(1,3,15)$ &  &  \\
		    &  & $(2,2,6)$ &  &  \\
		    &  & $(2,2,10)$ &  &  \\
		    &  & $(2,2,\overline{10})$ &  &  \\
		    \cline{2-5}
		    & $\mathbf{54}$ & $(1,1,1)$ &  &  \\
		    &  & $(2,2,6)_{GB}$ &  &  \\
		    &  & $(3,3,1)$ &  &  \\
		    &  & $(1,1,20')$ &  &  \\
			\hline
		\end{tabular}
		\caption{$SO(10)\to \mathcal{G}_{2_L2_R4_CD}\to\mathcal{G}_{2_L2_R3_C1_{B-L}D}\to\rm SM $. The bold multiplets contribute to the RGE and others are the heavy degrees of freedom which are integrated out. }\label{SO10-G224D-G2231D}
		
	\end{center}
	
\end{table}

\pagebreak
\begin{center}
\underline{Threshold corrections at $M_X$}
\end{center}
\input{"thrcr/thrcr_mx_SO10-G224D-G2231D.tex"}
\begin{center}
\underline{Threshold corrections at $M_{II}$}
\end{center}
\input{"thrcr/thrcr_mi_SO10-G224D-G2231D.tex"}
\begin{center}
\underline{Threshold corrections at $M_{II}$}
\end{center}
\input{"thrcr/thrcr_mi1_SO10-G224D-G2231D.tex"}
\subsection*{$\mathbf{SO(10)\to \mathcal{G}_{2_L 2_R 4_C D}\to \mathcal{G}_{2_L 1_R 4_C}\to {\rm \bf{SM}}}$}

The threshold corrections arise after integrating out the heavy scalar fields are tabulated in the Table~\ref{SO10-G224D-G214}.

\begin{table}[h!]
	
	\begin{center}
		
		\begin{tabular}{|c|c|c|c|c|}
			\hline
			& $SO(10)$ & $\mathcal{G}_{2_L2_R4_CD}$ & $\mathcal{G}_{2_L1_R4_C}$ & $\mathcal{G}_{2_L1_Y3_C}$ \\
			\hline
			Scalars & $\mathbf{10}$ & $\mathbf{(2,2,1)}$ & $\mathbf{(2,\pm{\frac{1}{2}},1)}$ & $\mathbf{(2,\pm{\frac{1}{2}},1)}$ \\
			&  & $(1,1,6)$ &  &  \\
			\cline{2-5}
		    & $\mathbf{\overline{126}}$ & $\mathbf{(1,3,\overline{10})}$ & $\mathbf{(1,-1,\overline{10})}$ & $(1,0,1)_{GB}$ \\
		    &  &  &  & $(1,-\frac{2}{3},\overline{3})_{GB}$ \\
		    &  &  &  & $(1,-\frac{4}{3},\overline{6})$ \\
		    &  &  & $(1,0,\overline{10})$ &  \\
		    &  &  & $(1,1,\overline{10})$ &  \\
		    \cline{3-5}
		    &  & $\mathbf{(3,1,10)_D}$ & $(3,0,10)$ &  \\
		    
		    \cline{3-5}
		    &  & $(2,2,15)$ &  &  \\
		    &  & $(1,1,6)$ &  &  \\
		    \cline{2-5}
		    & $\mathbf{45}$ & $\mathbf{(1,3,1)}$ & $(1,-1,1)_{GB}$ &  \\
		    &  &  & $(1,0,1)$ &  \\
		    &  &  & $(1,1,1)_{GB}$ &  \\
		    \cline{3-5}
		    &  & $\mathbf{(3,1,1)_D}$ & $(3,0,1)_D$ &  \\
		    \cline{3-5}
		    &  & $(1,1,15)$ &  &  \\
		    &  & $(2,2,6)$ &  &  \\
		    \cline{2-5}
		    & $\mathbf{54}$ & $(1,1,1)$ &  &  \\
		    &  & $(2,2,6)_{GB}$ &  &  \\
		    &  & $(3,3,1)$ &  &  \\
		    &  & $(1,1,20')$ &  &  \\
		    \cline{2-5}
			\hline
		\end{tabular}
		\caption{$SO(10)\to \mathcal{G}_{2_L2_R4_CD}\to\mathcal{G}_{2_L1_R4_C}\to\rm SM $. The bold multiplets contribute to the RGE and others are the heavy degrees of freedom which are integrated out. }\label{SO10-G224D-G214}
		
	\end{center}
	
\end{table}

\begin{center}
\underline{Threshold corrections at $M_X$}
\end{center}
\input{"thrcr/thrcr_mx_SO10-G224D-G214.tex"}
\pagebreak
\begin{center}
\underline{Threshold corrections at $M_{I}$}
\end{center}
\input{"thrcr/thrcr_mi_SO10-G224D-G214.tex"}
\begin{center}
\underline{Threshold corrections at $M_{II}$}
\end{center}
\input{"thrcr/thrcr_mi1_SO10-G224D-G214.tex"}
\pagebreak
\subsection*{$\mathbf{SO(10)\to \mathcal{G}_{2_L 2_R 4_C\slashed{D} }\to \mathcal{G}_{2_L 1_R 4_C}\to {\rm \bf{SM}}}$}
\begin{table}[h!]
	
	The threshold corrections arise after integrating out the heavy scalar fields are tabulated in the Table~\ref{SO10-G224-G214}.
	
	\begin{center}
		
		\begin{tabular}{|c|c|c|c|c|}
			\hline
			& $SO(10)$ & $\mathcal{G}_{2_L2_R4_C\slashed{D}}$ & $\mathcal{G}_{2_L1_R4_C}$ & $\mathcal{G}_{2_L1_Y3_C}$ \\
			\hline
			Scalars & $\mathbf{10}$ & $\mathbf{(2,2,1)}$ & $\mathbf{(2,\pm{\frac{1}{2}},1)}$ & $\mathbf{(2,\pm{\frac{1}{2}},1)}$ \\
			&  & $(1,1,6)$ &  &  \\
			\cline{2-5}
		    & $\mathbf{\overline{126}}$ & $\mathbf{(1,3,\overline{10})}$ & $\mathbf{(1,-1,\overline{10})}$ & $(1,0,1)_{GB}$ \\
		    &  &  &  & $(1,-\frac{2}{3},\overline{3})_{GB}$ \\
		    &  &  &  & $(1,-\frac{4}{3},\overline{6})$ \\
		    &  &  & $(1,0,\overline{10})$ &  \\
		    &  &  & $(1,1,\overline{10})$ &  \\
		    \cline{3-5}
		    &  & $(3,1,10)$ & &  \\

		    &  & $(2,2,15)$ &  &  \\
		    &  & $(1,1,6)$ &  &  \\
		    \cline{2-5}
		    & $\mathbf{45}$ & $\mathbf{(1,3,1)}$ & $(1,-1,1)_{GB}$ &  \\
		    &  &  & $(1,0,1)$ &  \\
		    &  &  & $(1,1,1)_{GB}$ &  \\
		    \cline{3-5}
		    &  & $(3,1,1)$ &  &  \\
		    &  & $(1,1,15)$ &  &  \\
		    &  & $(2,2,6)$ &  &  \\
		    \cline{2-5}
		    
		    & $\mathbf{210}$ & $(1,1,1)$ &  &  \\
		    &  & $(1,1,15)$ &  &  \\
		    
		    &  & $(3,1,15)$ &  &  \\
		    &  & $(1,3,15)$ &  &  \\
		    &  & $(2,2,6)_{GB}$ &  &  \\
		    &  & $(2,2,10)$ &  &  \\
		    &  & $(2,2,\overline{10})$ &  &  \\
		    \cline{2-5}
			\hline
		\end{tabular}
		\caption{$SO(10)\to \mathcal{G}_{2_L2_R4_C\slashed{D}}\to\mathcal{G}_{2_L1_R4_C}\to\rm SM $. The bold multiplets contribute to the RGE and others are the heavy degrees of freedom which are integrated out. }\label{SO10-G224-G214}
		
	\end{center}
	
\end{table}

\begin{center}
\underline{Threshold corrections at $M_X$}
\end{center}
\input{"thrcr/thrcr_mx_SO10-G224-G214.tex"}
\begin{center}
\underline{Threshold corrections at $M_{I}$}
\end{center}
\input{"thrcr/thrcr_mi_SO10-G224-G214.tex"}
\begin{center}
\underline{Threshold corrections at $M_{II}$}
\end{center}
\input{"thrcr/thrcr_mi1_SO10-G224-G214.tex"}

\subsection*{$\mathbf{E(6)\to \mathcal{G}_{3_L 3_R 3_C D }\to \mathcal{G}_{2_L 2_R 3_C 1_{B-L} D}\to {\rm \bf{SM}}}$}

The threshold corrections arise after integrating out the heavy scalar fields are tabulated in the Table~\ref{E6-G333D-G2231D}.

\begin{table}[h!]
	
	\begin{center}
		
		\begin{tabular}{|c|c|c|c|c|}
			\hline
			& $E(6)$ & $\mathcal{G}_{3_L3_R3_CD}$ & $\mathcal{G}_{2_L2_R3_C 1_{B-L}D}$ & $\mathcal{G}_{2_L1_Y3_C}$ \\
			\hline
			Fermions & $\mathbf{27}$ & $\mathbf{(\bar{3},3,1)}$ & $(1,1,1,0)$ &  \\
			&  &  & $\mathbf{(1,2,1,1)}$ & $(1,1,1)$ \\
			
			&  &  &  & $(1,0,1)$ \\
			&  &  & $\mathbf{(2,1,1,-1)}$ & $(2,-\frac{1}{2},1)$ \\
			&  &  & $(2,2,1,0)$ & \\
			\cline{3-5}
			&  & $\mathbf{(3,1,3)}$ & $\mathbf{(2,1,3,\frac{1}{3})}$ & $(2,\frac{1}{6},3)$ \\
			&  &  & $(1,1,3,-\frac{2}{3})$ & \\
			\cline{3-5}
			&  & $\mathbf{(1,\overline{3},\overline{3})}$ & $\mathbf{(1,2,\overline{3},-\frac{1}{3})}$ & $(1,-\frac{2}{3},\overline{3})$ \\
			&  &  &  & $(1,\frac{1}{3},\overline{3})$ \\
			&  &  & $(1,1,\overline{3},\frac{2}{3})$ & \\
			\hline
			Scalars & $\mathbf{27}$ & $\mathbf{(\bar{3},3,1)}$ & $\mathbf{(2,2,1,0)}$ & $\mathbf{(2,\pm{\frac{1}{2}},1)}$ \\
			&  &  & $(1,1,1,0)$ &  \\
			&  &  & $(1,2,1,1)$ &  \\
			&  &  & $(2,1,1,-1)$ &  \\
			\cline{3-5}
			&  & $(3,1,3)$ &  &  \\
			&  & $(1,\overline{3},\overline{3})$ &  &  \\
			\cline{2-5}
		    & $\mathbf{351'}$ & $\mathbf{(6,\overline{6},1)}$ & $\mathbf{(1,3,1,-2)}$ & $(1,0,1)_{GB}$ \\
		    &  &  &  & $(1,-1,1)_{GB}$ \\
		    &  &  &  & $(1,-2,1)$ \\
		    &  &  & $\mathbf{(3,1,1,2)_D}$ & $(3,1,1)$ \\
		    &  &  & $(1,1,1,0)$ &  \\
		    &  &  & $(1,2,1,-1)$ &  \\
		    &  &  & $(2,1,1,1)$ &  \\
		    &  &  & $(2,2,1,0)$ &  \\
		    &  &  & $(2,3,1,-1)$ &  \\
		    &  &  & $(3,2,1,1)$ &  \\
		    &  &  & $(3,3,1,0)$ &  \\
		   
		    \cline{3-5}
		    &  & $(\overline{6},1,\overline{6})$ & &  \\
		    &  & $(1,6,6)$ & &  \\
		    &  & $(\overline{3},3,1)$ & &  \\
		    &  & $(3,1,3)$ & &  \\
		    &  & $(1,\overline{3},\overline{3})$ & &  \\
			&  & $(3,8,3)$ & &  \\
		    &  & $(8,\overline{3},\overline{3})$ &  &  \\
		    &  & $(\overline{3},3,8)$ &  &  \\
		    \cline{2-5}
		    & $\mathbf{27}$ & $\mathbf{(\bar{3},3,1)}$ & $(1,1,1,0)$ &  \\
			&  &  & $(1,2,1,1)_{GB}$ &  \\
			&  &  & $(2,1,1,-1)_{GB}$ &  \\
			&  &  & $(2,2,1,0)$ & \\
		    &  & $(3,1,3)$ & &  \\
		    &  & $(1,\overline{3},\overline{3})$ & &  \\
		    \hline
\end{tabular}
\caption{(To be continued)}
	\end{center}
\end{table}
\begin{table}[h!]\ContinuedFloat
	\begin{center}
		
		\begin{tabular}{|c|c|c|c|c|}
			\hline
			& $E(6)$ & $\mathcal{G}_{3_L3_R3_CD}$ & $\mathcal{G}_{2_L2_R3_C 1_{B-L}D}$ & $\mathcal{G}_{2_L1_Y3_C}$ \\
			\hline
			Scalars & $\mathbf{650}$ & $(1,1,1)$ &  &  \\
			&  & $(1,1,1)$ &  &  \\
			
			&  & $(8,1,1)$ &  &  \\
			&  & $(1,8,1)$ &  &  \\
			&  & $(1,1,8)$ &  &  \\
			&  & $(3,3,\overline{3})_{GB}$ &  &  \\
			&  & $(3,3,\overline{3})_{GB}$ &  &  \\
			&  & $(\overline{3},\overline{3},3)_{GB}$ &  &  \\
			&  & $(\overline{3},\overline{3},3)_{GB}$ &  &  \\
			&  & $(6,\overline{3},3)$ &  &  \\
			&  & $(\overline{3},6,3)$ &  &  \\
			&  & $(\overline{6},3,\overline{3})$ &  &  \\
			&  & $(3,\overline{6},\overline{3})$ &  &  \\
			&  & $(\overline{3},\overline{3},\overline{6})$ &  &  \\
			&  & $(3,3,6)$ &  &  \\
			&  & $(8,8,1)$ &  &  \\
			&  & $(8,1,8)$ &  &  \\
			&  & $(1,8,8)$ &  &  \\
			\cline{2-5}
			\hline
		\end{tabular}
		\caption{$E(6)\to \mathcal{G}_{3_L 3_R 3_C D }\to \mathcal{G}_{2_L 2_R 3_C 1_{B-L} D}\to\rm SM $. The bold multiplets contribute to the RGE and others are the heavy degrees of freedom which are integrated out. }\label{E6-G333D-G2231D}		
	\end{center}
	\end{table}

\begin{center}
\underline{Threshold corrections at $M_X$}
\end{center}
\input{"thrcr/thrcr_mx_E6-G333D-G2231D.tex"}
\begin{center}
\underline{Threshold corrections at $M_{I}$}
\end{center}
\input{"thrcr/thrcr_mi_E6-G333D-G2231D.tex"}
\begin{center}
\underline{Threshold corrections at $M_{II}$}
\end{center}
\input{"thrcr/thrcr_mi1_E6-G333D-G2231D.tex"}

\section{Normalisation of the  representations}
\begin{table}[h!]
\begin{center}
\begin{tabular}{|c|c|c|}
	\hline
	Gauge group & Dimension of representation $(R)$ & Normalisation of representation $(\mathrm{Tr} \, t_R^2)$ \\
	\hline
	$SU(2)$ & $2$ & $\frac{1}{2}$ \\
	& $3$ & $2$ \\
	\hline
	$SU(3)$ & $3$ & $\frac{1}{2}$ \\
	& $6$ & $\frac{5}{2}$ \\
	& $8$ & $3$ \\
	& $10$ & $\frac{15}{2}$ \\
	\hline
	$SU(4)$ & $4$ & $\frac{1}{2}$ \\
	& $6$ & $1$ \\
	& $10$ & $3$ \\
	& $15$ & $4$ \\
	& $20$ & $\frac{13}{2}$ \\
	& $20'$ & $8$ \\
	& $20''$ & $\frac{21}{2}$ \\
	\hline
\end{tabular}
\caption{Normalisation of the representations of different $SU(N)$ groups.}\label{index_repn}
\end{center}
\end{table}
\section{GUT normalisation of  the abelian charges}
\begin{table}[h!]
\begin{center}
\begin{tabular}{|c|c|c|}
	\hline
	Breaking pattern & Branching rule & $U(1)$ charge normalization \\
	\hline 
	$SU(2)_R\to U(1)_R$& $2= (\frac{1}{2})\oplus (-\frac{1}{2})$ & $1$ \\
	\hline
	$SU(3)_{L,R}\to SU(2)_{L,R}\otimes U(1)_{L,R}$ & $3 = (1,-\frac{4}{3\sqrt{2}}) \oplus (2,\frac{2}{3\sqrt{2}})$ & $\sqrt{\frac{3}{8}}$ \\
	\hline
	$SU(4)_C\to SU(3)_C\otimes U(1)_{B-L}$ & $4 = (1,-1)\oplus (3, \frac{1}{3})$ & $\sqrt{\frac{3}{8}}$ \\
	\hline
	$E(6)\to SO(10)\otimes U(1)_X$ & $27 = (1,4)\oplus (10,-2)\oplus (16,1)$ & $\frac{1}{2\sqrt{6}}$ \\
	
	\hline
\end{tabular}
\caption{Normalizations of the abelian charges embedded in unified groups. In GUTs the hypercharge $(Y)$  is normalized by $\sqrt{(3/5)}$.}\label{abelian_normn}
\end{center}

\end{table}
\providecommand{\href}[2]{#2}
\addcontentsline*{toc}{section}{}
\bibliographystyle{JHEP}
\bibliography{Threshold-protondecay}

\end{document}

%% file: RGE/E6-G333-G2231D.tex
\begin{align*} 
 M_{II} \ {\rm to} \ M_{I} \ : & \;\;\;b_{2L} =-\frac{7}{3}, \; b_{2R} =-\frac{7}{3}, \; b_{3C} =-7, \; b_{1LR} =7;  \;\;\; b_{ij}= 
\begin{pmatrix}
\frac{80}{3} & 3 & 12 & \frac{27}{2} \\ 
3 & \frac{80}{3} & 12 & \frac{27}{2} \\ 
\frac{9}{2} & \frac{9}{2} & -26 & \frac{1}{2} \\ 
\frac{81}{2} & \frac{81}{2} & 4 & \frac{115}{2}
\end{pmatrix}. \\ 
M_{I} \ {\rm to} \ M_X \ : & \;\;\;b_{3L} =\frac{1}{2}, \; b_{3R} =\frac{1}{2}, \; b_{3C} =-5;  \;\;\; b_{ij}= 
\begin{pmatrix}
253 & 220 & 12 \\ 
220 & 253 & 12 \\ 
12 & 12 & 12
\end{pmatrix}.
 \end{align*}

%% file: RGE/E6-G333-G2231.tex
\begin{align*} 
 M_{II} \ {\rm to} \ M_{I} \ : & \;\;\;b_{2L} =-3, \; b_{2R} =-\frac{7}{3}, \; b_{3C} =-7, \; b_{1LR} =\frac{11}{2};  \;\;\; b_{ij}= 
\begin{pmatrix}
8 & 3 & 12 & \frac{3}{2} \\ 
3 & \frac{80}{3} & 12 & \frac{27}{2} \\ 
\frac{9}{2} & \frac{9}{2} & -26 & \frac{1}{2} \\ 
\frac{9}{2} & \frac{81}{2} & 4 & \frac{61}{2}
\end{pmatrix}. \\ 
M_{I} \ {\rm to} \ M_X \ : & \;\;\;b_{3L} =\frac{1}{2}, \; b_{3R} =\frac{1}{2}, \; b_{3C} =-5;  \;\;\; b_{ij}= 
\begin{pmatrix}
253 & 220 & 12 \\ 
220 & 253 & 12 \\ 
12 & 12 & 12
\end{pmatrix}.
 \end{align*}

%% file: RGE/E6-G2241D-G214.tex
\begin{align*} 
 M_{II} \ {\rm to} \ M_{I} \ : & \;\;\;b_{2L} =-\frac{13}{6}, \; b_{1X} =\frac{31}{6}, \; b_{4C} =-\frac{19}{2};  \;\;\; b_{ij}= 
\begin{pmatrix}
\frac{217}{12} & \frac{4}{3} & \frac{45}{2} \\ 
4 & \frac{35}{8} & \frac{175}{8} \\ 
\frac{9}{2} & \frac{35}{24} & \frac{-355}{8}
\end{pmatrix}. \\ 
M_{I} \ {\rm to} \ M_X \ : & \;\;\;b_{2L} =\frac{1}{3}, \; b_{2R} =\frac{1}{3}, \; b_{4C} =-8, \; b_{1X} =\frac{58}{9};  \;\;\; b_{ij}= 
\begin{pmatrix}
\frac{359}{6} & \frac{15}{2} & \frac{75}{2} & \frac{5}{2} \\ 
\frac{15}{2} & \frac{359}{6} & \frac{75}{2} & \frac{5}{2} \\ 
\frac{15}{2} & \frac{15}{2} & \frac{-7}{2} & \frac{11}{6} \\ 
\frac{15}{2} & \frac{15}{2} & \frac{55}{2} & \frac{91}{18}
\end{pmatrix}.
 \end{align*}

%% file: RGE/E6-G2241-G214.tex
\begin{align*} 
 M_{II} \ {\rm to} \ M_{I} \ : & \;\;\;b_{2L} =-\frac{13}{6}, \; b_{1X} =\frac{31}{6}, \; b_{4C} =-\frac{19}{2};  \;\;\; b_{ij}= 
\begin{pmatrix}
\frac{217}{12} & \frac{4}{3} & \frac{45}{2} \\ 
4 & \frac{35}{8} & \frac{175}{8} \\ 
\frac{9}{2} & \frac{35}{24} & \frac{-355}{8}
\end{pmatrix}. \\ 
M_{I} \ {\rm to} \ M_X \ : & \;\;\;b_{2L} =-1, \; b_{2R} =\frac{1}{3}, \; b_{4C} =-\frac{25}{3}, \; b_{1X} =\frac{19}{3};  \;\;\; b_{ij}= 
\begin{pmatrix}
\frac{65}{2} & \frac{15}{2} & \frac{45}{2} & \frac{13}{6} \\ 
\frac{15}{2} & \frac{359}{6} & \frac{75}{2} & \frac{5}{2} \\ 
\frac{9}{2} & \frac{15}{2} & \frac{-41}{3} & \frac{5}{3} \\ 
\frac{13}{2} & \frac{15}{2} & 25 & 5
\end{pmatrix}.
 \end{align*}

%% file: RGE/SO10-G224D-G2231D.tex
\begin{align*} 
 M_{II} \ {\rm to} \ M_{I} \ : & \;\;\;b_{2L} =-\frac{7}{3}, \; b_{2R} =-\frac{7}{3}, \; b_{3C} =-7, \; b_{1(B-L)} =7;  \;\;\; b_{ij}= 
\begin{pmatrix}
\frac{80}{3} & 3 & 12 & \frac{27}{2} \\ 
3 & \frac{80}{3} & 12 & \frac{27}{2} \\ 
\frac{9}{2} & \frac{9}{2} & -26 & \frac{1}{2} \\ 
\frac{81}{2} & \frac{81}{2} & 4 & \frac{115}{2}
\end{pmatrix}. \\ 
M_{I} \ {\rm to} \ M_X \ : & \;\;\;b_{2L} =\frac{11}{3}, \; b_{2R} =\frac{11}{3}, \; b_{4C} =-4;  \;\;\; b_{ij}= 
\begin{pmatrix}
\frac{584}{3} & 3 & \frac{765}{2} \\ 
3 & \frac{584}{3} & \frac{765}{2} \\ 
\frac{153}{2} & \frac{153}{2} & \frac{661}{2}
\end{pmatrix}.
 \end{align*}

%% file: RGE/SO10-G224D-G2231.tex
\begin{align*} 
 M_{II} \ {\rm to} \ M_{I} \ : & \;\;\;b_{2L} =-3, \; b_{2R} =-\frac{7}{3}, \; b_{3C} =-7, \; b_{1(B-L)} =\frac{11}{2};  \;\;\; b_{ij}= 
\begin{pmatrix}
8 & 3 & 12 & \frac{3}{2} \\ 
3 & \frac{80}{3} & 12 & \frac{27}{2} \\ 
\frac{9}{2} & \frac{9}{2} & -26 & \frac{1}{2} \\ 
\frac{9}{2} & \frac{81}{2} & 4 & \frac{61}{2}
\end{pmatrix}. \\ 
M_{I} \ {\rm to} \ M_X \ : & \;\;\;b_{2L} =\frac{11}{3}, \; b_{2R} =\frac{11}{3}, \; b_{4C} =-4;  \;\;\; b_{ij}= 
\begin{pmatrix}
\frac{584}{3} & 3 & \frac{765}{2} \\ 
3 & \frac{584}{3} & \frac{765}{2} \\ 
\frac{153}{2} & \frac{153}{2} & \frac{661}{2}
\end{pmatrix}.
 \end{align*}

%% file: RGE/SO10-G224D-G214.tex
\begin{align*} 
 M_{II} \ {\rm to} \ M_{I} \ : & \;\;\;b_{2L} =-3, \; b_{1R} =\frac{23}{3}, \; b_{4C} =-\frac{29}{3};  \;\;\; b_{ij}= 
\begin{pmatrix}
8 & 1 & \frac{45}{2} \\ 
3 & 44 & \frac{405}{2} \\ 
\frac{9}{2} & \frac{27}{2} & \frac{-101}{6}
\end{pmatrix}. \\ 
M_{I} \ {\rm to} \ M_X \ : & \;\;\;b_{2L} =4, \; b_{2R} =4, \; b_{4C} =-\frac{14}{3};  \;\;\; b_{ij}= 
\begin{pmatrix}
204 & 3 & \frac{765}{2} \\ 
3 & 204 & \frac{765}{2} \\ 
\frac{153}{2} & \frac{153}{2} & \frac{1759}{6}
\end{pmatrix}.
 \end{align*}

%% file: RGE/SO10-G224-G2231.tex
\begin{align*} 
 M_{II} \ {\rm to} \ M_{I} \ : & \;\;\;b_{2L} =-3, \; b_{2R} =-\frac{7}{3}, \; b_{3C} =-7, \; b_{1(B-L)} =\frac{11}{2};  \;\;\; b_{ij}= 
\begin{pmatrix}
8 & 3 & 12 & \frac{3}{2} \\ 
3 & \frac{80}{3} & 12 & \frac{27}{2} \\ 
\frac{9}{2} & \frac{9}{2} & -26 & \frac{1}{2} \\ 
\frac{9}{2} & \frac{81}{2} & 4 & \frac{61}{2}
\end{pmatrix}. \\ 
M_{I} \ {\rm to} \ M_X \ : & \;\;\;b_{2L} =-3, \; b_{2R} =\frac{11}{3}, \; b_{4C} =-7;  \;\;\; b_{ij}= 
\begin{pmatrix}
8 & 3 & \frac{45}{2} \\ 
3 & \frac{584}{3} & \frac{765}{2} \\ 
\frac{9}{2} & \frac{153}{2} & \frac{289}{2}
\end{pmatrix}.
 \end{align*}

%% file: RGE/SO10-G224-G214.tex
\begin{align*} 
 M_{II} \ {\rm to} \ M_{I} \ : & \;\;\;b_{2L} =-3, \; b_{1R} =\frac{23}{3}, \; b_{4C} =-\frac{29}{3};  \;\;\; b_{ij}= 
\begin{pmatrix}
8 & 1 & \frac{45}{2} \\ 
3 & 44 & \frac{405}{2} \\ 
\frac{9}{2} & \frac{27}{2} & \frac{-101}{6}
\end{pmatrix}. \\ 
M_{I} \ {\rm to} \ M_X \ : & \;\;\;b_{2L} =-3, \; b_{2R} =4, \; b_{4C} =-\frac{23}{3};  \;\;\; b_{ij}= 
\begin{pmatrix}
8 & 3 & \frac{45}{2} \\ 
3 & 204 & \frac{765}{2} \\ 
\frac{9}{2} & \frac{153}{2} & \frac{643}{6}
\end{pmatrix}.
 \end{align*}

%% file: thrcr_I/thrcr_mx_SO10-G224D.tex
\small{\begin{align} 
\Lambda_{2L}(M_X) & = 6 \log \frac{ M_S\left({3, 3, 1}\right)}{M_X}+ 30 \log \frac{ M_S\left({2, 2, 15}\right)}{M_X},\nonumber \\
 \Lambda_{2R}(M_X) & = 6 \log \frac{ M_S\left({3, 3, 1}\right)}{M_X}+ 30 \log \frac{ M_S\left({2, 2, 15}\right)}{M_X},\nonumber \\
 \Lambda_{4C}(M_X) & =\log \frac{ M_S\left({1, 1, 6}\right)}{M_X}+ 8 \log \frac{ M_S\left({1, 1, 20'}\right)}{M_X}+ 32 \log \frac{ M_S\left({2, 2, 15}\right)}{M_X}+ 2 \log \frac{ M_S\left({1, 1, 6}\right)}{M_X}. 
 \end{align}}

%% file: thrcr_I/thrcr_mr_SO10-G224D.tex
\begin{align} 
\Lambda_{2L}(M_I) & = 4 \log \frac{ M_S\left({3, -1, 1}\right)}{M_{I}}+ 12 \log \frac{ M_S\left({3, \frac{-1}{3}, 3}\right)}{M_{I}}+ 24 \log \frac{ M_S\left({3, \frac{1}{3}, 6}\right)}{M_{I}},\nonumber \\
 \Lambda_{1Y}(M_I) & = \frac{3}{5} \left(8 \log \frac{ M_S\left({1, 2, 1}\right)}{M_{I}}+ \frac{32}{3} \log \frac{ M_S\left({1, \frac{4}{3}, \overline{3}}\right)}{M_{I}}+ \frac{16}{3} \log \frac{ M_S\left({1, \frac{2}{3}, \overline{6}}\right)}{M_{I}}+ \frac{2}{3} \log \frac{ M_S\left({1, \frac{1}{3}, \overline{3}}\right)}{M_{I}}\right.\nonumber \\ &+ \frac{4}{3} \log \frac{ M_S\left({1, \frac{-1}{3}, \overline{6}}\right)}{M_{I}}+ \frac{64}{3} \log \frac{ M_S\left({1, \frac{-4}{3}, \overline{6}}\right)}{M_{I}}+ 6 \log \frac{ M_S\left({3, -1, 1}\right)}{M_{I}}+ 2 \log \frac{ M_S\left({3, \frac{-1}{3}, 3}\right)}{M_{I}}\nonumber \\ &+\left. 4 \log \frac{ M_S\left({3, \frac{1}{3}, 6}\right)}{M_{I}}\right),\nonumber \\
 \Lambda_{3C}(M_I) & =\log \frac{ M_S\left({1, \frac{4}{3}, \overline{3}}\right)}{M_{I}}+ 5 \log \frac{ M_S\left({1, \frac{2}{3}, \overline{6}}\right)}{M_{I}}+\log \frac{ M_S\left({1, \frac{1}{3}, \overline{3}}\right)}{M_{I}}+ 5 \log \frac{ M_S\left({1, \frac{-1}{3}, \overline{6}}\right)}{M_{I}}\nonumber \\ &+ 5 \log \frac{ M_S\left({1, \frac{-4}{3}, \overline{6}}\right)}{M_{I}}+ 3 \log \frac{ M_S\left({3, \frac{-1}{3}, 3}\right)}{M_{I}}+ 15 \log \frac{ M_S\left({3, \frac{1}{3}, 6}\right)}{M_{I}}. 
 \end{align} 

%% file: thrcr_I/thrcr_mx_SO10-G224.tex
\small{\begin{align} 
\Lambda_{2L}(M_X) & = 30 \log \frac{ M_S\left({3, 1, 15}\right)}{M_X}+ 30 \log \frac{ M_S\left({2, 2, 15}\right)}{M_X}+ 20 \log \frac{ M_S\left({2, 2, 10}\right)}{M_X}+ 40 \log \frac{ M_S\left({3, 10, 1}\right)}{M_X},\nonumber \\
 \Lambda_{2R}(M_X) & = 30 \log \frac{ M_S\left({1, 3, 15}\right)}{M_X}+ 30 \log \frac{ M_S\left({2, 2, 15}\right)}{M_X}+ 20 \log \frac{ M_S\left({2, 2, 10}\right)}{M_X},\nonumber \\
 \Lambda_{4C}(M_X) & =\log \frac{ M_S\left({1, 1, 6}\right)}{M_X}+ 4 \log \frac{ M_S\left({1, 1, 15}\right)}{M_X}+ 12 \log \frac{ M_S\left({3, 1, 15}\right)}{M_X}+ 12 \log \frac{ M_S\left({1, 3, 15}\right)}{M_X}\nonumber \\ & + 32 \log \frac{ M_S\left({2, 2, 15}\right)}{M_X}+ 2 \log \frac{ M_S\left({1, 1, 6}\right)}{M_X}+ 24 \log \frac{ M_S\left({2, 2, 10}\right)}{M_X}+ 18 \log \frac{ M_S\left({3, 10, 1}\right)}{M_X}. 
 \end{align}}

%% file: thrcr_I/thrcr_mr_SO10-G224.tex
\begin{align} 
\Lambda_{2L}(M_I) & =0,\nonumber \\
 \Lambda_{1Y}(M_I) & = \frac{3}{5}\left( 8 \log \frac{ M_S\left({1, 2, 1}\right)}{M_{I}}+ \frac{32}{3} \log \frac{ M_S\left({1, \frac{4}{3}, \overline{3}}\right)}{M_{I}}+ \frac{16}{3} \log \frac{ M_S\left({1, \frac{2}{3}, \overline{6}}\right)}{M_{I}}+ \frac{2}{3} \log \frac{ M_S\left({1, \frac{1}{3}, \overline{3}}\right)}{M_{I}}\right.\nonumber \\ & \left.+ \frac{4}{3} \log \frac{ M_S\left({1, \frac{-1}{3}, \overline{6}}\right)}{M_{I}}+ \frac{64}{3} \log \frac{ M_S\left({1, \frac{-4}{3}, \overline{6}}\right)}{M_{I}}\right),\nonumber \\
 \Lambda_{3C}(M_I) & =\log \frac{ M_S\left({1, \frac{4}{3}, \overline{3}}\right)}{M_{I}}+ 5 \log \frac{ M_S\left({1, \frac{2}{3}, \overline{6}}\right)}{M_{I}}+\log \frac{ M_S\left({1, \frac{1}{3}, \overline{3}}\right)}{M_{I}}+ 5 \log \frac{ M_S\left({1, \frac{-1}{3}, \overline{6}}\right)}{M_{I}}\nonumber \\ &+ 5 \log \frac{ M_S\left({1, \frac{-4}{3}, \overline{6}}\right)}{M_{I}}. 
 \end{align} 

%% file: thrcr_I/thrcr_mx_SO10-G2231D.tex
\small{\begin{align} 
\Lambda_{2L}(M_X) & = 2 \log \frac{ M_S\left({3, 1, 1, 0}\right)}{M_X}+ 16 \log \frac{ M_S\left({3, 1, 8, 0}\right)}{M_X}+ 12 \log \frac{ M_S\left({3, 1, 3, \frac{-2}{3}}\right)}{M_X}\nonumber \\ & + 24 \log \frac{ M_S\left({3, 1, 6, \frac{2}{3}}\right)}{M_X}+ 2 \log \frac{ M_S\left({2, 2, 1, 0}\right)}{M_X}+ 6 \log \frac{ M_S\left({2, 2, 3, \frac{4}{3}}\right)}{M_X}\nonumber \\ & + 6 \log \frac{ M_S\left({2, 2, 3, \frac{4}{3}}\right)}{M_X}+ 16 \log \frac{ M_S\left({2, 2, 8, 0}\right)}{M_X}+ 12 \log \frac{ M_S\left({3, 1, 3, \frac{4}{3}}\right)}{M_X}\nonumber \\ & + 2 \log \frac{ M_S\left({2, 2, 1, 2}\right)}{M_X}+ 6 \log \frac{ M_S\left({2, 2, 3, \frac{-2}{3}}\right)}{M_X}+ 12 \log \frac{ M_S\left({2, 2, 6, \frac{2}{3}}\right)}{M_X},\nonumber \\
 \Lambda_{2R}(M_X) & = 2 \log \frac{ M_S\left({1, 3, 1, 0}\right)}{M_X}+ 16 \log \frac{ M_S\left({1, 3, 8, 0}\right)}{M_X}+ 12 \log \frac{ M_S\left({1, 3, \overline{3}, \frac{2}{3}}\right)}{M_X}\nonumber \\ & + 24 \log \frac{ M_S\left({1, 3, \overline{6}, \frac{-2}{3}}\right)}{M_X}+ 2 \log \frac{ M_S\left({2, 2, 1, 0}\right)}{M_X}+ 6 \log \frac{ M_S\left({2, 2, 3, \frac{4}{3}}\right)}{M_X}\nonumber \\ & + 6 \log \frac{ M_S\left({2, 2, 3, \frac{4}{3}}\right)}{M_X}+ 16 \log \frac{ M_S\left({2, 2, 8, 0}\right)}{M_X}+ 12 \log \frac{ M_S\left({1, 3, 3, \frac{4}{3}}\right)}{M_X}\nonumber \\ & + 2 \log \frac{ M_S\left({2, 2, 1, 2}\right)}{M_X}+ 6 \log \frac{ M_S\left({2, 2, 3, \frac{-2}{3}}\right)}{M_X}+ 12 \log \frac{ M_S\left({2, 2, 6, \frac{2}{3}}\right)}{M_X},\nonumber \\
 \Lambda_{3C}(M_X) & = 3 \log \frac{ M_S\left({1, 1, 8, 0}\right)}{M_X}+ 9 \log \frac{ M_S\left({3, 1, 8, 0}\right)}{M_X}+ 9 \log \frac{ M_S\left({1, 3, 8, 0}\right)}{M_X}\nonumber \\ & +\log \frac{ M_S\left({1, 1, 3, \frac{-2}{3}}\right)}{M_X}+ 3 \log \frac{ M_S\left({1, 3, \overline{3}, \frac{2}{3}}\right)}{M_X}+ 15 \log \frac{ M_S\left({1, 3, \overline{6}, \frac{-2}{3}}\right)}{M_X}\nonumber \\ & + 3 \log \frac{ M_S\left({3, 1, 3, \frac{-2}{3}}\right)}{M_X}+ 15 \log \frac{ M_S\left({3, 1, 6, \frac{2}{3}}\right)}{M_X}+ 4 \log \frac{ M_S\left({2, 2, 3, \frac{4}{3}}\right)}{M_X}\nonumber \\ & + 4 \log \frac{ M_S\left({2, 2, 3, \frac{4}{3}}\right)}{M_X}+ 24 \log \frac{ M_S\left({2, 2, 8, 0}\right)}{M_X}+\log \frac{ M_S\left({1, 1, \overline{3}, \frac{2}{3}}\right)}{M_X}\nonumber \\ & +\log \frac{ M_S\left({1, 1, \overline{3}, \frac{2}{3}}\right)}{M_X}+ 3 \log \frac{ M_S\left({3, 1, 3, \frac{4}{3}}\right)}{M_X}+ 3 \log \frac{ M_S\left({1, 3, 3, \frac{4}{3}}\right)}{M_X}\nonumber \\ & + 4 \log \frac{ M_S\left({2, 2, 3, \frac{-2}{3}}\right)}{M_X}+ 20 \log \frac{ M_S\left({2, 2, 6, \frac{2}{3}}\right)}{M_X},\nonumber \\
 \Lambda_{1(B-L)}(M_X) & = \frac{3}{8}\left(\frac{8}{3} \log \frac{ M_S\left({1, 1, 3, \frac{-2}{3}}\right)}{M_X}+ 8 \log \frac{ M_S\left({1, 3, \overline{3}, \frac{2}{3}}\right)}{M_X}+ 16 \log \frac{ M_S\left({1, 3, \overline{6}, \frac{-2}{3}}\right)}{M_X}\right.\nonumber \\ & + 8 \log \frac{ M_S\left({3, 1, 3, \frac{-2}{3}}\right)}{M_X}+ 16 \log \frac{ M_S\left({3, 1, 6, \frac{2}{3}}\right)}{M_X}+ \frac{128}{3} \log \frac{ M_S\left({2, 2, 3, \frac{4}{3}}\right)}{M_X}\nonumber \\ & + \frac{128}{3} \log \frac{ M_S\left({2, 2, 3, \frac{4}{3}}\right)}{M_X}+ \frac{8}{3} \log \frac{ M_S\left({1, 1, \overline{3}, \frac{2}{3}}\right)}{M_X}+ \frac{8}{3} \log \frac{ M_S\left({1, 1, \overline{3}, \frac{2}{3}}\right)}{M_X}\nonumber \\ & + 32 \log \frac{ M_S\left({3, 1, 3, \frac{4}{3}}\right)}{M_X}+ 32 \log \frac{ M_S\left({1, 3, 3, \frac{4}{3}}\right)}{M_X}+ 32 \log \frac{ M_S\left({2, 2, 1, 2}\right)}{M_X}\nonumber \\ & \left. + \frac{32}{3} \log \frac{ M_S\left({2, 2, 3, \frac{-2}{3}}\right)}{M_X}+ \frac{64}{3} \log \frac{ M_S\left({2, 2, 6, \frac{2}{3}}\right)}{M_X} \right). 
 \end{align}}

%% file: thrcr_I/thrcr_mr_SO10-G2231D.tex
\small{\begin{align} 
\Lambda_{2L}(M_I) & = 4 \log \frac{ M_S\left({3, -1, 1}\right)}{M_{I}},\nonumber \\
 \Lambda_{1Y}(M_I) & = \frac{3}{5}\left( 8 \log \frac{ M_S\left({1, 2, 1}\right)}{M_{I}}+ 6 \log \frac{ M_S\left({3, -1, 1}\right)}{M_{I}} \right),\nonumber \\
 \Lambda_{3C}(M_I) & =0. 
 \end{align} }

%% file: thrcr_I/thrcr_mx_SO10-G2231.tex
\small{\begin{align} 
\Lambda_{2L}(M_X) & = 2 \log \frac{ M_S\left({3, 1, 1, 0}\right)}{M_X}+ 4 \log \frac{ M_S\left({3, 1, 1, -2}\right)}{M_X}+ 12 \log \frac{ M_S\left({3, 1, 3, \frac{-2}{3}}\right)}{M_X}\nonumber \\ & + 24 \log \frac{ M_S\left({3, 1, 6, \frac{2}{3}}\right)}{M_X}+ 2 \log \frac{ M_S\left({2, 2, 1, 0}\right)}{M_X}+ 6 \log \frac{ M_S\left({2, 2, 3, \frac{4}{3}}\right)}{M_X}\nonumber \\ & + 6 \log \frac{ M_S\left({2, 2, 3, \frac{4}{3}}\right)}{M_X}+ 16 \log \frac{ M_S\left({2, 2, 8, 0}\right)}{M_X},\nonumber \\
 \Lambda_{2R}(M_X) & = 2 \log \frac{ M_S\left({1, 3, 1, 0}\right)}{M_X}+ 12 \log \frac{ M_S\left({1, 3, \overline{3}, \frac{2}{3}}\right)}{M_X}+ 24 \log \frac{ M_S\left({1, 3, \overline{6}, \frac{-2}{3}}\right)}{M_X}\nonumber \\ & + 2 \log \frac{ M_S\left({2, 2, 1, 0}\right)}{M_X}+ 6 \log \frac{ M_S\left({2, 2, 3, \frac{4}{3}}\right)}{M_X}+ 6 \log \frac{ M_S\left({2, 2, 3, \frac{4}{3}}\right)}{M_X}\nonumber \\ & + 16 \log \frac{ M_S\left({2, 2, 8, 0}\right)}{M_X},\nonumber \\
 \Lambda_{3C}(M_X) & = 3 \log \frac{ M_S\left({1, 1, 8, 0}\right)}{M_X}+\log \frac{ M_S\left({1, 1, 3, \frac{-2}{3}}\right)}{M_X}+ 3 \log \frac{ M_S\left({1, 3, \overline{3}, \frac{2}{3}}\right)}{M_X}\nonumber \\ & + 15 \log \frac{ M_S\left({1, 3, \overline{6}, \frac{-2}{3}}\right)}{M_X}+ 3 \log \frac{ M_S\left({3, 1, 3, \frac{-2}{3}}\right)}{M_X}+ 15 \log \frac{ M_S\left({3, 1, 6, \frac{2}{3}}\right)}{M_X}\nonumber \\ & + 4 \log \frac{ M_S\left({2, 2, 3, \frac{4}{3}}\right)}{M_X}+ 4 \log \frac{ M_S\left({2, 2, 3, \frac{4}{3}}\right)}{M_X}+ 24 \log \frac{ M_S\left({2, 2, 8, 0}\right)}{M_X}\nonumber \\ & +\log \frac{ M_S\left({1, 1, \overline{3}, \frac{2}{3}}\right)}{M_X}+\log \frac{ M_S\left({1, 1, \overline{3}, \frac{2}{3}}\right)}{M_X},\nonumber \\
 \Lambda_{1(B-L)}(M_X) & = \frac{3}{8}\left(\frac{8}{3} \log \frac{ M_S\left({1, 1, 3, \frac{-2}{3}}\right)}{M_X}+ 8 \log \frac{ M_S\left({1, 3, \overline{3}, \frac{2}{3}}\right)}{M_X}+ 16 \log \frac{ M_S\left({1, 3, \overline{6}, \frac{-2}{3}}\right)}{M_X} \right. \nonumber \\ & + 24 \log \frac{ M_S\left({3, 1, 1, -2}\right)}{M_X}+ 8 \log \frac{ M_S\left({3, 1, 3, \frac{-2}{3}}\right)}{M_X}+ 16 \log \frac{ M_S\left({3, 1, 6, \frac{2}{3}}\right)}{M_X}\nonumber \\ & + \frac{128}{3} \log \frac{ M_S\left({2, 2, 3, \frac{4}{3}}\right)}{M_X}+ \frac{128}{3} \log \frac{ M_S\left({2, 2, 3, \frac{4}{3}}\right)}{M_X}+ \frac{8}{3} \log \frac{ M_S\left({1, 1, \overline{3}, \frac{2}{3}}\right)}{M_X}\nonumber \\ & \left. + \frac{8}{3} \log \frac{ M_S\left({1, 1, \overline{3}, \frac{2}{3}}\right)}{M_X}\right). 
 \end{align}}

%% file: thrcr_I/thrcr_mr_SO10-G2231.tex
\small{\begin{align} 
\Lambda_{2L}(M_I) & =0,\nonumber \\
 \Lambda_{1Y}(M_I) & = \frac{24}{5} \log \frac{ M_S\left({1, 2, 1}\right)}{M_{I}},\nonumber \\
 \Lambda_{3C}(M_I) & =0. 
 \end{align} }

%% file: thrcr_I/thrcr_mx_E6-G2241D.tex
\small{\begin{align} 
\Lambda_{2L}(M_X) & = 2 \log \frac{ M_S\left({3, 1, 1, 0}\right)}{M_X}+ 6 \log \frac{ M_S\left({2, 2, 6, 0}\right)}{M_X}+ 6 \log \frac{ M_S\left({3, 3, 1, 0}\right)}{M_X}\nonumber \\ & + 30 \log \frac{ M_S\left({3, 1, 15, 0}\right)}{M_X}+ 6 \log \frac{ M_S\left({2, 2, 6, 0}\right)}{M_X}+ 4 \log \frac{ M_S\left({2, 1, 4, 1}\right)}{M_X}\nonumber \\ & +\log \frac{ M_S\left({2, 2, 1, -2}\right)}{M_X}+\log \frac{ M_S\left({2, 2, 1, 6}\right)}{M_X}+ 4 \log \frac{ M_S\left({2, 1, 4, -3}\right)}{M_X}\nonumber \\ & + 12 \log \frac{ M_S\left({2, 3, 4, -3}\right)}{M_X}+ 32 \log \frac{ M_S\left({3, 2, \overline{4}, -3}\right)}{M_X}+ 20 \log \frac{ M_S\left({2, 1, 20, -3}\right)}{M_X}\nonumber \\ & + 20 \log \frac{ M_S\left({2, 2, 10, 0}\right)}{M_X},\nonumber \\
 \Lambda_{2R}(M_X) & = 6 \log \frac{ M_S\left({1, 3, 1, 0}\right)}{M_X}+ 6 \log \frac{ M_S\left({2, 2, 6, 0}\right)}{M_X}+ 6 \log \frac{ M_S\left({3, 3, 1, 0}\right)}{M_X}\nonumber \\ & + 30 \log \frac{ M_S\left({1, 3, 15, 0}\right)}{M_X}+ 6 \log \frac{ M_S\left({2, 2, 6, 0}\right)}{M_X}+ 4 \log \frac{ M_S\left({1, 2, \overline{4}, 1}\right)}{M_X}\nonumber \\ & +\log \frac{ M_S\left({2, 2, 1, -2}\right)}{M_X}+\log \frac{ M_S\left({2, 2, 1, 6}\right)}{M_X}+ 4 \log \frac{ M_S\left({1, 2, \overline{4}, -3}\right)}{M_X}\nonumber \\ & + 32 \log \frac{ M_S\left({2, 3, 4, -3}\right)}{M_X}+ 12 \log \frac{ M_S\left({3, 2, \overline{4}, -3}\right)}{M_X}+ 20 \log \frac{ M_S\left({1, 2, \overline{20}, -3}\right)}{M_X}\nonumber \\ & + 20 \log \frac{ M_S\left({2, 2, 10, 0}\right)}{M_X},\nonumber \\
 \Lambda_{4C}(M_X) & = 4 \log \frac{ M_S\left({1, 1, 15, 0}\right)}{M_X}+ 4 \log \frac{ M_S\left({2, 2, 6, 0}\right)}{M_X}+ 8 \log \frac{ M_S\left({1, 1, 20', 0}\right)}{M_X}\nonumber \\ & + 12 \log \frac{ M_S\left({3, 1, 15, 0}\right)}{M_X}+ 12 \log \frac{ M_S\left({1, 3, 15, 0}\right)}{M_X}+ 4 \log \frac{ M_S\left({2, 2, 6, 0}\right)}{M_X}\nonumber \\ & + 2 \log \frac{ M_S\left({1, 1, 6, -2}\right)}{M_X}+ 2 \log \frac{ M_S\left({2, 1, 4, 1}\right)}{M_X}+ 2 \log \frac{ M_S\left({1, 2, \overline{4}, 1}\right)}{M_X}\nonumber \\ & + 2 \log \frac{ M_S\left({1, 1, 6, -2}\right)}{M_X}+ 2 \log \frac{ M_S\left({1, 1, 6, 6}\right)}{M_X}+ 2 \log \frac{ M_S\left({2, 1, 4, -3}\right)}{M_X}\nonumber \\ & + 2 \log \frac{ M_S\left({1, 2, \overline{4}, -3}\right)}{M_X}+ 6 \log \frac{ M_S\left({2, 3, 4, -3}\right)}{M_X}+ 6 \log \frac{ M_S\left({3, 2, \overline{4}, -3}\right)}{M_X}\nonumber \\ & + 26 \log \frac{ M_S\left({2, 1, 20, -3}\right)}{M_X}+ 26 \log \frac{ M_S\left({1, 2, \overline{20}, -3}\right)}{M_X}+ 24 \log \frac{ M_S\left({2, 2, 10, 0}\right)}{M_X},\nonumber \\
 \Lambda_{1X}(M_X) & = \frac{1}{24}\left( 48 \log \frac{ M_S\left({1, 1, 6, -2}\right)}{M_X}+ 32 \log \frac{ M_S\left({1, 1, 1, 4}\right)}{M_X}+ 16 \log \frac{ M_S\left({2, 1, 4, 1}\right)}{M_X}\right. \nonumber \\ & + 16 \log \frac{ M_S\left({1, 2, \overline{4}, 1}\right)}{M_X}+ 48 \log \frac{ M_S\left({1, 1, 6, -2}\right)}{M_X}+ 32 \log \frac{ M_S\left({1, 1, 1, 4}\right)}{M_X}\nonumber \\ & + 32 \log \frac{ M_S\left({2, 2, 1, -2}\right)}{M_X}+ 288 \log \frac{ M_S\left({2, 2, 1, 6}\right)}{M_X}+ 432 \log \frac{ M_S\left({1, 1, 6, 6}\right)}{M_X}\nonumber \\ & + 48 \log \frac{ M_S\left({2, 1, 4, -3}\right)}{M_X}+ 48 \log \frac{ M_S\left({1, 2, \overline{4}, -3}\right)}{M_X}+ 432 \log \frac{ M_S\left({2, 3, 4, -3}\right)}{M_X}\nonumber \\ & \left. + 432 \log \frac{ M_S\left({3, 2, \overline{4}, -3}\right)}{M_X}+ 720 \log \frac{ M_S\left({2, 1, 20, -3}\right)}{M_X}+ 720 \log \frac{ M_S\left({1, 2, \overline{20}, -3}\right)}{M_X}\right). 
 \end{align}}

%% file: thrcr_I/thrcr_mr_E6-G2241D.tex
\small{\begin{align} 
\Lambda_{2L}(M_I) & =\log \frac{ M_S\left({2, \frac{1}{2}, 1}\right)}{M_{I}}+ 3 \log \frac{ M_S\left({2, \frac{1}{6}, 3}\right)}{M_{I}}+ 4 \log \frac{ M_F\left({2, \frac{1}{2}, 1}\right)}{M_{I}},\nonumber \\
 \Lambda_{1Y}(M_I) & =\frac{3}{5}\left(\log \frac{ M_S\left({2, \frac{1}{2}, 1}\right)}{M_{I}}+ \frac{1}{3} \log \frac{ M_S\left({2, \frac{1}{6}, 3}\right)}{M_{I}}+ \frac{8}{3} \log \frac{ M_F\left({1, \frac{-1}{3}, 3}\right)}{M_{I}}+ 4 \log \frac{ M_F\left({2, \frac{1}{2}, 1}\right)}{M_{I}}\right),\nonumber \\
 \Lambda_{3C}(M_I) & = 2 \log \frac{ M_S\left({2, \frac{1}{6}, 3}\right)}{M_{I}}+ 4 \log \frac{ M_F\left({1, \frac{-1}{3}, 3}\right)}{M_{I}}. 
 \end{align} }

%% file: thrcr_I/thrcr_mx_E6-G2241.tex
\small{\begin{align} 
\Lambda_{2L}(M_X) & = 2 \log \frac{ M_S\left({3, 1, 1, 0}\right)}{M_X}+ 6 \log \frac{ M_S\left({2, 2, 6, 0}\right)}{M_X}+ 6 \log \frac{ M_S\left({3, 3, 1, 0}\right)}{M_X}\nonumber \\ & + 6 \log \frac{ M_S\left({2, 2, 6, 0}\right)}{M_X}+ 30 \log \frac{ M_S\left({3, 1, 15, 0}\right)}{M_X}+ 4 \log \frac{ M_S\left({2, 1, 4, 1}\right)}{M_X}\nonumber \\ & + 4 \log \frac{ M_S\left({2, 1, 4, 1}\right)}{M_X}+\log \frac{ M_S\left({2, 2, 1, -2}\right)}{M_X}+\log \frac{ M_S\left({2, 2, 1, 6}\right)}{M_X}\nonumber \\ & + 4 \log \frac{ M_S\left({2, 1, 4, -3}\right)}{M_X}+ 12 \log \frac{ M_S\left({2, 3, 4, -3}\right)}{M_X}+ 32 \log \frac{ M_S\left({3, 2, \overline{4}, -3}\right)}{M_X}\nonumber \\ & + 20 \log \frac{ M_S\left({2, 1, 20, -3}\right)}{M_X}+ 20 \log \frac{ M_S\left({2, 2, 10, 0}\right)}{M_X},\nonumber \\
 \Lambda_{2R}(M_X) & = 6 \log \frac{ M_S\left({1, 3, 1, 0}\right)}{M_X}+ 6 \log \frac{ M_S\left({2, 2, 6, 0}\right)}{M_X}+ 6 \log \frac{ M_S\left({3, 3, 1, 0}\right)}{M_X}\nonumber \\ & + 6 \log \frac{ M_S\left({2, 2, 6, 0}\right)}{M_X}+ 30 \log \frac{ M_S\left({1, 3, 15, 0}\right)}{M_X}+ 4 \log \frac{ M_S\left({1, 2, \overline{4}, 1}\right)}{M_X}\nonumber \\ & +\log \frac{ M_S\left({2, 2, 1, -2}\right)}{M_X}+\log \frac{ M_S\left({2, 2, 1, 6}\right)}{M_X}+ 4 \log \frac{ M_S\left({1, 2, \overline{4}, -3}\right)}{M_X}\nonumber \\ & + 32 \log \frac{ M_S\left({2, 3, 4, -3}\right)}{M_X}+ 12 \log \frac{ M_S\left({3, 2, \overline{4}, -3}\right)}{M_X}+ 20 \log \frac{ M_S\left({1, 2, \overline{20}, -3}\right)}{M_X}\nonumber \\ & + 20 \log \frac{ M_S\left({2, 2, 10, 0}\right)}{M_X},\nonumber \\
 \Lambda_{4C}(M_X) & = 4 \log \frac{ M_S\left({1, 1, 15, 0}\right)}{M_X}+ 4 \log \frac{ M_S\left({2, 2, 6, 0}\right)}{M_X}+ 8 \log \frac{ M_S\left({1, 1, 20', 0}\right)}{M_X}\nonumber \\ & + 4 \log \frac{ M_S\left({2, 2, 6, 0}\right)}{M_X}+ 12 \log \frac{ M_S\left({3, 1, 15, 0}\right)}{M_X}+ 12 \log \frac{ M_S\left({1, 3, 15, 0}\right)}{M_X}\nonumber \\ & + 2 \log \frac{ M_S\left({1, 1, 6, -2}\right)}{M_X}+ 2 \log \frac{ M_S\left({2, 1, 4, 1}\right)}{M_X}+ 2 \log \frac{ M_S\left({1, 2, \overline{4}, 1}\right)}{M_X}\nonumber \\ & + 2 \log \frac{ M_S\left({1, 1, 6, -2}\right)}{M_X}+ 2 \log \frac{ M_S\left({2, 1, 4, 1}\right)}{M_X}+ 2 \log \frac{ M_S\left({1, 1, 6, 6}\right)}{M_X}\nonumber \\ & + 2 \log \frac{ M_S\left({2, 1, 4, -3}\right)}{M_X}+ 2 \log \frac{ M_S\left({1, 2, \overline{4}, -3}\right)}{M_X}+ 6 \log \frac{ M_S\left({2, 3, 4, -3}\right)}{M_X}\nonumber \\ & + 6 \log \frac{ M_S\left({3, 2, \overline{4}, -3}\right)}{M_X}+ 26 \log \frac{ M_S\left({2, 1, 20, -3}\right)}{M_X}+ 26 \log \frac{ M_S\left({1, 2, \overline{20}, -3}\right)}{M_X}\nonumber \\ & + 24 \log \frac{ M_S\left({2, 2, 10, 0}\right)}{M_X},\nonumber \\
 \Lambda_{1X}(M_X) & = \frac{1}{24}\left(48 \log \frac{ M_S\left({1, 1, 6, -2}\right)}{M_X}+ 32 \log \frac{ M_S\left({1, 1, 1, 4}\right)}{M_X}+ 16 \log \frac{ M_S\left({2, 1, 4, 1}\right)}{M_X}\right. \nonumber \\ & + 16 \log \frac{ M_S\left({1, 2, \overline{4}, 1}\right)}{M_X}+ 48 \log \frac{ M_S\left({1, 1, 6, -2}\right)}{M_X}+ 32 \log \frac{ M_S\left({1, 1, 1, 4}\right)}{M_X}\nonumber \\ & + 16 \log \frac{ M_S\left({2, 1, 4, 1}\right)}{M_X}+ 32 \log \frac{ M_S\left({2, 2, 1, -2}\right)}{M_X}+ 288 \log \frac{ M_S\left({2, 2, 1, 6}\right)}{M_X}\nonumber \\ & + 432 \log \frac{ M_S\left({1, 1, 6, 6}\right)}{M_X}+ 48 \log \frac{ M_S\left({2, 1, 4, -3}\right)}{M_X}+ 48 \log \frac{ M_S\left({1, 2, \overline{4}, -3}\right)}{M_X}\nonumber \\ & + 432 \log \frac{ M_S\left({2, 3, 4, -3}\right)}{M_X}+ 432 \log \frac{ M_S\left({3, 2, \overline{4}, -3}\right)}{M_X}+ 720 \log \frac{ M_S\left({2, 1, 20, -3}\right)}{M_X}\nonumber \\ & \left. + 720 \log \frac{ M_S\left({1, 2, \overline{20}, -3}\right)}{M_X}\right) . 
 \end{align}}

%% file: thrcr_I/thrcr_mr_E6-G2241.tex
\begin{align} 
\Lambda_{2L}(M_I) & = 4 \log \frac{ M_F\left({2, \frac{1}{2}, 1}\right)}{M_{I}},\nonumber \\
 \Lambda_{1Y}(M_I) & =\frac{3}{5}\left( \frac{8}{3} \log \frac{ M_F\left({1, \frac{-1}{3}, 3}\right)}{M_{I}}+ 4 \log \frac{ M_F\left({2, \frac{1}{2}, 1}\right)}{M_{I}}\right),\nonumber \\
 \Lambda_{3C}(M_I) & = 4 \log \frac{ M_F\left({1, \frac{-1}{3}, 3}\right)}{M_{I}}. 
 \end{align} 

%% file: thrcr/thrcr_mx_SO10-G224D-G2231D.tex
\small{\begin{align} 
\Lambda_{2L}(M_X) & = 30 \log \frac{ M_S\left({3, 1, 15}\right)}{M_X}+ 6 \log \frac{ M_S\left({2, 2, 6}\right)}{M_X}+ 6 \log \frac{ M_S\left({3, 3, 1}\right)}{M_X}\nonumber \\ & + 30 \log \frac{ M_S\left({2, 2, 15}\right)}{M_X}+ 20 \log \frac{ M_S\left({2, 2, 10}\right)}{M_X},\nonumber \\
 \Lambda_{2R}(M_X) & = 30 \log \frac{ M_S\left({1, 3, 15}\right)}{M_X}+ 6 \log \frac{ M_S\left({2, 2, 6}\right)}{M_X}+ 6 \log \frac{ M_S\left({3, 3, 1}\right)}{M_X}\nonumber \\ & + 30 \log \frac{ M_S\left({2, 2, 15}\right)}{M_X}+ 20 \log \frac{ M_S\left({2, 2, 10}\right)}{M_X},\nonumber \\
 \Lambda_{4C}(M_X) & =\log \frac{ M_S\left({1, 1, 6}\right)}{M_X}+ 12 \log \frac{ M_S\left({3, 1, 15}\right)}{M_X}+ 12 \log \frac{ M_S\left({1, 3, 15}\right)}{M_X}\nonumber \\ & + 4 \log \frac{ M_S\left({2, 2, 6}\right)}{M_X}+ 32 \log \frac{ M_S\left({2, 2, 15}\right)}{M_X}+ 2 \log \frac{ M_S\left({1, 1, 6}\right)}{M_X}\nonumber \\ & + 24 \log \frac{ M_S\left({2, 2, 10}\right)}{M_X}. 
 \end{align}}

%% file: thrcr/thrcr_mi_SO10-G224D-G2231D.tex
\begin{align} 
\Lambda_{2L}(M_{I}) & = 12 \log \frac{ M_S\left({3, 1, 3, \frac{-2}{3}}\right)}{M_{I}}+ 24 \log \frac{ M_S\left({3, 1, 6, \frac{2}{3}}\right)}{M_{I}},\nonumber \\
 \Lambda_{2R}(M_{I}) & =\log \frac{ M_S\left({1, 1, 8, 0}\right)}{M_{I}}+ 12 \log \frac{ M_S\left({1, 3, \overline{3}, \frac{2}{3}}\right)}{M_{I}}+ 24 \log \frac{ M_S\left({1, 3, \overline{6}, \frac{-2}{3}}\right)}{M_{I}},\nonumber \\
 \Lambda_{3C}(M_{I}) & = 3 \log \frac{ M_S\left({1, 1, 8, 0}\right)}{M_{I}}+ 3 \log \frac{ M_S\left({1, 3, \overline{3}, \frac{2}{3}}\right)}{M_{I}}+ 15 \log \frac{ M_S\left({1, 3, \overline{6}, \frac{-2}{3}}\right)}{M_{I}}\nonumber \\ & + 3 \log \frac{ M_S\left({3, 1, 3, \frac{-2}{3}}\right)}{M_{I}}+ 15 \log \frac{ M_S\left({3, 1, 6, \frac{2}{3}}\right)}{M_{I}},\nonumber \\
 \Lambda_{1(B-L)}(M_{I}) & =\frac{3}{8}\left( 8 \log \frac{ M_S\left({1, 3, \overline{3}, \frac{2}{3}}\right)}{M_{I}}+ 16 \log \frac{ M_S\left({1, 3, \overline{6}, \frac{-2}{3}}\right)}{M_{I}}+ 8 \log \frac{ M_S\left({3, 1, 3, \frac{-2}{3}}\right)}{M_{I}}\right.\nonumber \\ & \left. + 16 \log \frac{ M_S\left({3, 1, 6, \frac{2}{3}}\right)}{M_{I}}\right). 
 \end{align} 

%% file: thrcr/thrcr_mi1_SO10-G224D-G2231D.tex
\begin{align} 
\Lambda_{2L}(M_{II}) & = 4 \log \frac{ M_S\left({3, -1, 1}\right)}{M_{II}},\nonumber \\
 \Lambda_{1Y}(M_{II}) & = \frac{3}{5}\left( 8 \log \frac{ M_S\left({1, 2, 1}\right)}{M_{II}}+ 6 \log \frac{ M_S\left({3, -1, 1}\right)}{M_{II}}\right),\nonumber \\
 \Lambda_{3C}(M_{II}) & =0. 
 \end{align} 

%% file: thrcr/thrcr_mx_SO10-G224D-G214.tex
\small{\begin{align} 
\Lambda_{2L}(M_X) & = 6 \log \frac{ M_{S}\left({2, 2, 6}\right)}{M_X}+ 6 \log \frac{ M_{S}\left({3, 3, 1}\right)}{M_X}+ 30 \log \frac{ M_{S}\left({2, 2, 15}\right)}{M_X},\nonumber \\
 \Lambda_{2R}(M_X) & = 6 \log \frac{ M_{S}\left({2, 2, 6}\right)}{M_X}+ 6 \log \frac{ M_{S}\left({3, 3, 1}\right)}{M_X}+ 30 \log \frac{ M_{S}\left({2, 2, 15}\right)}{M_X},\nonumber \\
 \Lambda_{4C}(M_X) & =\log \frac{ M_{S}\left({1, 1, 6}\right)}{M_X}+ 4 \log \frac{ M_{S}\left({1, 1, 15}\right)}{M_X}+ 4 \log \frac{ M_{S}\left({2, 2, 6}\right)}{M_X}\nonumber \\ & + 8 \log \frac{ M_{S}\left({1, 1, 20'}\right)}{M_X}+ 32 \log \frac{ M_{S}\left({2, 2, 15}\right)}{M_X}+ 2 \log \frac{ M_{S}\left({1, 1, 6}\right)}{M_X}. 
 \end{align}}

%% file: thrcr/thrcr_mi_SO10-G224D-G214.tex
\begin{align} 
\Lambda_{2L}(M_{I}) & = 2 \log \frac{ M_{S}\left({3, 0, 1}\right)}{M_{I}}+ 40 \log \frac{ M_{S}\left({3, 0, 10}\right)}{M_{I}},\nonumber \\
 \Lambda_{1R}(M_{I}) & = 20 \log \frac{ M_{S}\left({1, 1, \overline{10}}\right)}{M_{I}},\nonumber \\
 \Lambda_{4C}(M_{I}) & = 6 \log \frac{ M_{S}\left({1, 1, \overline{10}}\right)}{M_{I}}+ 6 \log \frac{ M_{S}\left({1, 0, \overline{10}}\right)}{M_{I}}+ 18 \log \frac{ M_{S}\left({3, 0, 10}\right)}{M_{I}}. 
 \end{align} 

%% file: thrcr/thrcr_mi1_SO10-G224D-G214.tex
\begin{align} 
\Lambda_{2L}(M_{II}) & =0,\nonumber \\
 \Lambda_{1Y}(M_{II}) & = \frac{64}{5} \log \frac{ M_{S}\left({1, \frac{-4}{3}, \overline{6}}\right)}{M_{II}},\nonumber \\
 \Lambda_{3C}(M_{II}) & = 5 \log \frac{ M_{S}\left({1, \frac{-4}{3}, \overline{6}}\right)}{M_{II}}. 
 \end{align} 

%% file: thrcr/thrcr_mx_SO10-G224-G214.tex
\small{\begin{align} 
\Lambda_{2L}(M_X) & = 6 \log \frac{ M_S\left({2, 2, 6}\right)}{M_X}+ 2 \log \frac{ M_S\left({3, 1, 1}\right)}{M_X}+ 30 \log \frac{ M_S\left({3, 1, 15}\right)}{M_X}\nonumber \\ & + 30 \log \frac{ M_S\left({2, 2, 15}\right)}{M_X}+ 40 \log \frac{ M_S\left({3, 1, 10}\right)}{M_X}+ 20 \log \frac{ M_S\left({2, 2, 10}\right)}{M_X},\nonumber \\
 \Lambda_{2R}(M_X) & = 6 \log \frac{ M_S\left({2, 2, 6}\right)}{M_X}+ 30 \log \frac{ M_S\left({1, 3, 15}\right)}{M_X}+ 30 \log \frac{ M_S\left({2, 2, 15}\right)}{M_X}\nonumber \\ & + 20 \log \frac{ M_S\left({2, 2, 10}\right)}{M_X},\nonumber \\
 \Lambda_{4C}(M_X) & =\log \frac{ M_S\left({1, 1, 6}\right)}{M_X}+ 4 \log \frac{ M_S\left({1, 1, 15}\right)}{M_X}+ 4 \log \frac{ M_S\left({2, 2, 6}\right)}{M_X}\nonumber \\ & + 4 \log \frac{ M_S\left({1, 1, 15}\right)}{M_X}+ 12 \log \frac{ M_S\left({3, 1, 15}\right)}{M_X}+ 12 \log \frac{ M_S\left({1, 3, 15}\right)}{M_X}\nonumber \\ & + 32 \log \frac{ M_S\left({2, 2, 15}\right)}{M_X}+ 2 \log \frac{ M_S\left({1, 1, 6}\right)}{M_X}+ 18 \log \frac{ M_S\left({3, 1, 10}\right)}{M_X}\nonumber \\ & + 24 \log \frac{ M_S\left({2, 2, 10}\right)}{M_X}. 
 \end{align}}

%% file: thrcr/thrcr_mi_SO10-G224-G214.tex
\begin{align} 
\Lambda_{2L}(M_{I}) & =0,\nonumber \\
 \Lambda_{1R}(M_{I}) & = 20 \log \frac{ M_S\left({1, 1, \overline{10}}\right)}{M_{I}},\nonumber \\
 \Lambda_{4C}(M_{I}) & = 6 \log \frac{ M_S\left({1, 1, \overline{10}}\right)}{M_{I}}+ 6 \log \frac{ M_S\left({1, 0, \overline{10}}\right)}{M_{I}}. 
 \end{align} 

%% file: thrcr/thrcr_mi1_SO10-G224-G214.tex
\begin{align} 
\Lambda_{2L}(M_{II}) & =0,\nonumber \\
 \Lambda_{1Y}(M_{II}) & = \frac{64}{5} \log \frac{ M_S\left({1, \frac{-4}{3}, \overline{6}}\right)}{M_{II}},\nonumber \\
 \Lambda_{3C}(M_{II}) & = 5 \log \frac{ M_S\left({1, \frac{-4}{3}, \overline{6}}\right)}{M_{II}}. 
 \end{align} 

%% file: thrcr/thrcr_mx_E6-G333D-G2231D.tex
\small{\begin{align} 
\Lambda_{3L}(M_X) & =3 \log \frac{ M_S\left({8, 1, 1}\right)}{M_X}+24 \log \frac{ M_S\left({8, 8, 1}\right)}{M_X}+24 \log \frac{ M_S\left({8, 1, 8}\right)}{M_X}\nonumber \\ &+3 \log \frac{ M_S\left({3, 1, 3}\right)}{M_X}+3 \log \frac{ M_S\left({3, 1, 3}\right)}{M_X}+24 \log \frac{ M_S\left({3, 8, 3}\right)}{M_X}\nonumber \\ &+54 \log \frac{ M_S\left({8, \overline{3}, \overline{3}}\right)}{M_X}+30 \log \frac{ M_S\left({\overline{6}, 1, \overline{6}}\right)}{M_X}+24 \log \frac{ M_S\left({\overline{3}, 3, 8}\right)}{M_X}\nonumber \\ &+3 \log \frac{ M_S\left({3, 1, 3}\right)}{M_X}+45 \log \frac{ M_S\left({6, \overline{3}, 3}\right)}{M_X}+18 \log \frac{ M_S\left({\overline{3}, 6, 3}\right)}{M_X}\nonumber \\ &+18 \log \frac{ M_S\left({3, 3, 6}\right)}{M_X},\nonumber \\
 \Lambda_{3R}(M_X) & =3 \log \frac{ M_S\left({1, 8, 1}\right)}{M_X}+24 \log \frac{ M_S\left({8, 8, 1}\right)}{M_X}+24 \log \frac{ M_S\left({1, 8, 8}\right)}{M_X}\nonumber \\ &+3 \log \frac{ M_S\left({1, \overline{3}, \overline{3}}\right)}{M_X}+54 \log \frac{ M_S\left({3, 8, 3}\right)}{M_X}+3 \log \frac{ M_S\left({1, \overline{3}, \overline{3}}\right)}{M_X}\nonumber \\ &+24 \log \frac{ M_S\left({8, \overline{3}, \overline{3}}\right)}{M_X}+30 \log \frac{ M_S\left({1, 6, 6}\right)}{M_X}+24 \log \frac{ M_S\left({\overline{3}, 3, 8}\right)}{M_X}\nonumber \\ &+3 \log \frac{ M_S\left({1, \overline{3}, \overline{3}}\right)}{M_X}+18 \log \frac{ M_S\left({6, \overline{3}, 3}\right)}{M_X}+45 \log \frac{ M_S\left({\overline{3}, 6, 3}\right)}{M_X}\nonumber \\ &+18 \log \frac{ M_S\left({3, 3, 6}\right)}{M_X},\nonumber \\
 \Lambda_{3C}(M_X) & =3 \log \frac{ M_S\left({1, 1, 8}\right)}{M_X}+24 \log \frac{ M_S\left({8, 1, 8}\right)}{M_X}+24 \log \frac{ M_S\left({1, 8, 8}\right)}{M_X}\nonumber \\ &+3 \log \frac{ M_S\left({3, 1, 3}\right)}{M_X}+3 \log \frac{ M_S\left({1, \overline{3}, \overline{3}}\right)}{M_X}+3 \log \frac{ M_S\left({3, 1, 3}\right)}{M_X}\nonumber \\ &+24 \log \frac{ M_S\left({3, 8, 3}\right)}{M_X}+3 \log \frac{ M_S\left({1, \overline{3}, \overline{3}}\right)}{M_X}+24 \log \frac{ M_S\left({8, \overline{3}, \overline{3}}\right)}{M_X}\nonumber \\ &+30 \log \frac{ M_S\left({\overline{6}, 1, \overline{6}}\right)}{M_X}+30 \log \frac{ M_S\left({1, 6, 6}\right)}{M_X}+54 \log \frac{ M_S\left({\overline{3}, 3, 8}\right)}{M_X}\nonumber \\ &+3 \log \frac{ M_S\left({3, 1, 3}\right)}{M_X}+3 \log \frac{ M_S\left({1, \overline{3}, \overline{3}}\right)}{M_X}+18 \log \frac{ M_S\left({6, \overline{3}, 3}\right)}{M_X}\nonumber \\ &+18 \log \frac{ M_S\left({\overline{3}, 6, 3}\right)}{M_X}+45 \log \frac{ M_S\left({3, 3, 6}\right)}{M_X}. 
 \end{align}}

%% file: thrcr/thrcr_mi_E6-G333D-G2231D.tex
\small{\begin{align} 
\Lambda_{2L}(M_{I}) & =1 \log \frac{ M_S\left({2, 1, 1, 1}\right)}{M_{I}}+1 \log \frac{ M_S\left({2, 1, 1, 1}\right)}{M_{I}}+2 \log \frac{ M_S\left({2, 2, 1, 0}\right)}{M_{I}}\nonumber \\ &+3 \log \frac{ M_S\left({2, 3, 1, -1}\right)}{M_{I}}+4 \log \frac{ M_S\left({3, 1, 1, 2}\right)}{M_{I}}+12 \log \frac{ M_S\left({3, 3, 1, 0}\right)}{M_{I}}\nonumber \\ &+2 \log \frac{ M_S\left({2, 2, 1, 0}\right)}{M_{I}}+4 \log \frac{ M_F\left({2, 2, 1, 0}\right)}{M_{I}},\nonumber \\
 \Lambda_{2R}(M_{I}) & =1 \log \frac{ M_S\left({1, 2, 1, -1}\right)}{M_{I}}+1 \log \frac{ M_S\left({1, 2, 1, -1}\right)}{M_{I}}+2 \log \frac{ M_S\left({2, 2, 1, 0}\right)}{M_{I}}\nonumber \\ &+8 \log \frac{ M_S\left({2, 3, 1, -1}\right)}{M_{I}}+12 \log \frac{ M_S\left({3, 3, 1, 0}\right)}{M_{I}}+2 \log \frac{ M_S\left({2, 2, 1, 0}\right)}{M_{I}}\nonumber \\ &+4 \log \frac{ M_F\left({2, 2, 1, 0}\right)}{M_{I}},\nonumber \\
 \Lambda_{3C}(M_{I}) & =4 \log \frac{ M_F\left({1, 1, 3, \frac{-2}{3}}\right)}{M_{I}},\nonumber \\
 \Lambda_{1LR}(M_{I}) & =\frac{3}{8}\left( 4 \log \frac{ M_S\left({2, 1, 1, 1}\right)}{M_{I}}+4 \log \frac{ M_S\left({1, 2, 1, -1}\right)}{M_{I}}+4 \log \frac{ M_S\left({2, 1, 1, 1}\right)}{M_{I}} \right. \nonumber \\ &  +4 \log \frac{ M_S\left({1, 2, 1, -1}\right)}{M_{I}}+12 \log \frac{ M_S\left({2, 3, 1, -1}\right)}{M_{I}}+24 \log \frac{ M_S\left({3, 1, 1, 2}\right)}{M_{I}}\nonumber \\ & \left. +\frac{32}{3} \log \frac{ M_F\left({1, 1, 3, \frac{-2}{3}}\right)}{M_{I}}\right). 
 \end{align}}

%% file: thrcr/thrcr_mi1_E6-G333D-G2231D.tex
\begin{align} 
\Lambda_{2L}(M_{II}) & =4 \log \frac{ M_S\left({3, 1, 1}\right)}{M_{II}},\nonumber \\
 \Lambda_{1Y}(M_{II}) & =\frac{3}{5}\left( 8 \log \frac{ M_S\left({1, -2, 1}\right)}{M_{II}}+6 \log \frac{ M_S\left({3, 1, 1}\right)}{M_{II}}\right),\nonumber \\
 \Lambda_{3C}(M_{II}) & =0. 
 \end{align} 